\documentclass[aps,prb,twocolumn,shortbibliography,superscriptaddress,article]{revtex4-1}
\usepackage{epsfig}
\usepackage{epstopdf}
\usepackage{amsmath}
\usepackage{amsfonts}
\usepackage{amssymb}

\usepackage{hyperref}
\usepackage{bm}
\usepackage{makecell}
\usepackage{rotating}
\usepackage{hyperref}
\usepackage{multirow}
\usepackage{graphicx}
\usepackage{array} 
\usepackage{tabularx} % make sure this is in your preamble
\newcolumntype{Y}{>{\centering\arraybackslash}X}

\usepackage{chemformula} 
\usepackage[T1]{fontenc} 
\usepackage[percent]{overpic}
\usepackage{academicons} 
\usepackage{xcolor}
\usepackage{ragged2e}

\usepackage{graphicx}% Include figure files
\usepackage{dcolumn}% Align table columns on the decimal point
\usepackage{bm}% bold math
\usepackage{color}
\usepackage{comment}

\usepackage{tikz,xcolor,hyperref}
\DeclareMathAlphabet\mathbfcal{OMS}{cmsy}{b}{n}

\usepackage{physics}

\definecolor{lime}{HTML}{A6CE39}
\DeclareRobustCommand{\orcidicon}{%
	\begin{tikzpicture}
	\draw[lime, fill=lime] (0,0)
	circle [radius=0.16]
	node[white] {{\fontfamily{qag}\selectfont \tiny ID}};
	\draw[white, fill=white] (-0.0625,0.095)
	circle [radius=0.007];
	\end{tikzpicture}
	\hspace{-2mm}
}

\foreach \x in {A, ..., Z}{%
	\expandafter\xdef\csname orcid\x\endcsname{\noexpand\href{https://orcid.org/\csname orcidauthor\x\endcsname}{\noexpand\orcidicon}}
}

\begin{document}

%\title{How the Relativistic Spin-Momentum Locking \\ influences the spin photocurrents in altermagnets}
%\title{Relativistic and Non-Relativistic Contributions to Spin-Momentum Locking and Nonlinear Spin Photocurrents in an Altermagnetic Multiferroic}
\title{Nonrelativistic and Relativistic Contributions to Spin-Momentum Locking and Spin Photogalvanic Effect in an Altermagnetic Multiferroic}

\author{Giuseppe Cuono\orcidA}
\email{giuseppe.cuono@unimib.it}
\affiliation{Department of Materials Science, University of Milan-Bicocca, Via Roberto Cozzi 55, 20125 Milan, Italy}

\author{Subhadeep Bandyopadhyay\orcidB}
\affiliation{Consiglio Nazionale delle Ricerche (CNR-SPIN), Unit\'a di Ricerca presso Terzi c/o Universit\'a “G. D’Annunzio”, 66100 Chieti, Italy}

\author{Amar Fakhredine\orcidC}
\affiliation{Institute of Physics, Polish Academy of Sciences, Aleja Lotnik\'ow 32/46, 02668 Warsaw, Poland}

\author{Mathews Benny\orcidD}
\affiliation{International Research Centre Magtop, Institute of Physics, Polish Academy of Sciences, Aleja Lotnik\'ow 32/46, PL-02668 Warsaw, Poland}

\author{Xujia Gong\orcidX}
\affiliation{International Research Centre Magtop, Institute of Physics, Polish Academy of Sciences, Aleja Lotnik\'ow 32/46, PL-02668 Warsaw, Poland}

\author{Paolo Barone\orcidE}
\affiliation{Consiglio Nazionale delle Ricerche CNR-SPIN, Area della Ricerca di Tor Vergata, Via del Fosso del Cavaliere, 100, I-00133 Rome, Italy}

\author{Andrea Droghetti\orcidF}
\email{andrea.droghetti@unive.it}
\affiliation{Department of Molecular Sciences and Nanosystems, Ca’ Foscari University of Venice, via Torino 155, 30170, Venice-Mestre, Italy}

\author{Carmine Autieri\orcidG}
\email{autieri@magtop.ifpan.edu.pl}
\affiliation{International Research Centre Magtop, Institute of Physics, Polish Academy of Sciences,
Aleja Lotnik\'ow 32/46, PL-02668 Warsaw, Poland}

\author{Silvia Picozzi\orcidH}
\affiliation{Department of Materials Science, University of Milan-Bicocca, Via Roberto Cozzi 55, 20125 Milan, Italy}
\affiliation{Consiglio Nazionale delle Ricerche (CNR-SPIN), Unit\'a di Ricerca presso Terzi c/o Universit\'a “G. D’Annunzio”, 66100 Chieti, Italy}

\date{\today}
\begin{abstract}
Altermagnetic multiferroics provide a platform where non-relativistic spin-momentum locking (SML) coexists with relativistic spin textures, yet disentangling their different origins remains challenging. Here, using first-principles density-functional-theory calculations and symmetry analysis of the multipolar expansion of the spin polarization, we investigate the altermagnetic multiferroic BaCuF$_4$. Its dominant non-relativistic altermagnetic $d$-wave SML originates primarily from a CuF$_6$ octahedral rotation coupled to the polar distortion and can therefore be reversed by ferroelectric switching. We show that spin-orbit coupling generates additional spin multipoles and, depending on the Néel-vector orientation, induces spin canting and weak ferromagnetism. By exploiting the group–subgroup relations between spin groups and magnetic space groups, we then distinguish non-relativistic altermagnetic, canting-associated, and purely relativistic contributions to the SML. We further show that this hierarchy is encoded in the nonlinear spin-photogalvanic response, whose spin projection selectively probes contributions of different origins. These results establish a direct connection between the microscopic spin texture and nonlinear spin transport in altermagnetic multiferroics.
\end{abstract}


\maketitle

\section{Introduction}

Altermagnets are a recently identified class of compensated collinear magnets in which the interplay between crystal and magnetic symmetries generates a momentum-dependent spin polarization of the electronic states, known as spin-momentum locking (SML) \cite{Smejkal22,hayami2019momentum,yuan2023degeneracy,Smejkal22beyond}. In contrast to conventional spin textures, which arise from spin-orbit coupling (SOC), altermagnetic SML is purely nonrelativistic in origin. In the presence of SOC, however, additional symmetry-allowed components of the spin polarization can emerge \cite{Fakhredine25a,Fakhredine25b,Fernandes2024,PhysRevB.110.144412}, giving rise to more complex spin textures that can be naturally described within the framework of spin multipoles \cite{hirakida2025multipoleanalysisspincurrents,Hayami2024}.

Altermagnetic multiferroics \cite{SunNatureMater26,smejkal2024altermagneticmultiferroicsaltermagnetoelectriceffect,Gu25,Bezzerga25,D4MH01619J,Khan25b,Sun25Adv,Guo25altII,Cao24,Urru25}, in which altermagnetism coexists with ferroelectricity, constitute a particularly interesting class of emerging materials in this context. The ferroelectric polarization originates from a polar distortion that breaks inversion symmetry, creating a noncentrosymmetric environment. In the presence of SOC, this can give rise to Rashba or Weyl-like spin splitting \cite{Fakhredine25b}, and even persistent spin textures \cite{tenzin2025persistentspintexturesaltermagnetism}, coexisting with the nonrelativistic altermagnetic SML of the dominant spin component. SOC may also induce weak deviations from the collinear magnetic state, leading to weak ferromagnetism\cite{autieri2024staggereddzyaloshinskiimoriyainducingweak}. Although these relativistic effects have been shown to be significant in several altermagnetic multiferroics \cite{Fakhredine25b,tenzin2025persistentspintexturesaltermagnetism}, a quantitative understanding of their microscopic origins and relative contributions to the spin texture remains limited.

Probing these distinct contributions requires response functions that are sensitive to both inversion-symmetry breaking and the spin-dependent electronic structure. Nonlinear photocurrents have emerged as powerful probes in ferroelectrics and multiferroics \cite{Dai23,Zhao26,Sipe00,Dai21,Azpiroz18,Young12,Young12b,Cuono25SbSI,Roy26,stavric2025,Tiwari22,Xiao22}, topological materials \cite{Ma23,Puente23,Rees2020}, and, more recently, systems with unconventional magnetism, including altermagnets \cite{Yoshida26,Yang26,Ezawa25altermagnets,Ezawa25pwavemagnets,Jiang25,Dong25,Yang25,Blatter26,Sivianes25,Song2025Nature,Li26Altermagnets,Cuono26}. Among these responses, the spin photogalvanic effect -- the generation of a dc spin current by light --directly probes the spin-dependent electronic structure \cite{Lihm22,li2026purespinphotocurrentaltermagnetic,Xiao23spin,Xu21,Young13,Song21,Jiang25,Dong25,Yang25,Fei21,Ivchenko08,Zhou07,Bhat05}. However, how the different microscopic contributions to SML are encoded in the spin photogalvanic response of altermagnetic multiferroics has yet to be established.

Here, we address these issues in the prototypical altermagnetic multiferroic BaCuF$_4$ by combining density functional theory (DFT) calculations and symmetry analysis. We identify three distinct contributions to the SML, classified in terms of spin multipoles: an even-wave, nonrelativistic altermagnetic SML of the dominant spin component; a relativistic even-wave SML of the subdominant spin components; and an odd-wave SML that, in the present case, takes the form of Rashba-like contributions. For specific orientations of the N\'eel vector, SOC also induces spin canting and a weak-ferromagnetic moment, which further modify the components of the spin texture. We then establish that the spin photogalvanic tensors reflect the distinct microscopic contributions to SML through their characteristic spin-polarization directions and symmetry properties.

In particular, spin currents polarized along the N\'eel vector are associated with the nonrelativistic altermagnetic quadrupole, those polarized along the weak-ferromagnetic moment with the canting-induced monopole, and those polarized perpendicular to both with relativistic SOC multipoles. Furthermore, the calculated magnitudes reveal a corresponding hierarchy: the non-relativistic contributions produce the largest photocurrents, followed by those associated with the additional spin polarization emerging in the weak-ferromagnetic state, whereas the purely SOC-induced contributions are generally substantially smaller.

The paper is organized as follows. Section II describes the computational methods. Section III presents the DFT and symmetry analysis of the nonrelativistic and relativistic contributions to SML, together with the resulting spin photocurrents. Finally, Section IV summarizes the main conclusions.

\section{Computational details}

The DFT calculations were performed using the projector-augmented wave (PAW) method, as implemented in the \textsc{VASP} code~\cite{Kresse93,Kresse96,Kresse96b}. The exchange-correlation functional was described within the generalized gradient approximation (GGA) using the Perdew--Burke--Ernzerhof functional for solids (PBEsol)~\cite{Perdew08,Perdew96}.

The ground-state $Cmc2_1$ structure and the hypothetical nonpolar $Cmcm$ reference structure~\cite{DANCE_ACuF4,Claude_2006_BaMF4} were optimized by relaxing both the ionic positions and lattice parameters until the Hellmann--Feynman forces on all atoms were below 0.005~eV/\AA. The structural optimizations were performed assuming the A-AFM ordering of the Cu magnetic moments~\cite{Garcia18}, which is experimentally observed at high temperature~\cite{DANCE_ACuF4}. Symmetry-adapted mode (SAM) analysis was performed using the ISODISTORT software~\cite{Isodistort,Isodistort1}.

To account for the localized nature of the Cu $d$ electrons, the DFT+$U$ correction was applied within the Liechtenstein \cite{Liechtenstein95} formalism, using on-site Coulomb and exchange parameters of $U=7.0$~eV and $J=0.9$~eV, respectively. These values were chosen consistently with previous work~\cite{Garcia18}, where they provide an accurate description of the electronic structure. A plane-wave kinetic-energy cutoff of 500~eV was adopted, and the Brillouin zone was sampled using an $8\times4\times8$ Monkhorst--Pack $k$-point mesh for band structure calculations and $22\times7\times1$ for constructing the two-dimensional (2D) energy surface with $k_x = 0$.

The second-order photocurrent responses were evaluated following the methodology of Refs.~\cite{Azpiroz18,Lihm22,Puente23}, based on interpolation of the DFT band structure using maximally localized Wannier functions (MLWFs)~\cite{Marzari97,Mostofi08}, as implemented in the \textsc{Wannier90} package~\cite{Pizzi2020}. Cu $d$ and F $p$ orbitals were included in the Wannierization, resulting in 136 spinor bands in the presence of SOC. The Wannier-interpolated band structures are reported in the Supplemental Material (SM). The photoresponses were evaluated on a dense $100\times100\times100$ $k$-point grid, for which the photoconductivities are well converged, with no appreciable changes upon further increasing the grid density.

We use the conventional spin-current definition based on the symmetrized product of the spin and velocity operators, following Ref.~\cite{Lihm22}. The calculated response is rescaled by $2e/\hbar$ and reported in charge-current-equivalent units.

%section II
\section{Results}

\begin{figure}[t!]
\centering
\includegraphics[width=\columnwidth]{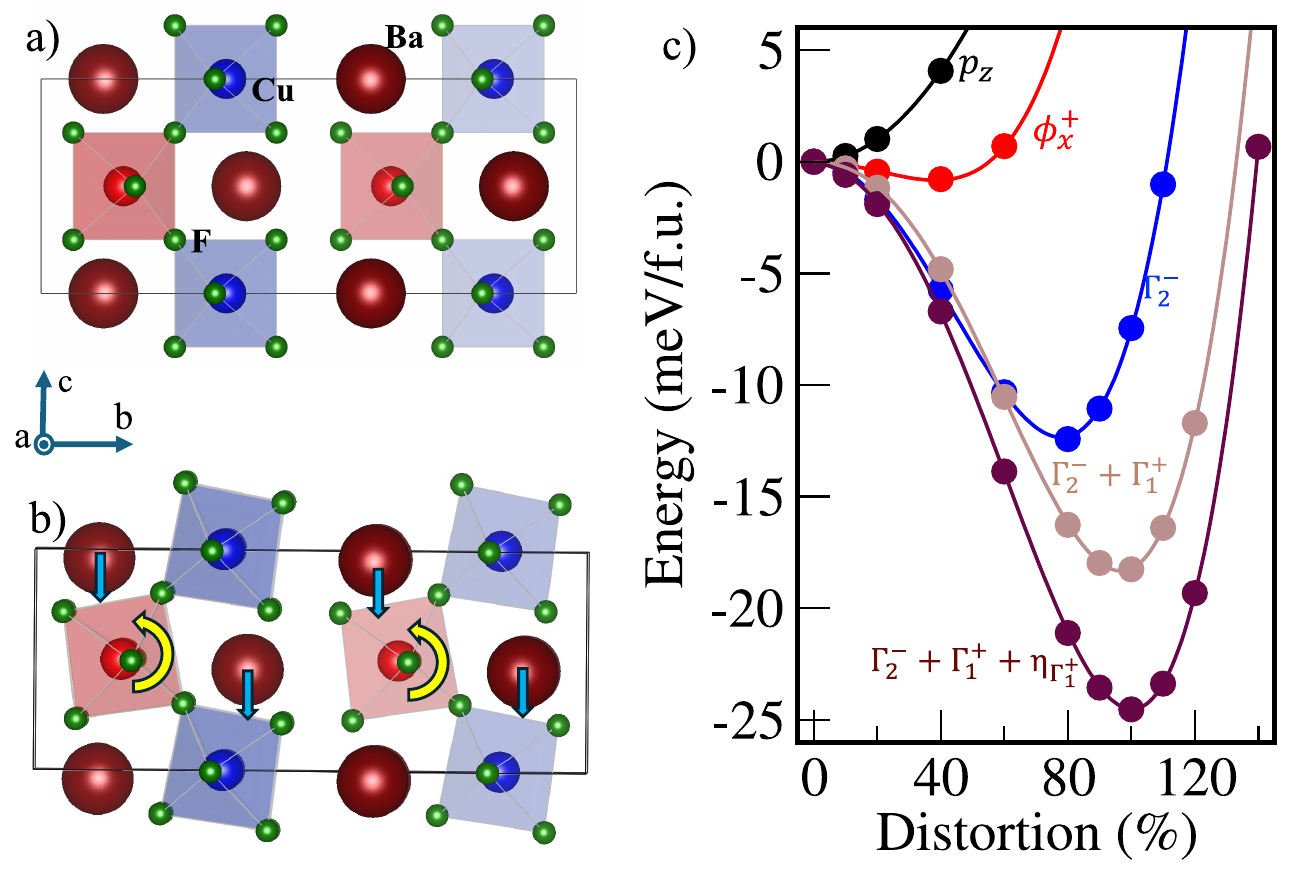}
\caption{Schematic representations of (a) the centrosymmetric $Cmcm$
reference structure and (b) the polar $Cmc2_1$ ground-state structure
of BaCuF$_4$. Brown, red/blue, and green spheres denote Ba, Cu, and F
atoms, respectively. Red and blue Cu atoms indicate oppositely oriented
magnetic moments in the A-type antiferromagnetic configuration. Yellow
curved arrows indicate the rotations of the CuF$_6$ octahedra about the
$x$ axis, while blue arrows indicate the polar displacements of the Ba
atoms along $z$. (c) Total energy as a function of the amplitudes of
the different structural distortions connecting the $Cmcm$ reference
structure to the $Cmc2_1$ ground state. The energy of the undistorted
$Cmcm$ structure is set to zero.}
\label{structure}
\end{figure} 

\subsection{Structural properties and electronic structure}

BaCuF$_4$ belongs to the BaMF$_4$ family of compounds \cite{Garcia18,Ibarra24} and crystallizes in the polar orthorhombic space group $Cmc2_1$ (No. 36). The crystal structure consists of layers of corner-sharing CuF$_6$ octahedra separated by Ba sheets and stacked along the $y$ direction, as shown in Fig.~\ref{structure}(a) and (b). The system is therefore quasi-two-dimensional in the $xz$ plane and exhibits a spontaneous electric polarization along the $z$ direction ($\mathbf{P}\parallel z$).

To understand the origin of the polar ground state, we performed a symmetry-mode analysis with respect to the hypothetical centrosymmetric $Cmcm$ reference structure \cite{Garcia18}. The corresponding energy landscape is shown in Fig.~\ref{structure}(c). The distortion from $Cmcm$ to $Cmc2_1$ can be decomposed into the irreducible representations (IRs) $\Gamma_2^-$ and $\Gamma_1^+$, together with the associated strain $\eta_{\Gamma_1^+}$. The latter comprises tensile strains of $0.15\%$
and $0.97\%$ along the $a$ and $b$ directions, respectively, and a
compressive strain of $-0.12\%$ along $c$. 

The $\Gamma_2^-$ mode comprises two symmetry-adapted distortions: a polar displacement $p_z$, mainly involving the Ba atoms and responsible for the electric polarization along $z$ \cite{Garcia18}, and an in-phase rotation $\phi_x^+$ of the CuF$_6$ octahedra about the $x$ axis. 
The two distortions make opposite contributions to the total energy, as shown by the black and red curves, respectively. While the polar mode $p_z$ alone increases the energy of the reference structure, the octahedral rotation $\phi_x^+$ lowers it.\footnote{The distortions are introduced in a frozen $Cmcm$ lattice. Both $p_z$ and $\phi_x^+$ individually lower the symmetry from $Cmcm$ to $Cmc2_1$.} Their combination (blue curve) yields a further energy gain and drives the instability of the $Cmcm$ phase, consistent with the unstable $\Gamma$-point phonon reported in Ref.~\cite{Garcia18}.

The $\Gamma_1^+$ mode is a fully symmetric distortion involving
displacements of Ba, Cu, and F atoms, with the latter two occurring
primarily within the CuF$_6$ layers. It shortens the in-plane Cu--F
bonds, resulting in a $2.6\%$ reduction in the volume of the CuF$_6$
octahedra. Since it does not lower the symmetry of the $Cmcm$ phase, it
emerges as a secondary structural response during relaxation of the
primary $\Gamma_2^-$ distortion. Nevertheless, it provides an additional energetic stabilization, as evidenced by the further reduction of the total energy, as shown by the brown curve. Finally, upon inclusion of the strain $\eta_{\Gamma_1^+}$, the system reaches the fully relaxed $Cmc2_1$ ground state, which lies approximately $25$ meV/f.u. below the centrosymmetric $Cmcm$ reference phase.

Besides these collective distortions, the CuF$_6$ octahedra exhibit a
local Jahn--Teller (JT) distortion associated with the nominal $d^9$
configuration of Cu$^{2+}$. In the approximately octahedral crystal
field generated by the surrounding F ligands, the Cu $d$ states split
into lower-energy $t_{2g}$ and higher-energy $e_g$ manifolds. The
$t_{2g}$ manifold is fully occupied, whereas the $e_g$ manifold hosts a single hole and is therefore JT active. The resulting distortion is
already present in the centrosymmetric $Cmcm$ reference structure and
persists in the polar $Cmc2_1$ phase; it is therefore distinct from the
collective symmetry-breaking distortions driving the
$Cmcm\rightarrow Cmc2_1$ transition.

\begin{figure}[t!]
    \centering
    \begin{overpic}[width=0.68\columnwidth,angle=270]{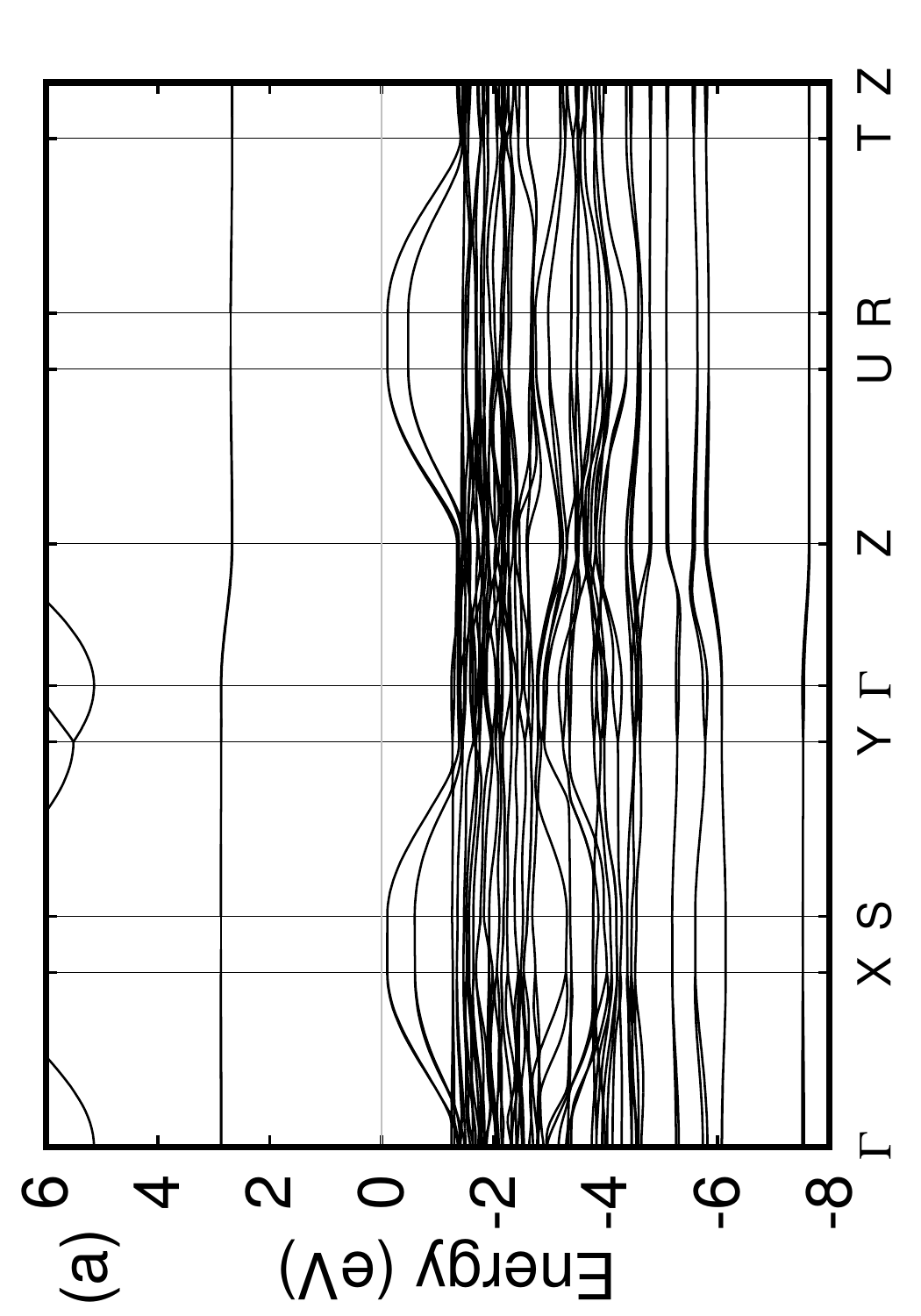}
        \put(60,60){\includegraphics[height=2.4cm]{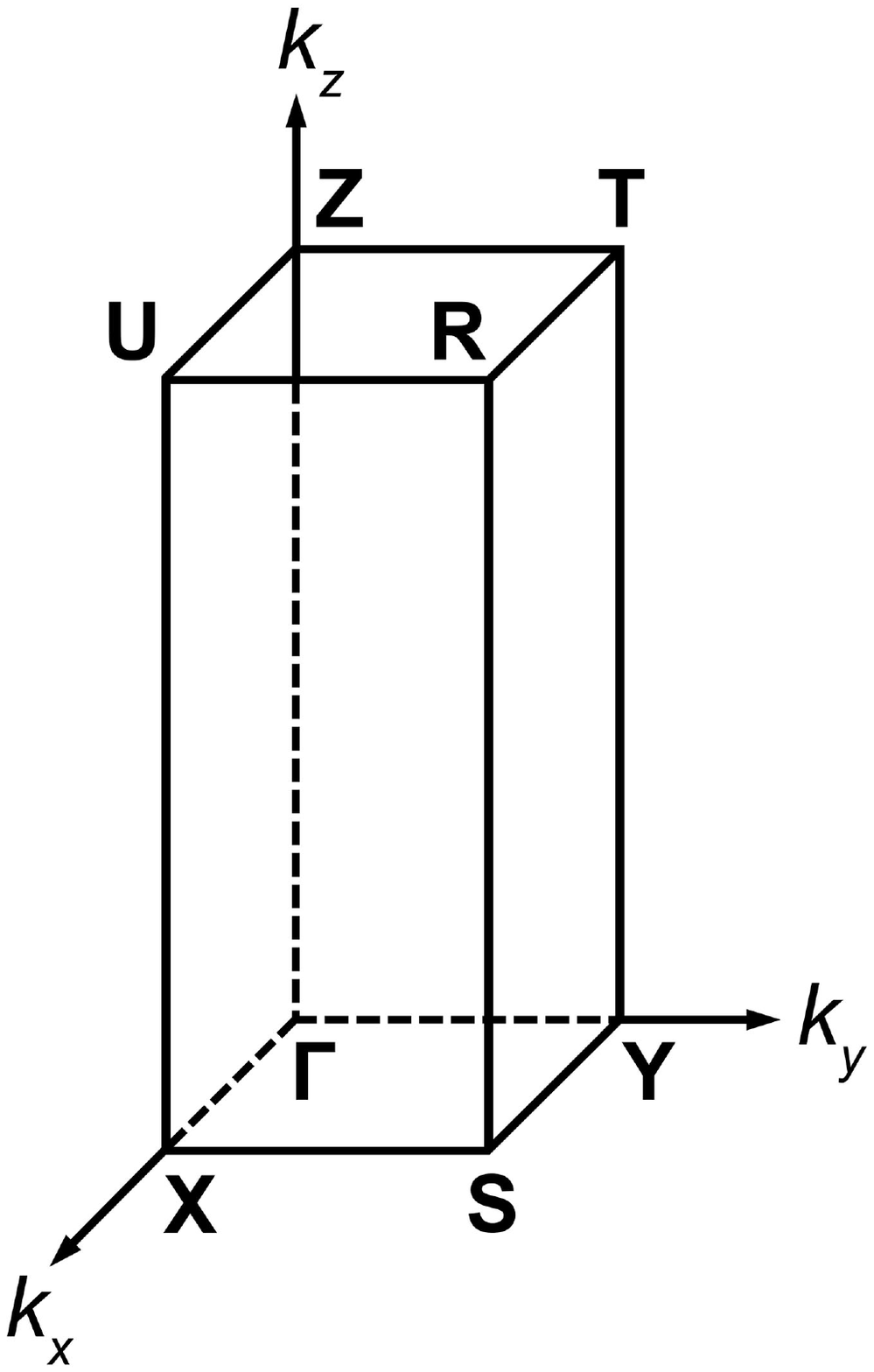}}
    \end{overpic}
    \vspace{1cm}
    \includegraphics[width=0.68\columnwidth,angle=270]{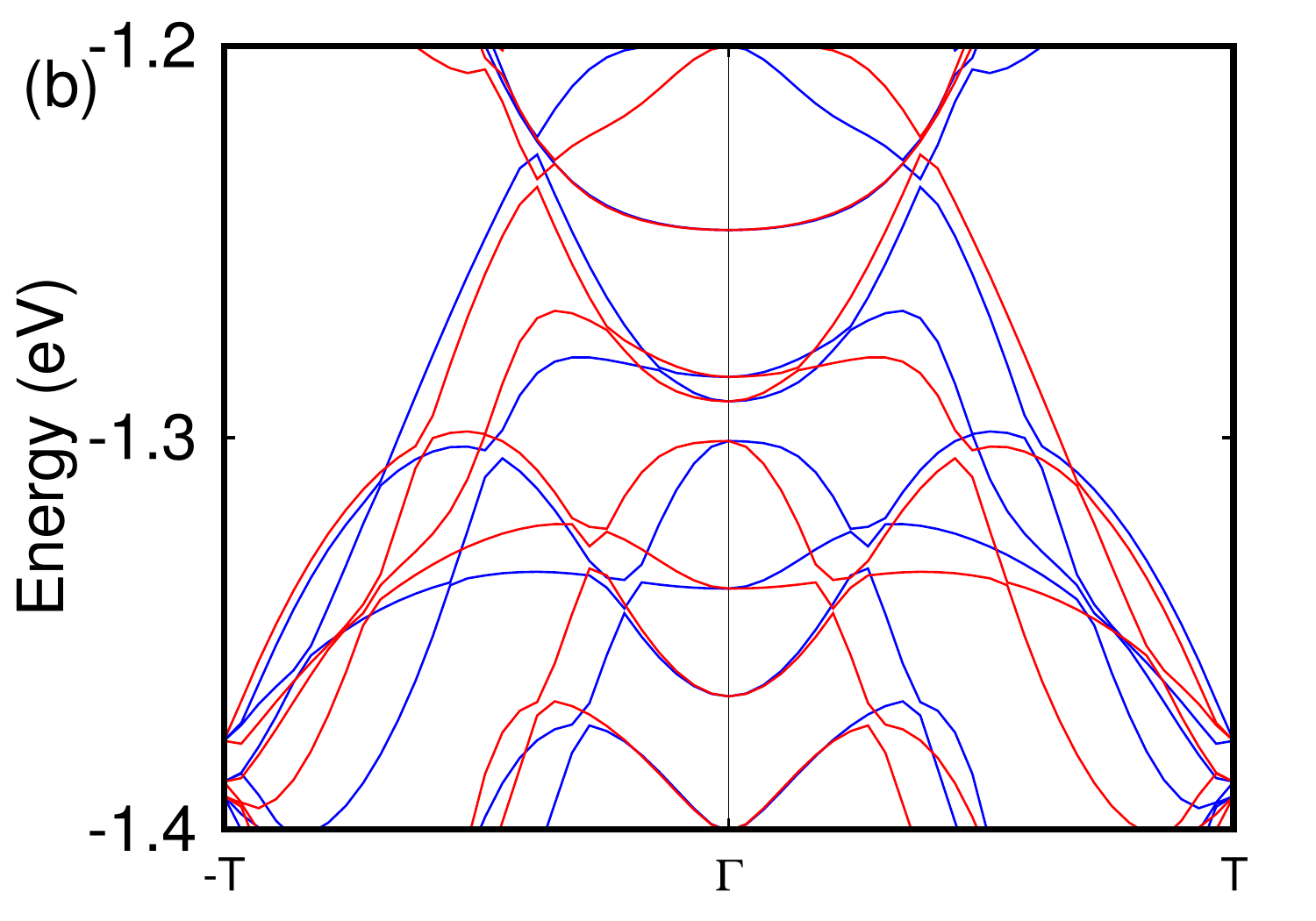}
    \footnotesize
    \caption{(a) Nonrelativistic band structure of BaCuF$_4$ along the high-symmetry path of the orthorhombic BZ shown in the inset. The isolated conduction band, approximately $2.7$~eV above the valence-band maximum, originates from the unoccupied Cu $e_g$ states. (b) Nonrelativistic spin splitting along the $-T\leftarrow\Gamma\rightarrow T$ direction of the BZ. Red and blue indicate opposite signs of the spin splitting.}
    \label{Bands_BaCuF4}
\end{figure}

In the relaxed $Cmc2_1$ structure, the JT distortion elongates the
Cu--F bonds along the crystallographic $a$ direction to $2.26$~\AA,
compared with an average length of $1.89$~\AA{} for the remaining
Cu--F bonds. The resulting lifting of the $e_g$ degeneracy places the unoccupied minority-spin state at the bottom of the conduction manifold. It forms the isolated, weakly dispersive band visible approximately $2.7$~eV above the valence-band
maximum in Fig.~\ref{Bands_BaCuF4}(a), thereby defining the insulating
gap. As shown in the SM, the inclusion of SOC leaves the band gap essentially
unaffected.

%%%%%%%  BEGINNING OF CARMINE AND AMAR PART %%%%%%%%%%%%

\begin{table*}[]
\centering
\begin{tabular}{|c|c|c|c|c|c|c|c|c|}
\hline
Magnetic Configuration & M$_1$ & M$_2$ & M$_3$ & M$_4$ & M$_{\text{tot}}$ & m$_a$ & m$_b$ & m$_c$  \\
\hline
$\mathbf{N} \parallel x$ & (m$_a$,0,0)   & (-m$_a$,0,0)  & (m$_a$,0,0) & (-m$_a$,0,0) & (0,0,0) & 0.796 & 0 & 0  \\
\hline
$\mathbf{N} \parallel y$ & (0,m$_b$,m$_c$)  & (0,-m$_b$,m$_c$)  & (0,m$_b$,m$_c$)  & (0,-m$_b$,m$_c$)  & (0,0,4m$_c$) & 0 & 0.794 & 0.020  \\
\hline
$\mathbf{N} \parallel z$ &    (0,-m$_b$,m$_c$) & (0,-m$_b$,-m$_c$)  & (0,-m$_b$,m$_c$) & (0,-m$_b$,-m$_c$) & (0,-4m$_b$,0) & 0 & 0.046 & 0.785   \\
\hline
\end{tabular}
\caption{Magnetic moments along different directions of the N\'eel vector ($\mathbf{N}$). m$_a$, m$_b$ and m$_c$ are reported in $\mu_B$. When the N\'eel vector points along the $y$-axis, weak ferromagnetism appears along $z$, and when it is along the $z$-axis, weak ferromagnetism occurs along $y$.}
\label{tab:table1}
\end{table*}

\begin{figure*}[!t]
\centering
\includegraphics[width=0.23\textwidth,angle=270]{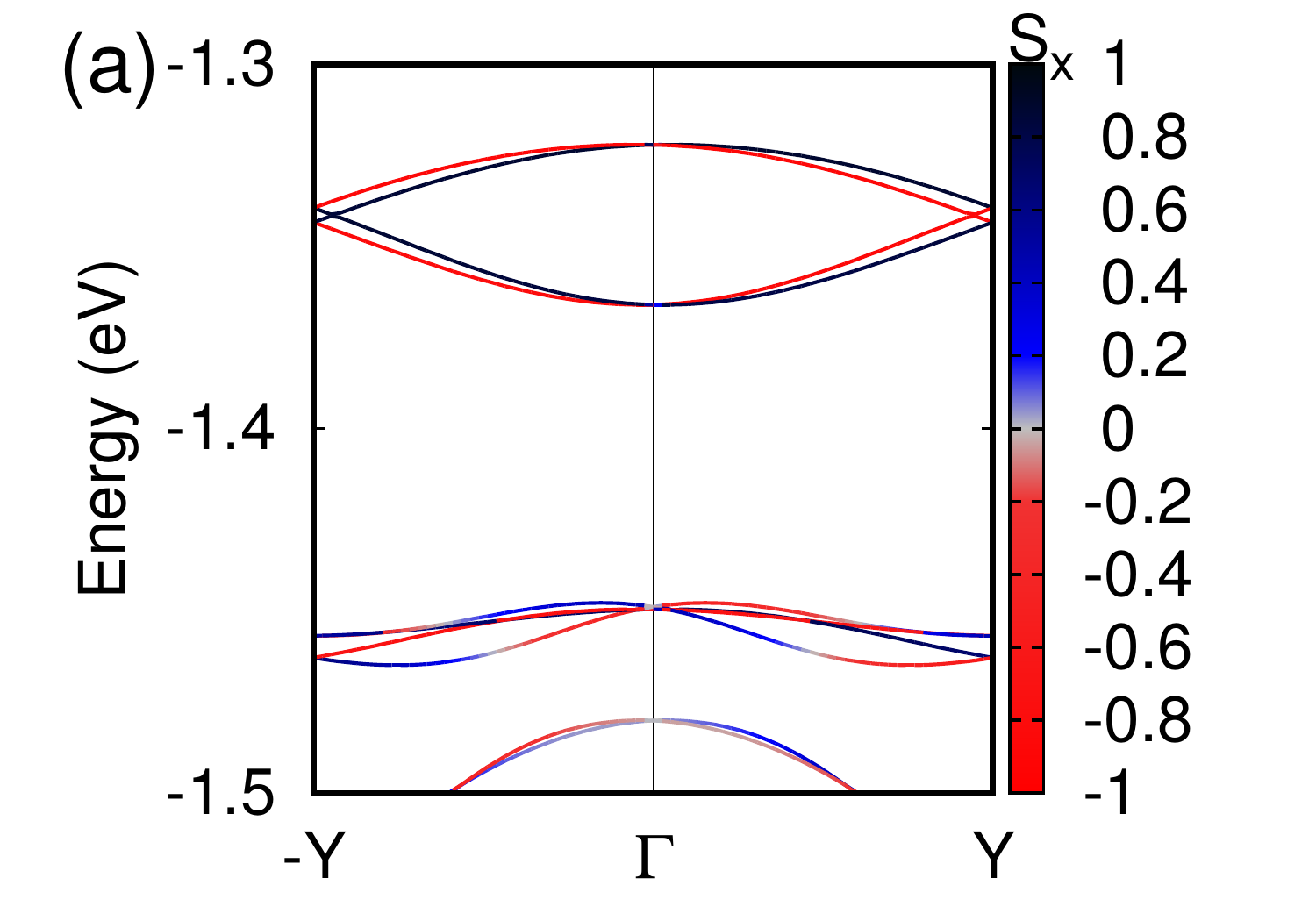}
\includegraphics[width=0.23\textwidth,angle=270]{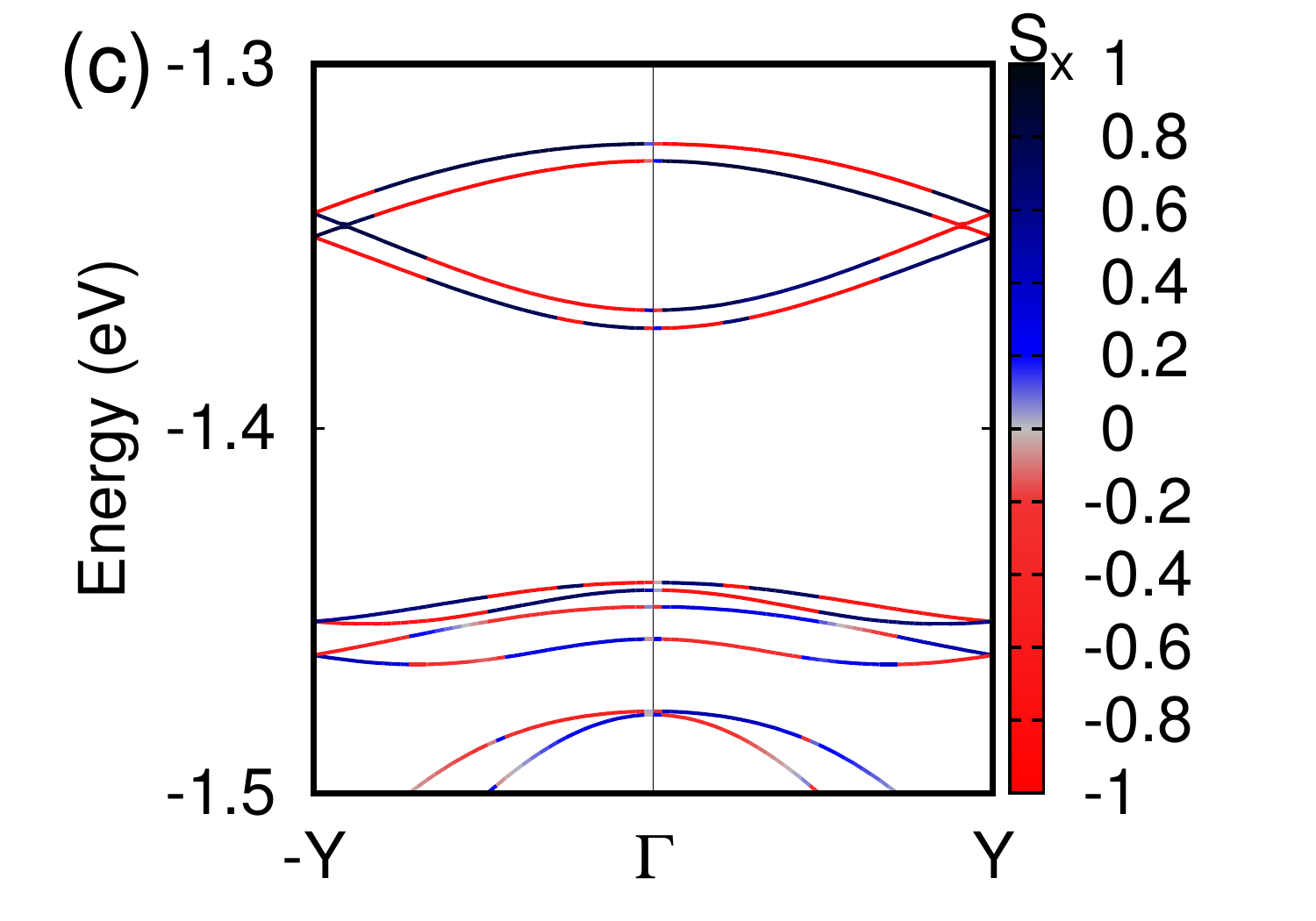}
\includegraphics[width=0.23\textwidth,angle=270]{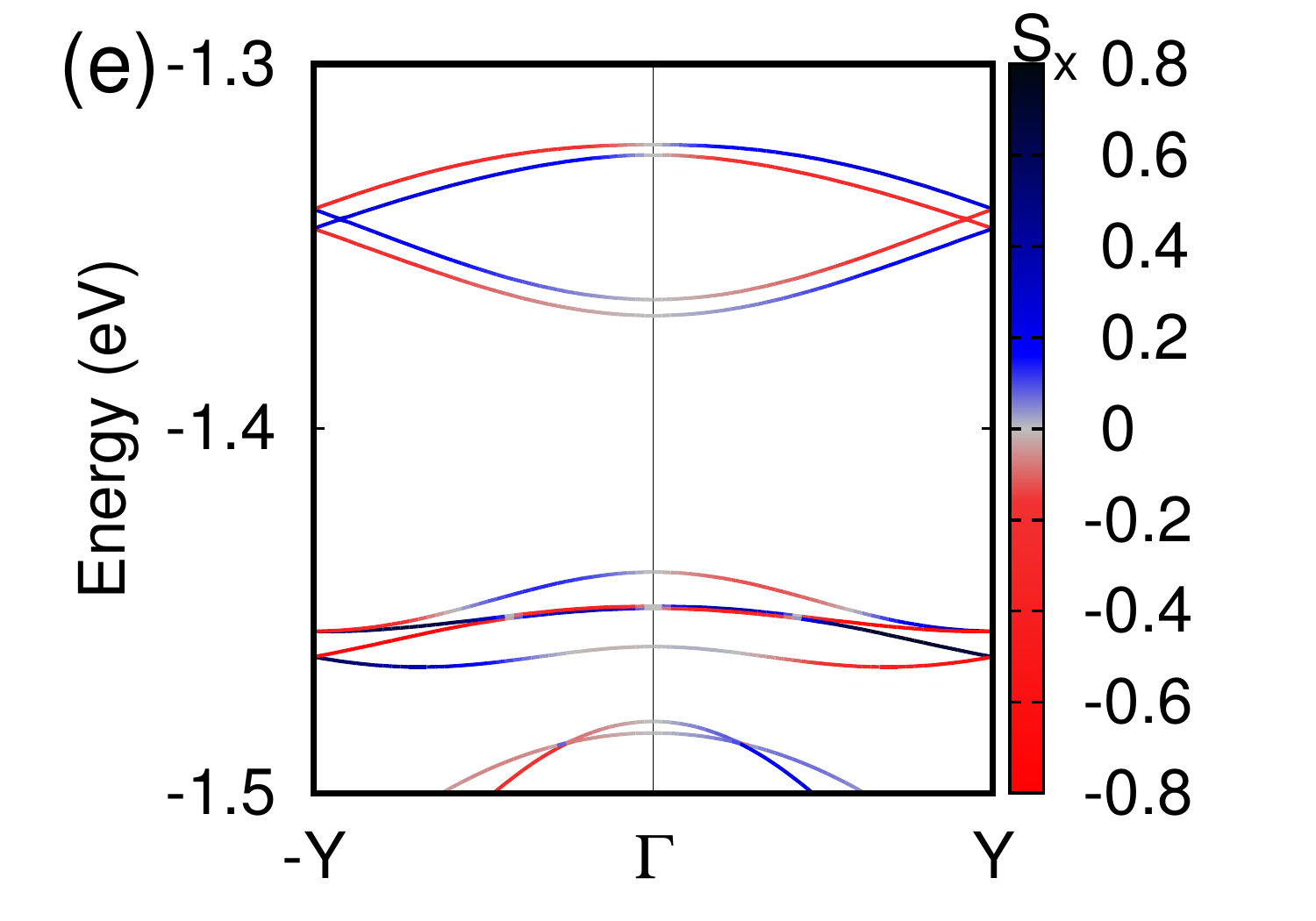}
\includegraphics[width=0.23\textwidth,angle=270]{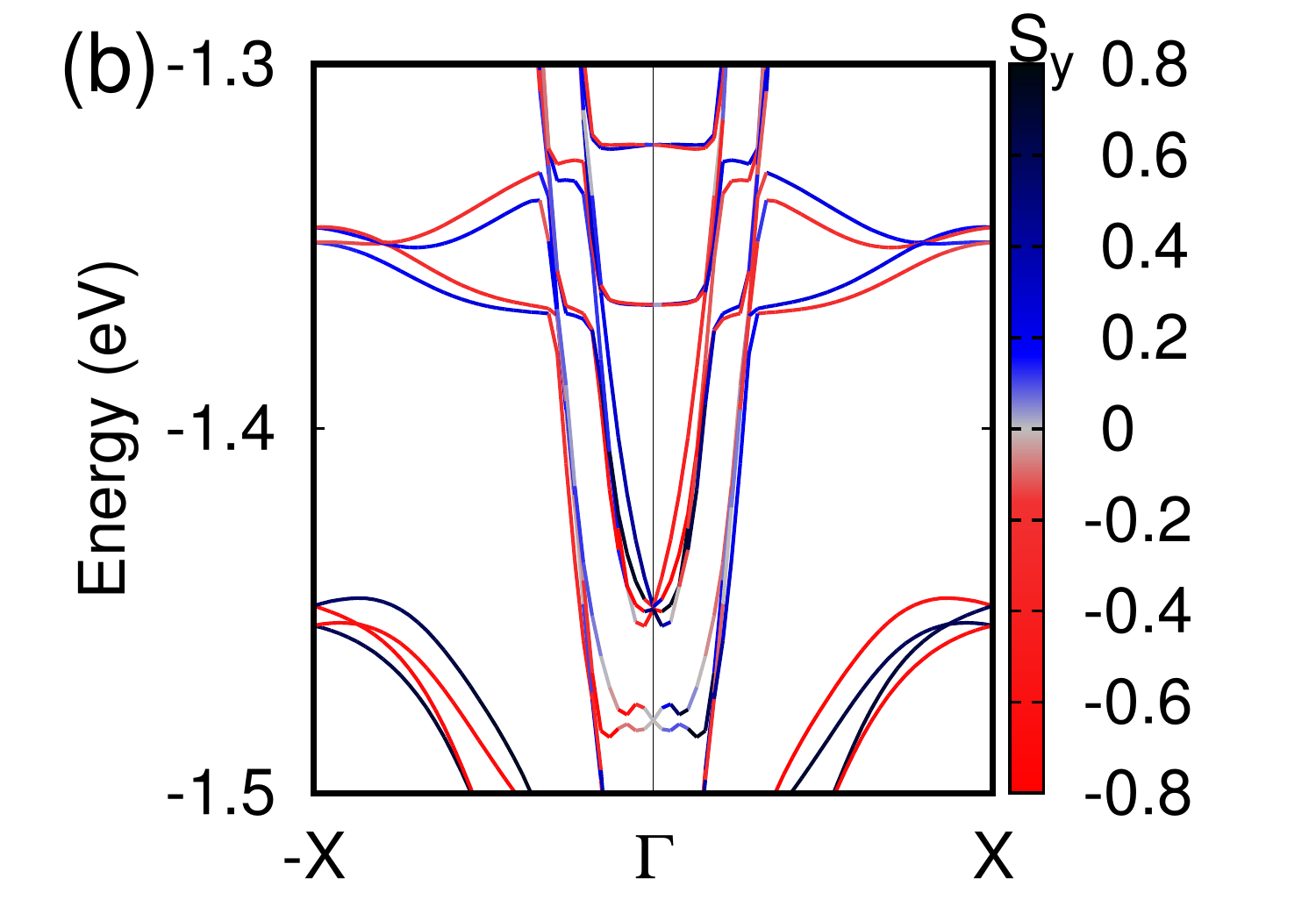}
\includegraphics[width=0.23\textwidth,angle=270]{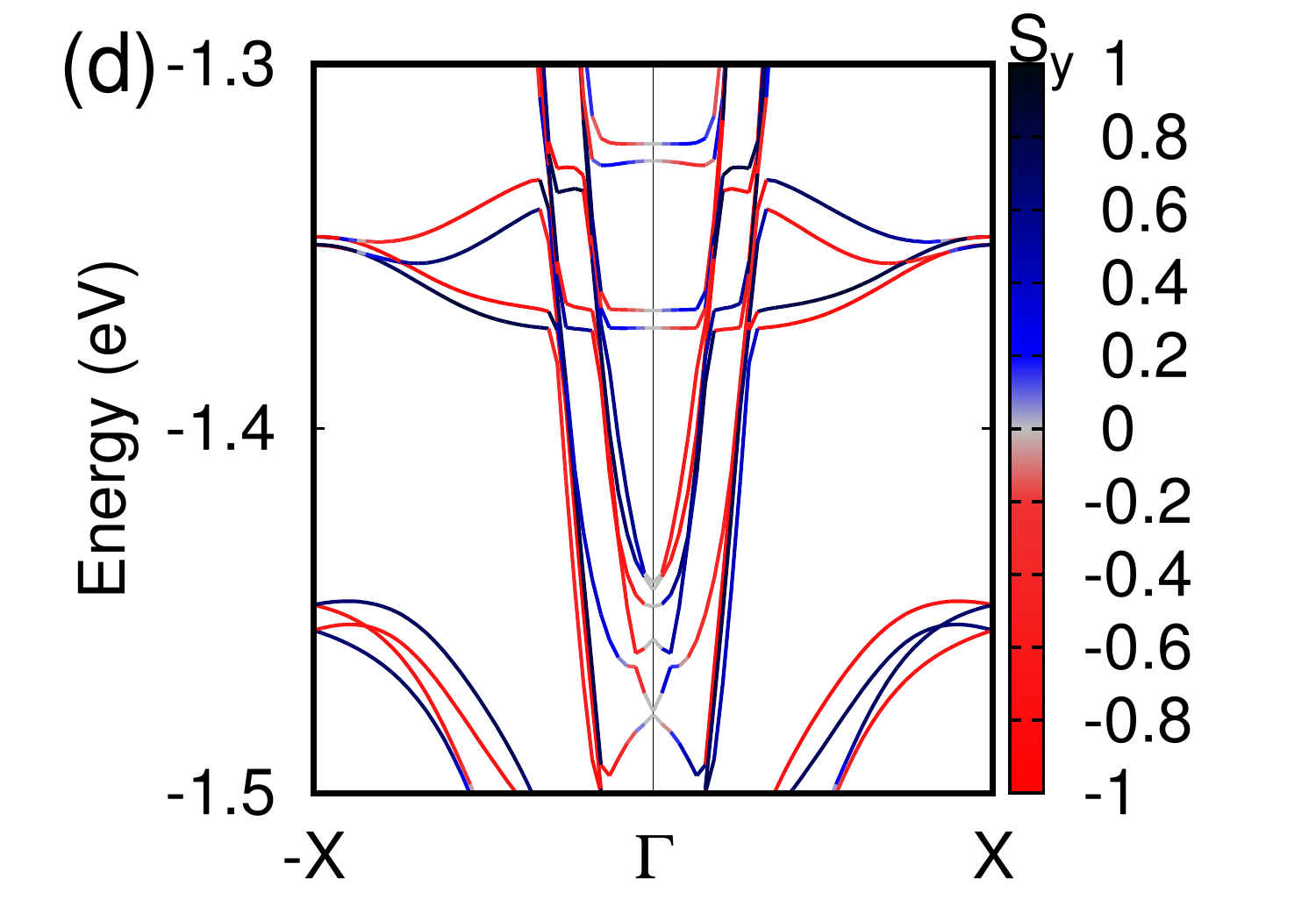}
\includegraphics[width=0.23\textwidth,angle=270]{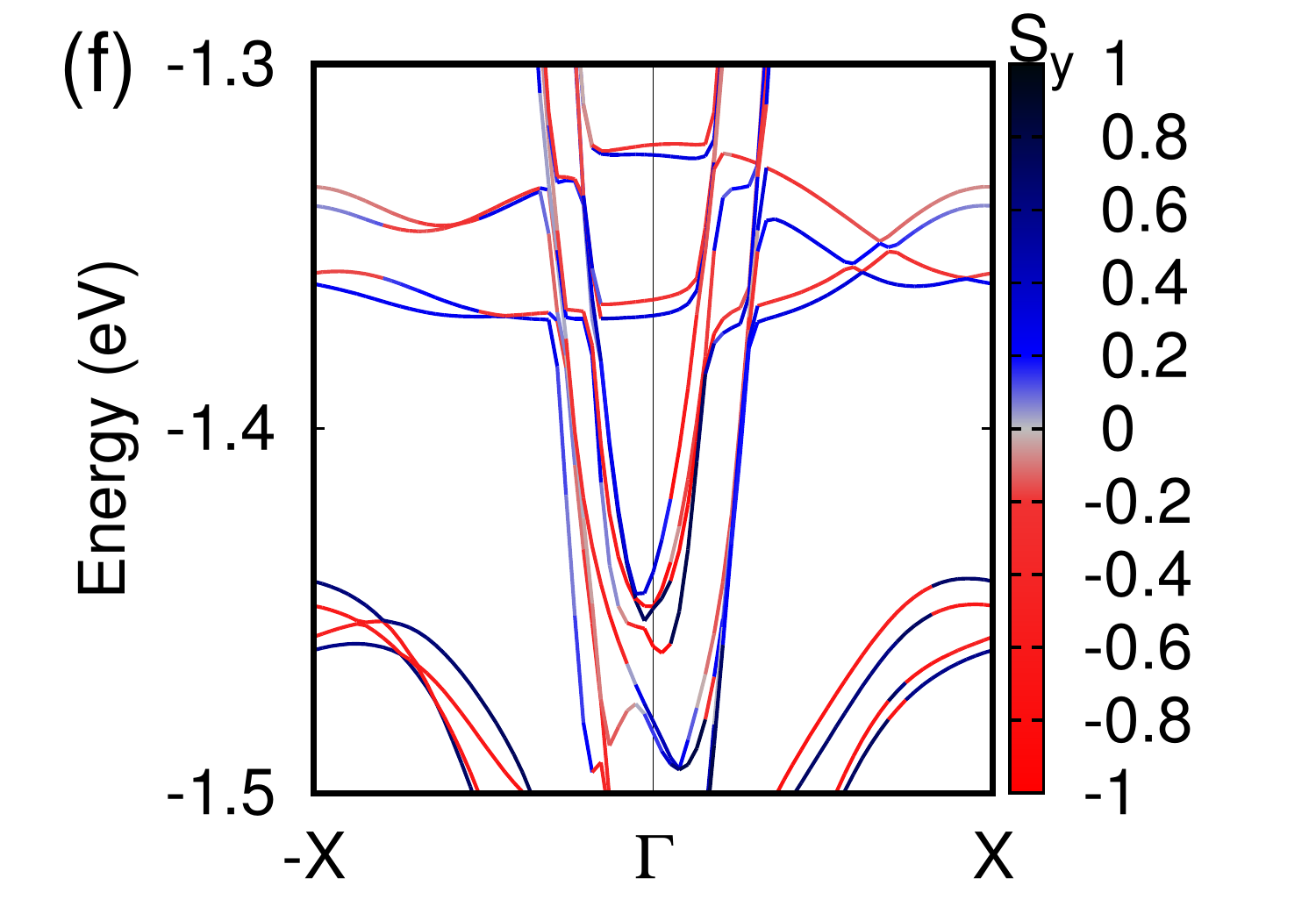}
\vspace{2mm}
\includegraphics[width=0.99\textwidth]{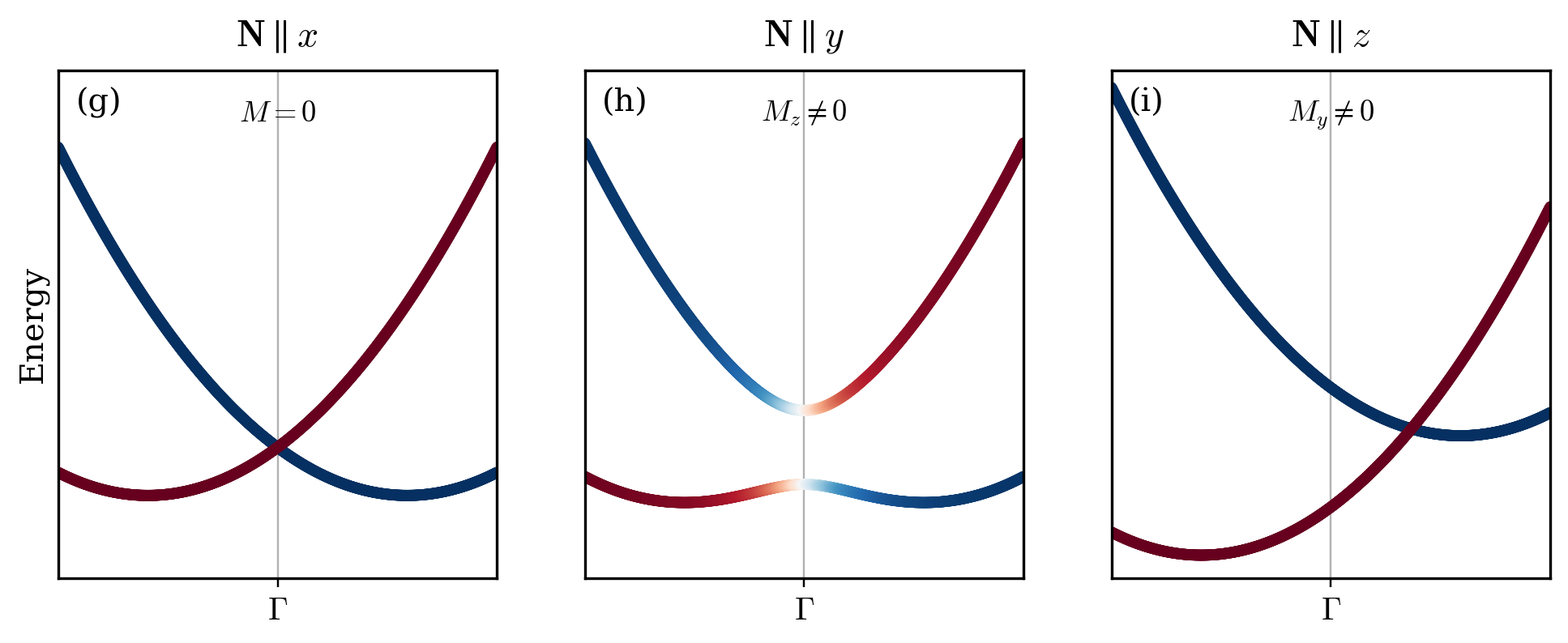}
\caption{Rashba-like contributions to the SML for the three N\'eel-vector orientations: (a),(b) $\mathbf N\parallel x$; (c),(d) $\mathbf N\parallel y$; and (e),(f) $\mathbf N\parallel z$. For $\mathbf N\parallel x$, the collinear magnetic state preserves the twofold degeneracy at $\Gamma$, whereas for $\mathbf N\parallel y,z$, the SOC-induced weak-ferromagnetic moment lifts it. For $\mathbf N\parallel y$, $\mathbf M\parallel\mathbf P\parallel z$, and a gap opens at the crossing. For $\mathbf N\parallel z$, $\mathbf M\parallel y$ is perpendicular to $\mathbf P$, shifting the crossing away from $\Gamma$ and producing a nonreciprocal dispersion. Panels (g)--(i) schematically illustrate the corresponding evolution of the Rashba-split bands. Red and blue denote opposite signs of the spin polarization.}
\label{figure_Rashba}
\end{figure*}

\subsection{Nonrelativistic spin-momentum locking}
The centrosymmetric $Cmcm$ reference structure does not exhibit altermagnetism in any of the A-, C-, or G-type magnetic configurations compatible with the unit cell (within the unit cell)\cite{Garcia18}. We therefore focus on the polar $Cmc2_1$ phase. We consider throughout this work the A-type antiferromagnetic configuration, which realizes the altermagnetic phase of BaCuF$_4$. The A-, G- and C-type orderings are nearly degenerate in energy \cite{Garcia18}.
In the absence of SOC, the band structure in Fig.~\ref{Bands_BaCuF4}(b) displays a spin splitting along the $\Gamma=(0,0,0)\rightarrow T=(0,\frac{1}{2},\frac{1}{2})$ direction of the Brillouin zone (BZ), shown in the inset of Fig. \ref{Bands_BaCuF4}. The splitting is odd in both $k_y$ and $k_z$, consistently with previous studies\cite{smejkal2024altermagneticmultiferroicsaltermagnetoelectriceffect}. More generally, it vanishes on the $k_y =0$ and $k_z
=0$ planes, whose intersection defines the nodal line along $k_x$. The resulting nonrelativistic altermagnetic spin texture therefore has $d$-wave character.

The emergence of altermagnetism is closely tied to the structural distortions stabilizing the $Cmc2_1$ phase. The nonrelativistic spin splitting originates primarily from the octahedral rotation mode $\phi_x^+$, whereas the ferroelectric polarization is associated with the polar displacement $p_z$. These two phenomena are nevertheless coupled: reversing the polarization reverses the sense of the CuF$_6$ octahedral rotations and, consequently, the sign of the altermagnetic spin splitting in the oppositely polarized $Cmc2_1$ state, in agreement with previous reports \cite{smejkal2024altermagneticmultiferroicsaltermagnetoelectriceffect,ray2025_arxiv}.

This coupling between ferroelectricity and altermagnetism is known as the altermagnetoelectric effect \cite{smejkal2024altermagneticmultiferroicsaltermagnetoelectriceffect}. Similar mechanisms have also been proposed in perovskite superlattices \cite{Sbandyopadhyay_2025,ray2025_arxiv} and rutile compounds \cite{SBandyopadhyay_MnF2}. By enabling electrical control of spin polarization, the altermagnetoelectric effect offers a promising platform for reconfigurable spintronic devices.

\subsection{Relativistic contributions to the spin-momentum locking}
%SOC preserves the nonrelativistic altermagnetic SML of the dominant spin component while generating additional spin-texture contributions that depend on the orientation of the N\'eel vector, $\mathbf N$.

SOC preserves the nonrelativistic altermagnetic SML of the dominant spin component while generating additional spin-texture contributions. It also determines the magnetic anisotropy: consistent with previous work, the magnetic easy axis is predicted to lie along $y$. Nevertheless, to understand how relativistic effects reshape the electronic and magnetic structure, we consider all three principal orientations of the Néel vector, $\mathbf N$, while retaining the A-type magnetic order throughout.

For $\mathbf{N}\parallel x$, the magnetic order remains collinear, and the system therefore retains a purely altermagnetic character. The dominant spin-polarization component, $S_x$, is most clearly visible along the $\Gamma$--T direction in the BZ, but vanishes along the $\Gamma$--X and $\Gamma$--Y, where the relativistic contributions can therefore be isolated.

Along these directions, the bands split away from the doubly degenerate $\Gamma$ point and develop a characteristic camel-back dispersion. The SOC-induced in-plane spin texture has finite $S_x$ along $\bar{\mathrm Y}$--$\Gamma$--$\mathrm Y$ and $S_y$ along the orthogonal $\bar{\mathrm X}$--$\Gamma$--$\mathrm X$ direction, as shown in Fig.~\ref{figure_Rashba}(a),(b). Both components reverse sign across $\Gamma$, consistent with a spin splitting linear in momentum, and exhibit nodal planes at $k_y=0$ for $S_x$ and $k_x=0$ for $S_y$. These features identify a Rashba-like spin texture arising from the interplay of SOC and the inversion-symmetry breaking associated with the ferroelectric polarization.

Beyond this Rashba-like in-plane spin texture, the DFT calculations additionally reveal an out-of-plane component $S_z$, with nodal planes at $k_x=0$ and $k_y=0$. This therefore constitutes a relativistic $d$-wave contribution, distinct from the dominant nonrelativistic altermagnetic $d$-wave texture.

%The SOC-induced spin texture is not limited to this in-plane contribution. Along the $(0,0,0)\rightarrow(\frac{1}{2},\frac{1}{2},0)$ direction, the spin-resolved bands also show a finite $S_z$ component. Unlike the in-plane spin-polarization, $S^z$ vanishes on both the $k_x =0$ and $k_y=0$ planes. This coexistence of in-plane and out-of-plane spin-polarization components demonstrates that SOC generates multiple relativistic contributions to the spin texture beyond the Rashba one. 

%For $\mathbf N\parallel y$ and $\mathbf N\parallel z$, the non-relativistic contribution to the spin polarization is likewise retained and remains parallel to the Néel vector, with $S_y$ and $S_z$ as the dominant components, respectively. However, in both cases, SOC induces a canting of the magnetic moments, giving rise to a small net magnetization $\mathbf M$ along the $z$ and $y$ directions, respectively. The system is therefore no longer purely altermagnetic, but also exhibits weak ferromagnetism. The calculated atomic magnetic moments and the magnetization per magnetic unit cell are reported in Table~\ref{tab:table1}. Consistent with previous work \cite{smejkal2024altermagneticmultiferroicsaltermagnetoelectriceffect}, the predicted magnetic easy axis lies along the $y$ direction.

For $\mathbf N\parallel y$ and $\mathbf N\parallel z$, the nonrelativistic spin polarization is likewise preserved and remains parallel to the Néel vector, with $S_y$ and $S_z$ as the respective dominant components. In both configurations, however, SOC induces spin canting and a weak-ferromagnetic moment $\mathbf M$, directed along $z$ for $\mathbf N\parallel y$ and along $y$ for $\mathbf N\parallel z$. The system is therefore no longer purely altermagnetic. The calculated atomic magnetic moments and net magnetization per magnetic unit cell are reported in Table~\ref{tab:table1}.% Consistent with previous work \cite{smejkal2024altermagneticmultiferroicsaltermagnetoelectriceffect}, the magnetic easy axis is predicted to lie along $y$.

The Rashba-like contribution to the spin-polarization persists for both these Néel-vector orientations, producing $S_x$ and $S_y$ components analogous to those found for $\mathbf N\parallel x$.
However, weak ferromagnetism lifts the double degeneracy of the band crossing at $\Gamma$, as expected for a Rashba system subject to an exchange field \cite{Bihlmayer2022}. The resulting modification of the band structure depends on the orientation of the weak-ferromagnetic moment relative to the electric polarization. The evolution of the Rashba-split bands for the three orientations of the Néel vector is shown in the sketch of Fig.~\ref{figure_Rashba}(g)-(i).

For $\mathbf N\parallel y$, when $\mathbf M\parallel\mathbf P\parallel z$, a finite $S_z$ develops and a gap opens at the band crossing, as shown in Fig.~\ref{figure_Rashba} (c), (d). In contrast, for $\mathbf N\parallel z$, $\mathbf M\parallel y$ is perpendicular to $\mathbf P$, and the band crossing shifts away from $\Gamma$. This gives rise to a non-reciprocal band structure, recently identified as a general phenomenon in multiferroics \cite{stavric2025}, whereby the electronic dispersion becomes asymmetric under momentum reversal about certain high-symmetry points of the BZ. For the present $\mathbf N\parallel z$ configuration, $\varepsilon_n(\mathbf{k}) \neq \varepsilon_n(-\mathbf{k})$ for $\mathbf{k}\parallel\mathbf M\times\mathbf P$, as shown in Fig.~\ref{figure_Rashba}(e), (f). This behaviour is analogous to that found for EuO in the MF-I configuration discussed in Ref.~\cite{stavric2025}. %The evolution of the Rashba-split bands for the three orientations of the Néel vector is shown in the sketch of Fig.~\ref{figure_Rashba}(g)-(i).  %These two behaviors are characteristic signatures of Rashba systems subject to an exchange field \cite{Bihlmayer2022}.  

Finally, these Néel-vector configurations exhibit additional $S_x$ contributions, which can be distinguished from the Rashba-like component by their nodal-plane structure. Specifically, for $\mathbf N\parallel y$, their nodal planes are located at $k_x=0$ and $k_z=0$, whereas for $\mathbf N\parallel z$, they occur at $k_x=0$ and $k_y=0$. These contributions thus have $d$-wave symmetry but are relativistic in origin.

In summary, the DFT calculations reveal several SOC-induced contributions to the spin texture beyond the nonrelativistic altermagnetic component. They can be classified according to their symmetry, as discussed next.

\begin{figure*}[t]
    \centering
        \centering
         \includegraphics[width=\textwidth]{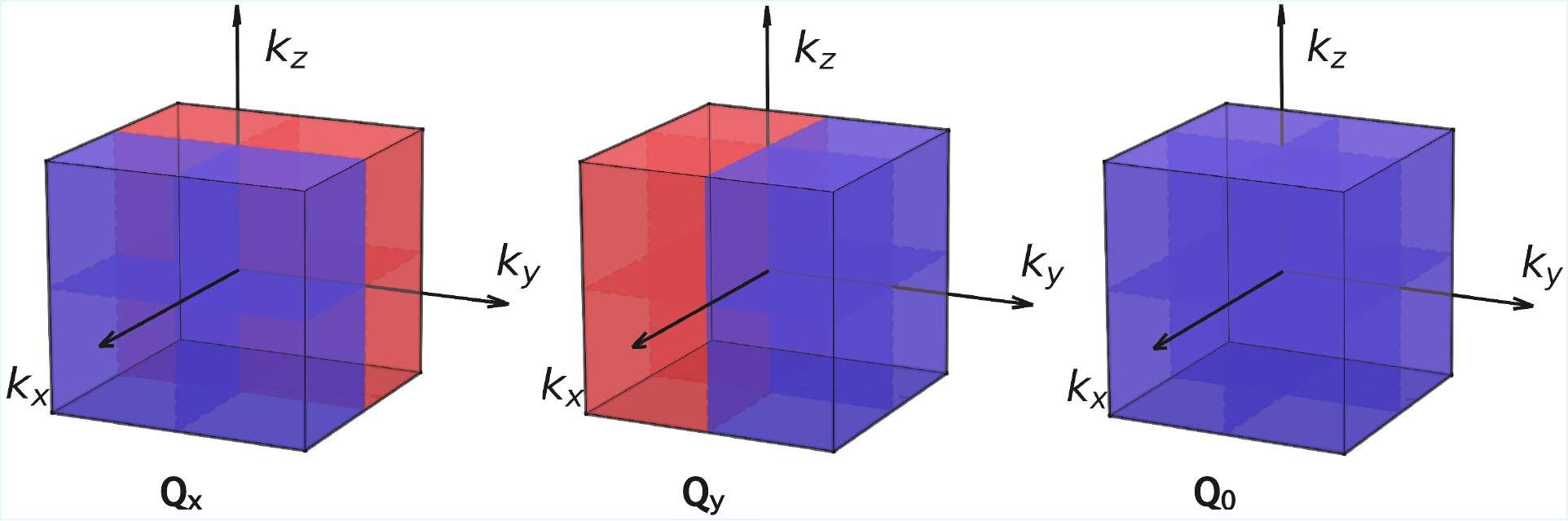}
        \includegraphics[width=\textwidth]{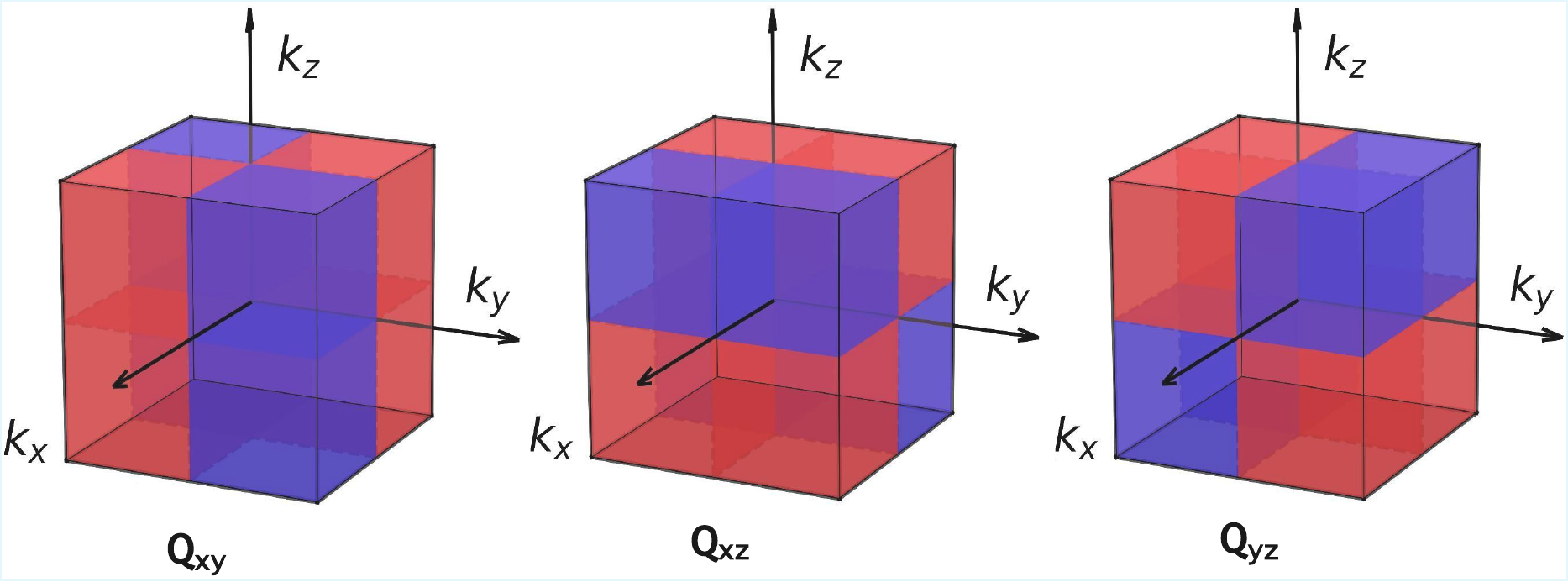}
       \caption{Momentum-space structures of the  monopole $Q_0$, dipoles $Q_x$ and $Q_y$, and quadrupoles $Q_{xy}$, $Q_{xz}$, and $Q_{yz}$ relevant to the spin-momentum locking in orthorhombic BaCuF$_4$. Red and blue denote regions with opposite signs of the spin polarization.}
        \label{figure2_SML}
\end{figure*}

\begin{table*}[ht!]
\centering
\resizebox{1\textwidth}{!}{%
\begin{tabular}{|c|c|c|c|}
\hline
 & \multicolumn{3}{|c|}{Spin components $S_i(\mathbf{k})$} \\
\hline
N\'eel vector & $S_x$ & $S_y$ & $S_z$ \\
\hline
$\mathbf{N} \parallel x$ &  $Q_y$,\; \textcolor{purple}{$Q_{yz}$} & $Q_x$,\; $Q_{xz}$ & $Q_{xy}$ \\
\hline
$\mathbf{N} \parallel y$ &  \textcolor{orange}{$Q_y$},\; $Q_{xz}$ & $Q_x$,\; \textcolor{purple}{$Q_{yz}$} & \textcolor{orange}{$Q_0$},\; [\textcolor{orange}{$Q_{xx}$},\; \textcolor{orange}{$Q_{yy}$},\; \textcolor{orange}{$Q_{zz}$}] \\
\hline
$\mathbf{N} \parallel z$  & \textcolor{orange}{$Q_y$},\; $Q_{xy}$ & \textcolor{orange}{$Q_0$},\; $Q_x$,\; [\textcolor{orange}{$Q_{xx}$},\; \textcolor{orange}{$Q_{yy}$},\; \textcolor{orange}{$Q_{zz}$}] & \textcolor{purple}{$Q_{yz}$} \\
\hline
%Paramagnetic &   $Q_y$ & $Q_x$ & 0 \\
%\hline
\end{tabular}
}
\caption{Allowed terms of the expansion of band spin-polarization. Purple-coloured terms are those
that would survive for collinear spin groups, while orange-coloured ones are those allowed by coplanar
spin groups - relevant for the second and third rows. When SOC is included and MSG applies,
then all terms appear. Without SOC and for the collinear phases, only the purple-coloured terms survive,
while both orange-coloured and purple-coloured terms are allowed for the coplanar (weak-ferromagnetic) cases without
SOC. Terms enclosed in square brackets cannot be directly identified from the DFT calculations since they cannot be easily disentangled by Q$_0$.  %In the paramagnetic phase, only odd-order terms can be nonzero, since even-order terms are
%zero under time reversal. 
}
\label{tab:spinmomentumlockingallowed}
\end{table*}

\subsection{Symmetry analysis of the relativistic spin-momentum locking}\label{sec.symmetry}

The spin polarization near the $\Gamma$ point can be expanded in powers of the crystal momentum as
\begin{equation}
S_i(\mathbf{k}) =
\sum_{n=0}^{\infty}
Q^{(n)}_{i,\alpha\beta\gamma\ldots}
k_\alpha k_\beta k_\gamma \ldots
\label{eq:Sk_expansion}
\end{equation}
where $i,\alpha,\beta,\gamma=x,y,z$ and summation over repeated indices is implied\cite{radaelli2025spintexture}. The coefficients $Q^{(n)}_{i,\alpha\beta\gamma\ldots}$ are the components of rank-$(n+1)$ spin multipole tensors,  illustrated in Fig.~\ref{figure2_SML}. %This expansion provides a systematic framework for classifying the momentum-dependent spin textures obtained from the DFT calculations. 
Specifically, the zeroth-order term ($Q^{(0)}$) corresponds to a spin monopole and represents a momentum-independent spin polarization associated with a net magnetization or weak ferromagnetism. The first-order terms ($Q^{(1)}_x$, $Q^{(1)}_y$, $Q^{(1)}_z$) are spin dipoles, characterized by a linear dependence on crystal momentum and a single nodal plane, encompassing Rashba- and Dresselhaus-like spin textures. The second-order terms ($Q^{(2)}_{\alpha\beta}$) are spin quadrupoles, characterized by a quadratic momentum dependence; the off-diagonal components exhibit two orthogonal nodal planes.

The allowed spin multipoles depend on the symmetries of both the spin and spatial degrees of freedom. Without SOC, these are described by spin groups; with SOC, spin and real-space transformations are coupled and the relevant symmetries are the magnetic space groups (MSGs). For $\mathbf N\parallel x,y,z$, the MSGs are,
respectively, $Cmc2_1$ (No.~36.172), $Cm'c'2_1$ (No.~36.176), and
$Cm'c2'_1$ (No.~36.174). The first forbids a net magnetization, whereas
the latter two allow weak ferromagnetism while preserving coplanarity,
in agreement with the DFT results.
In these latter two cases, each MSG is therefore associated with two oriented spin (super)groups\cite{Etxebarria.stensor.25} corresponding to the collinear and coplanar magnetic states:
$C^{1}m^{\bar{1}}c^{\bar{1}}2_1^{1}\,{}^{\infty_{\mathbf{N}}m}1$ and
$C^1m^{2_{\mathbf{M}}}c^{2_{\mathbf{M}}}2_1^{m_{\mathbf{M}\times\mathbf{N}}}1$, respectively. Here, $\mathbf N$ defines the spin quantization axis in the collinear state, while $\mathbf M$ denotes the weak-ferromagnetic moment in the coplanar state, whose quantization axis is normal to the spin plane. These groups were identified using FINDSPINGROUP \cite{Yu2026FINDSPINGROUP}.

The group--subgroup hierarchy provides the classification summarized in
Table~\ref{tab:spinmomentumlockingallowed}: purple multipoles are
allowed already by the collinear spin group and describe the
nonrelativistic altermagnetic SML; orange multipoles are additionally
allowed by the coplanar spin group and are associated with the
weak-ferromagnetic state; and black multipoles require the full MSG and
are purely relativistic. The complete transformation properties are
given in Appendix~\ref{app.spin_mult}.

The only multipole allowed in the collinear spin-group limit is the quadrupole $Q^{(2)}_{yz}$, which accounts for the dominant spin polarization parallel to the Néel vector, $\mathbf S\parallel\mathbf N$, for all three Néel-vector orientations. Its characteristic momentum-space structure, illustrated in Fig.~\ref{figure2_SML}, exhibits nodal planes at $k_y=0$ and $k_z=0$, accounting for the $d$-wave character observed in the DFT spin textures.

For $\mathbf N\parallel x$, weak ferromagnetism is not symmetry-allowed, and therefore there are no orange terms. The remaining black multipoles require the full MSG and therefore represent SOC-induced contributions. The Rashba-like in-plane spin texture is described by the dipoles $Q^{(1)}_x$ and $Q^{(1)}_y$, corresponding to $S_y$ and $S_x$, respectively. In addition, the out-of-plane spin-polarization component $S_z$  identified in the DFT calculations is described by the symmetry-allowed quadrupole $Q^{(2)}_{xy}$, with nodal planes at $k_x=0$ and $k_y=0$ as illustrated in Fig.~\ref{figure2_SML}.

For $\mathbf N\parallel y$ and $\mathbf N\parallel z$, weak ferromagnetism is symmetry allowed and is described by the orange monopole $Q^{(0)}$, corresponding to the spin-polarization components, $S_z$ and $S_y$ respectively. Interestingly, the reduced symmetry of the coplanar state also admits additional momentum-dependent contributions. In particular, the orange dipole $Q^{(1)}_y$, contributing to  $S_x$, is already permitted by the coplanar spin groups. Therefore, although $Q^{(1)}_y$ produces a linear-in-momentum spin polarization, it is not a purely relativistic contribution: once the coplanar magnetic configuration is established, this dipolar component can exist even in the absence of SOC. Nevertheless, SOC is required to stabilize the weakly ferromagnetic configuration, and in this sense  $Q^{(1)}_y$ can still be regarded as a relativistic contribution.
The same symmetry reduction also admits the orange diagonal quadrupoles $Q^{(2)}_{xx}$, $Q^{(2)}_{yy}$, and $Q^{(2)}_{zz}$, corresponding to spin-polarization components with quadratic momentum dependence proportional to $k_x^2$, $k_y^2$, and $k_z^2$, respectively. 

Finally, for these two Néel-vector orientations, the remaining black multipoles require the full MSG and therefore represent purely SOC-induced contributions. The quadrupoles $Q^{(2)}_{xz}$ and $Q^{(2)}_{xy}$ account for the additional in-plane spin-polarization component $S_x$ found for $\mathbf N\parallel y$ and $\mathbf N\parallel z$, respectively.

Not all symmetry-allowed multipoles in Table~\ref{tab:spinmomentumlockingallowed} can be unambiguously identified in the DFT calculations of the previous section. Some may be too small to be resolved numerically, while others may be masked by larger contributions with similar momentum dependence. Nevertheless, the DFT spin textures are consistent with the multipolar structure predicted by the symmetry analysis.
%Nevertheless, the excellent overall agreement demonstrates that combining symmetry analysis with first-principles calculations provides a powerful framework for disentangling the different microscopic contributions to complex spin textures.

\begin{figure}[t!]
    \centering
    \includegraphics[width=0.9\columnwidth,angle=0]{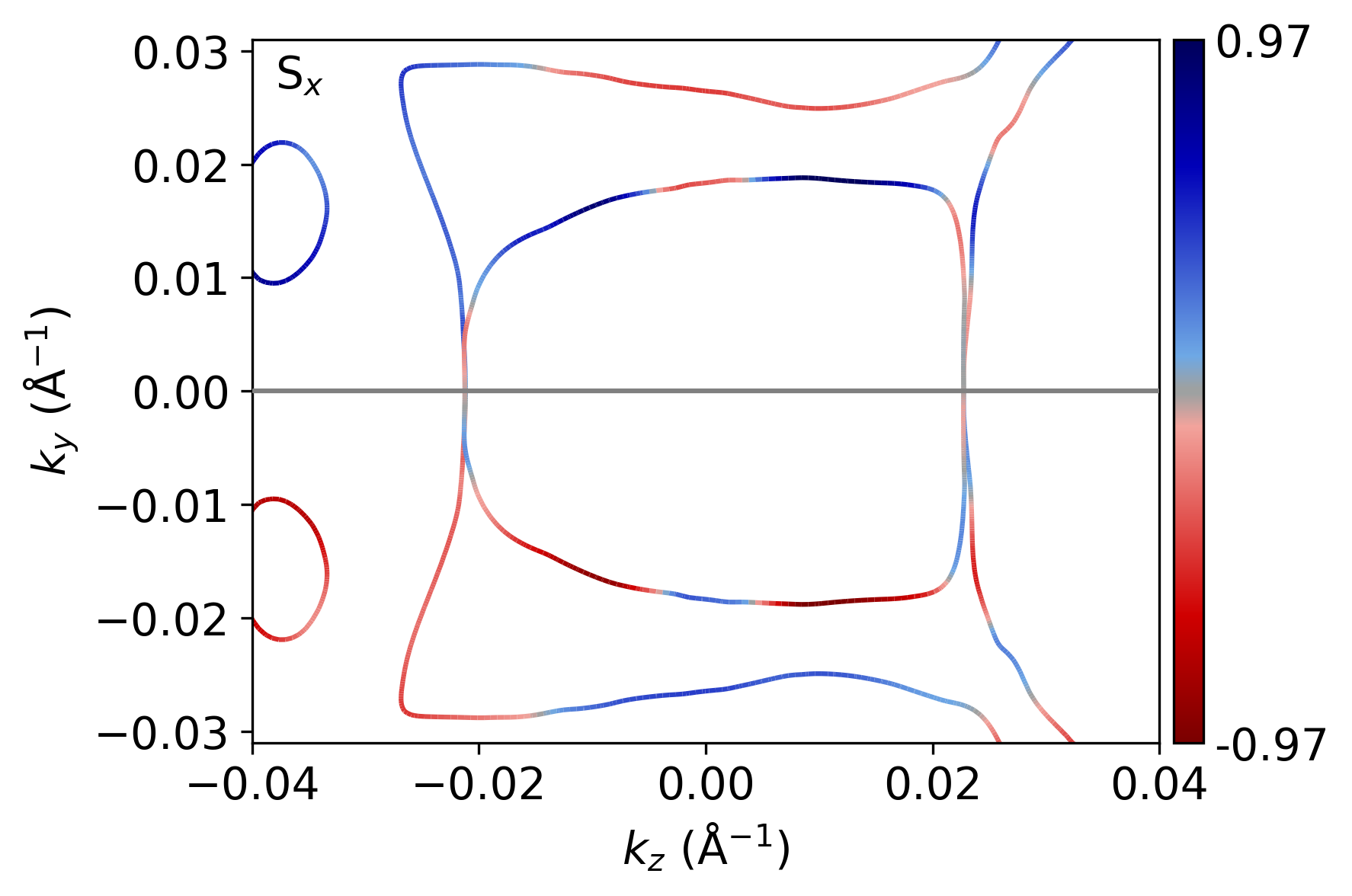}
    \caption{Two dimensional constant-energy contours for $\mathbf N \parallel x$ and spin-polarization component $S_x$, in the $k_x$=0 plane and at -1.7 eV from the top of the valence band, showing the $p_y$-wave spin-momentum locking character.}
    \label{fig:pwave}
\end{figure}

Beyond identifying the individual multipolar contributions, the symmetry analysis also provides a direct interpretation of the topology of the relativistic spin texture. For the dominant spin-polarization component $S_x$ for $\mathbf N \parallel x$, the SML is given by the superposition $Q^{(2)}_{yz}+Q^{(1)}_y$.
Consequently, the nodal plane $k_y=0$, common to both multipoles, is preserved, whereas the nodal plane $k_z=0$, belonging only to $Q^{(2)}_{yz}$, is no longer symmetry-protected. The resulting spin texture retains a single symmetry-protected nodal plane and therefore exhibits an effective $p_y$-wave character\cite{Fakhredine25b}, shown in Fig. \ref{fig:pwave}, also referred to as a hybrid $p/d$-wave magnet\cite{leon2025strainenhancedaltermagnetismca3ru2o7,luo2026unconventionalmagnetismsymmetryclassification}.

% S2

By contrast, for the dominant spin-polarization component $S_y$ for $\mathbf N \parallel y$, the SML takes the form $Q^{(2)}_{yz}+Q^{(1)}_x$. In this case, neither of the nodal planes of the nonrelativistic contribution remains symmetry-protected, resulting in a spin texture without symmetry-enforced nodal planes \cite{gong2026symmetryprotectednodalplanesaccidental,luo2026unconventionalmagnetismsymmetryclassification}.

Overall, these results demonstrate that combining symmetry analysis with first-principles calculations provides a systematic framework for disentangling the different microscopic contributions to the complex spin textures. As shown next, the same framework can be extended to spin-dependent response functions, establishing a connection with experiment.

\begin{table*}[t]
\centering

\begin{minipage}{0.48\textwidth}
\centering
\textbf{(a) LP shift conductivity, $\sigma_{\mathrm{LP}}$}

\vspace{0.15cm}

\scalebox{0.72}{%
\begin{tabular}{|c|c|c|c|}
\hline
 & \multicolumn{3}{|c|}{Spin projection} \\
\cline{2-4}
 & $S_x$ & $S_y$ & $S_z$ \\
\hline
$\mathbf N\parallel x$
& \textcolor{purple}{xxy, yxx, yyy, yzz, zyz}
& xxx, xyy, xzz, yxy, zxz
& xyz, yxz, zxy \\
\hline
$\mathbf N\parallel y$
& xxx, xyy, xzz, yxy, zxz
& \textcolor{purple}{xxy, yxx, yyy, yzz, zyz}
& \textcolor{orange}{xxz, yyz, zxx, zyy, zzz} \\
\hline
$\mathbf N\parallel z$
& xyz, yxz, zxy
& \textcolor{orange}{xxz, yyz, zxx, zyy, zzz}
& \textcolor{purple}{xxy, yxx, yyy, yzz, zyz} \\
\hline
\end{tabular}%
}
\end{minipage}
\hfill
\begin{minipage}{0.48\textwidth}
\centering
\textbf{(b) CP injection conductivity, $\eta_{\mathrm{CP}}$}

\vspace{0.15cm}

\scalebox{0.72}{%
\begin{tabular}{|c|c|c|c|}
\hline
 & \multicolumn{3}{|c|}{Spin projection} \\
\cline{2-4}
 & $S_x$ & $S_y$ & $S_z$ \\
\hline
$\mathbf N\parallel x$
& \textcolor{purple}{xxy, zyz}
& yxy, zxz
& xyz, yxz, zxy \\
\hline
$\mathbf N\parallel y$
& yxy, zxz
& \textcolor{purple}{xxy, zyz}
& \textcolor{orange}{xxz, yyz} \\
\hline
$\mathbf N\parallel z$
& xyz, yxz, zxy
& \textcolor{orange}{xxz, yyz}
& \textcolor{purple}{xxy, zyz} \\
\hline
\end{tabular}%

}
\end{minipage}

\vspace{0.15cm}

%\caption{
%Symmetry-allowed components of the time-reversal-odd spin photoconductivities:
%(a) LP shift and (b) CP injection.
%Purple entries survive in the collinear spin-group limit, orange entries are additionally allowed in the coplanar spin-group limit, and black entries require the full MSG.
%The TR-odd tensors vanish in the paramagnetic phase.
%}
%\label{tab:TRodd_spinphotoconductivities}

%\end{table*}
%\begin{table*}[t]
%\centering

\begin{minipage}{0.48\textwidth}
\centering
\textbf{(c) CP shift conductivity, $\sigma_{\mathrm{CP}}$}

\vspace{0.15cm}

\scalebox{0.72}{%
\begin{tabular}{|c|c|c|c|}
\hline
 & \multicolumn{3}{|c|}{Spin projection} \\
\cline{2-4}
 & $S_x$ & $S_y$ & $S_z$ \\
\hline
$\mathbf{N} \parallel x$ & xxy,zyz & yxy, zxz & xyz, yxz, zxy \\
\hline
$\mathbf{N} \parallel y$ & \textcolor{orange}{xxy, zyz} & yxy, zxz & xyz, yxz, zxy \\
\hline
$\mathbf{N} \parallel z$ & \textcolor{orange}{xxy, zyz} & yxy, zxz & xyz, yxz, zxy \\
\hline
%Paramagnetic & xxy,zyz & yxy, zxz & xyz, yxz, zxy \\
%\hline
\end{tabular}
}
\end{minipage}
\hfill
\begin{minipage}{0.48\textwidth}
\centering
\textbf{(d) LP injection conductivity, $\eta_{\mathrm{LP}}$}

\vspace{0.15cm}

\scalebox{0.72}{%
\begin{tabular}{|c|c|c|c|}
\hline
 & \multicolumn{3}{|c|}{Spin projection} \\
\cline{2-4}
 & $S_x$ & $S_y$ & $S_z$ \\
\hline
$\mathbf{N} \parallel x$ & xxy, yxx, yyy, yzz, zyz & xxx, xyy, xzz, yxy, zxz & xyz, yxz, zxy \\
\hline
 $\mathbf{N} \parallel y$ & \textcolor{orange}{xxy, yxx, yyy, yzz, zyz} & xxx, xyy, xzz, yxy, zxz & xyz, yxz, zxy \\
\hline
$\mathbf{N} \parallel z$ & \textcolor{orange}{xxy, yxx, yyy, yzz, zyz} & xxx, xyy, xzz, yxy, zxz & xyz, yxz, zxy \\
\hline
%Paramagnetic & \textcolor{orange}{xxy, yxx, yyy, yzz, zyz} & xxx, xyy, xzz, yxy, zxz & xyz, yxz, zxy \\
%\hline
\end{tabular}
}
\end{minipage}

\vspace{0.15cm}

\caption{
Symmetry-allowed elements of spin photoconductivities:
(a) LP shift, (b) CP injection (c) CP shift, and (d) LP injection.
Purple entries survive in the collinear spin-group limit, orange entries are additionally allowed in the coplanar spin-group limit, and black entries require the full MSG.
%The TR-odd tensors vanish in the paramagnetic phase.
}
\label{tab:TRodd_spinphotoconductivities}

\end{table*}

\subsection{Spin photogalvanic effect}

The second-order dc spin photocurrent density driven by an optical electric field 
$\mathbfcal{E}(\omega)$ of frequency $\omega$ is

\begin{equation}\begin{split}
j^{s,a}= 2[ \sigma^{s,abc}_\mathrm{LP}(\omega)+ \tau \eta^{s,abc}_\mathrm{LP}(\omega)]\Re[\mathcal{E}^b(\omega)\mathcal{E}^c(-\omega)]\\
+2i[ \sigma^{s,abc}_\mathrm{CP}(\omega)+ \tau \eta^{s,abc}_\mathrm{CP}(\omega)]\Im[\mathcal{E}^b(\omega)\mathcal{E}^c(-\omega)].
\end{split}\label{eq.j}
\end{equation}

Here, $s,a,b,c =x,y,z$, where $s$ denotes the spin component of the current, $a$ its propagation direction, and $b,c$ the Cartesian components of the electric field. The real and imaginary terms in Eq. (\ref{eq.j}) describe the photogalvanic effect under linearly and circularly polarized (LP and CP) light, respectively. The photoconductivity, a rank-four tensor, is decomposed into two contributions, the shift and injection photoconductivities, $\sigma_\mathrm{LP(CP)}$ and $\eta_\mathrm{LP(CP)}$. The shift contribution is intrinsic, and therefore independent of scattering, whereas the injection contribution is proportional to the spin lifetime, denoted by $\tau$. Physically, the shift current originates from interband coherence and corresponds to the spin-dependent displacement of the electronic wavepacket during an optical transition. By contrast, the injection current arises from an asymmetric photoexcitation of carriers in momentum space, resulting in a net spin current.

The symmetry-allowed photoconductivity elements are reported in Table~\ref{tab:TRodd_spinphotoconductivities}. They can be classified using the group--subgroup relations between the spin groups and MSGs established in Sec.~\ref{sec.symmetry}. As for the spin texture, this hierarchy separates components that survive in the collinear spin-group limit and are therefore of nonrelativistic altermagnetic origin, components allowed by the coplanar spin group and associated with weak ferromagnetism, and purely relativistic components that require the full MSG. Thus, the same color coding introduced for the spin multipoles is used in the table.

The four spin photoresponses can be further classified according to their behavior under time reversal. The spin LP shift and CP injection conductivities are odd under time reversal, whereas the spin LP injection and CP shift conductivities are even (see Appendix~\ref{app.spin_cond} for their transformation properties in Jahn notation). %This classification is particularly relevant when comparing the different N\'eel-vector orientations. 
The MSGs corresponding to the different N\'eel-vector orientations share the same spatial symmetry operations, but differ in whether these enter as unitary operations $g$ or as antiunitary operations $g\mathcal{T}$, where $\mathcal{T}$ denotes time reversal. For the time-reversal-even responses, $\mathcal{T}$ introduces no additional sign in the transformation law, so that $g$ and $g\mathcal{T}$ impose the same invariance condition. Consequently, changing the N\'eel-vector orientation does not change the set of symmetry-allowed tensor components. For the time-reversal-odd responses, instead, $\mathcal{T}$ introduces an additional sign, so that $g$ and $g\mathcal{T}$ impose different invariance conditions. Consequently, the set of symmetry-allowed tensor components can depend on the N\'eel-vector orientation.

The two time-reversal classes also differ in their spin-group decomposition: the time-reversal-odd responses retain contributions in the nonrelativistic collinear spin-group limit, whereas the time-reversal-even responses contain only contributions associated with the weak-ferromagnetic coplanar state and the full MSG. We examine the four photoresponses individually below.

\begin{figure*}[t!]
\centering
\parbox{\textwidth}{%
  \centering
  \includegraphics[width=0.32\textwidth]{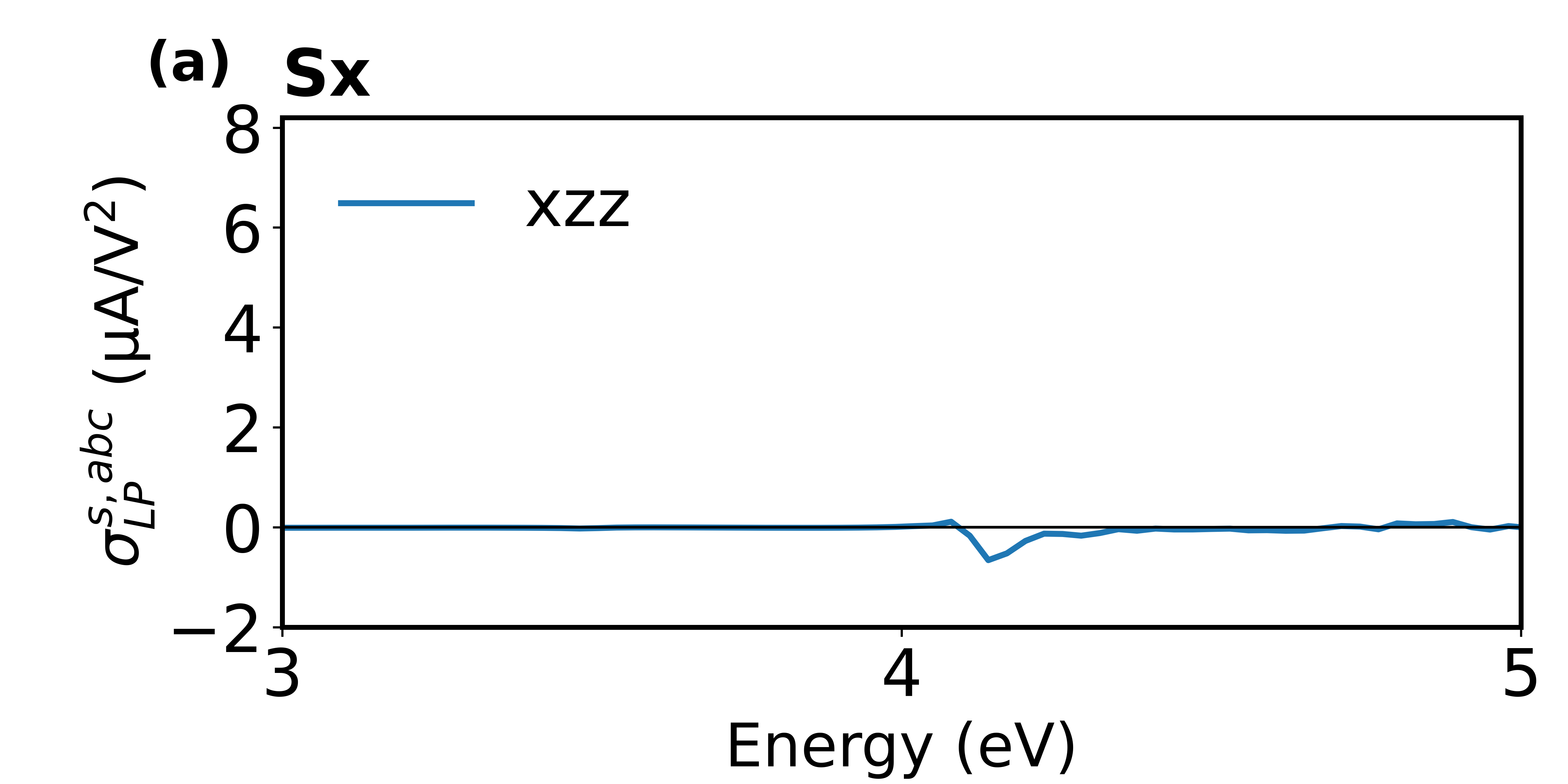}\hfill
  \includegraphics[width=0.32\textwidth]{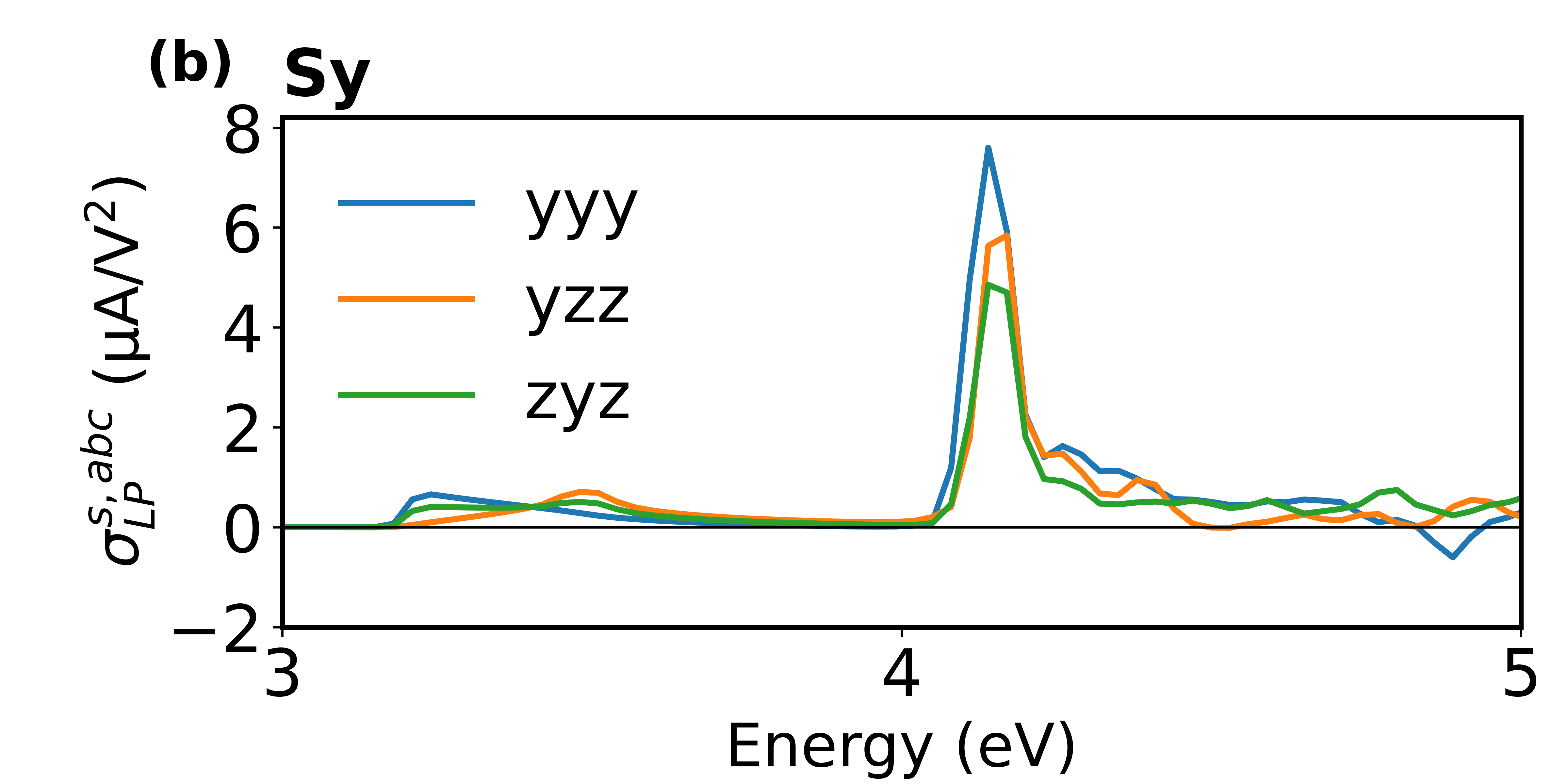}\hfill
  \includegraphics[width=0.32\textwidth]{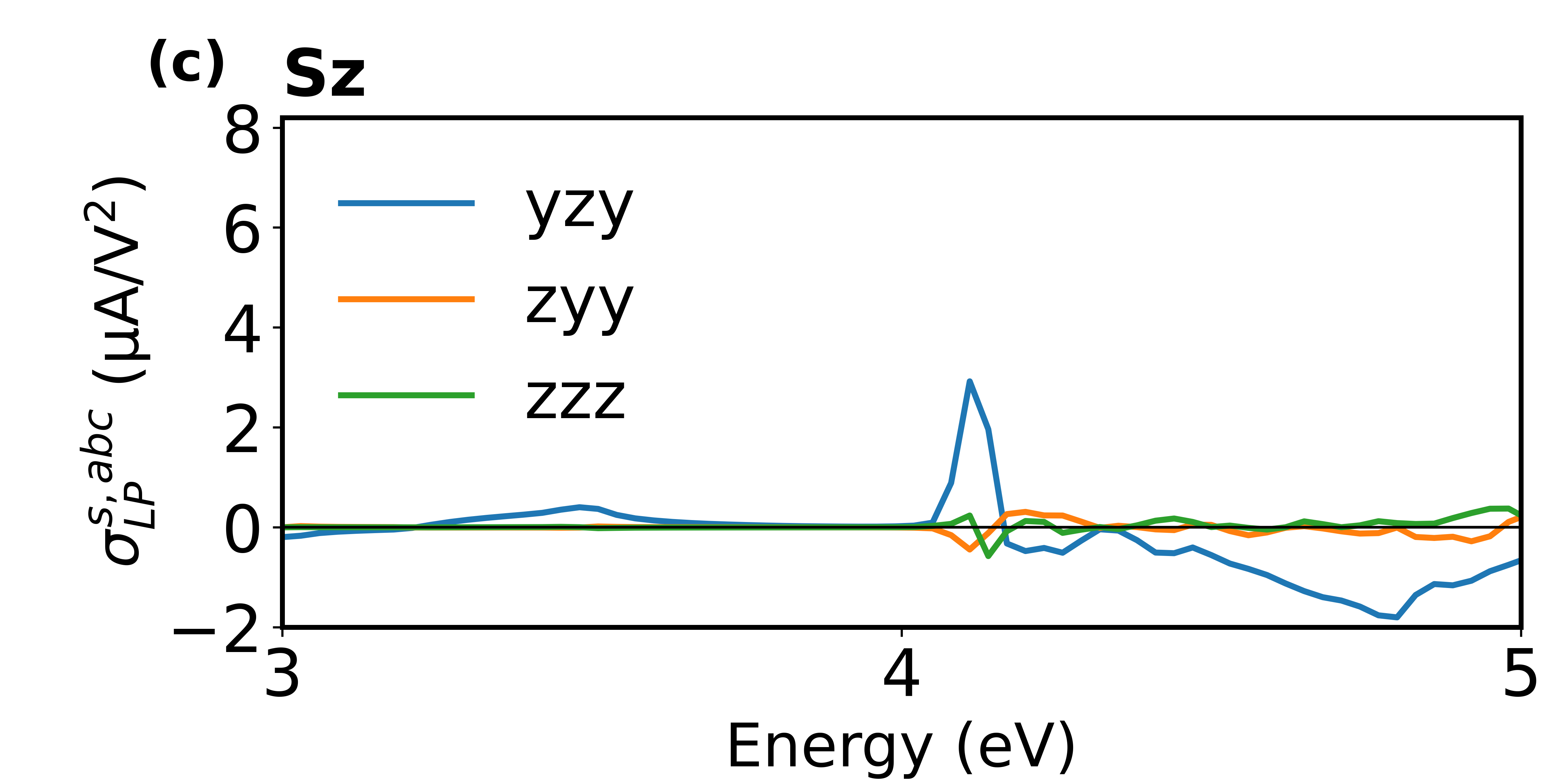}%
}
\vspace{1mm}
\parbox{\textwidth}{%
  \centering
  \includegraphics[width=0.32\textwidth]{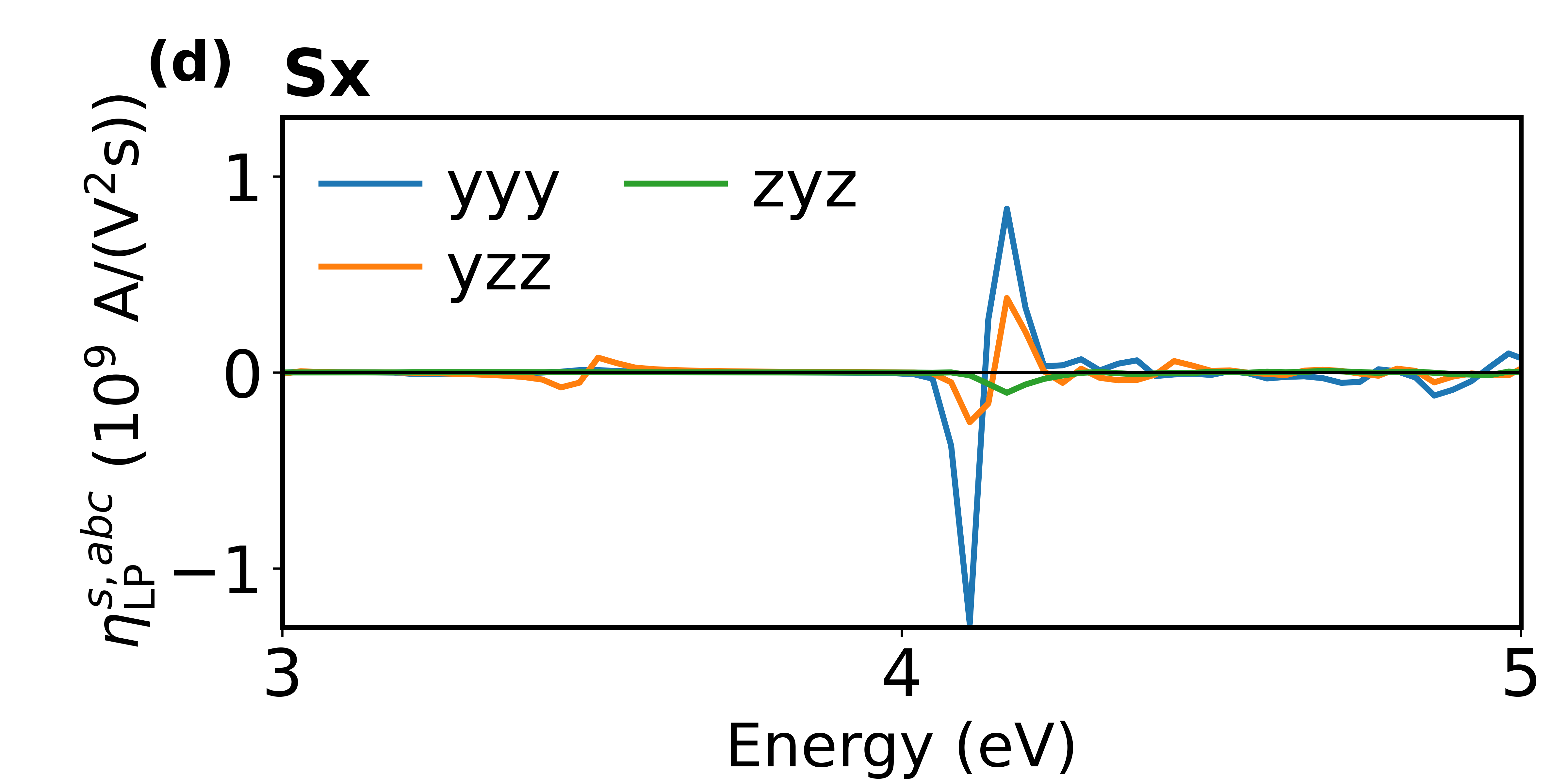}\hfill
  \includegraphics[width=0.32\textwidth]{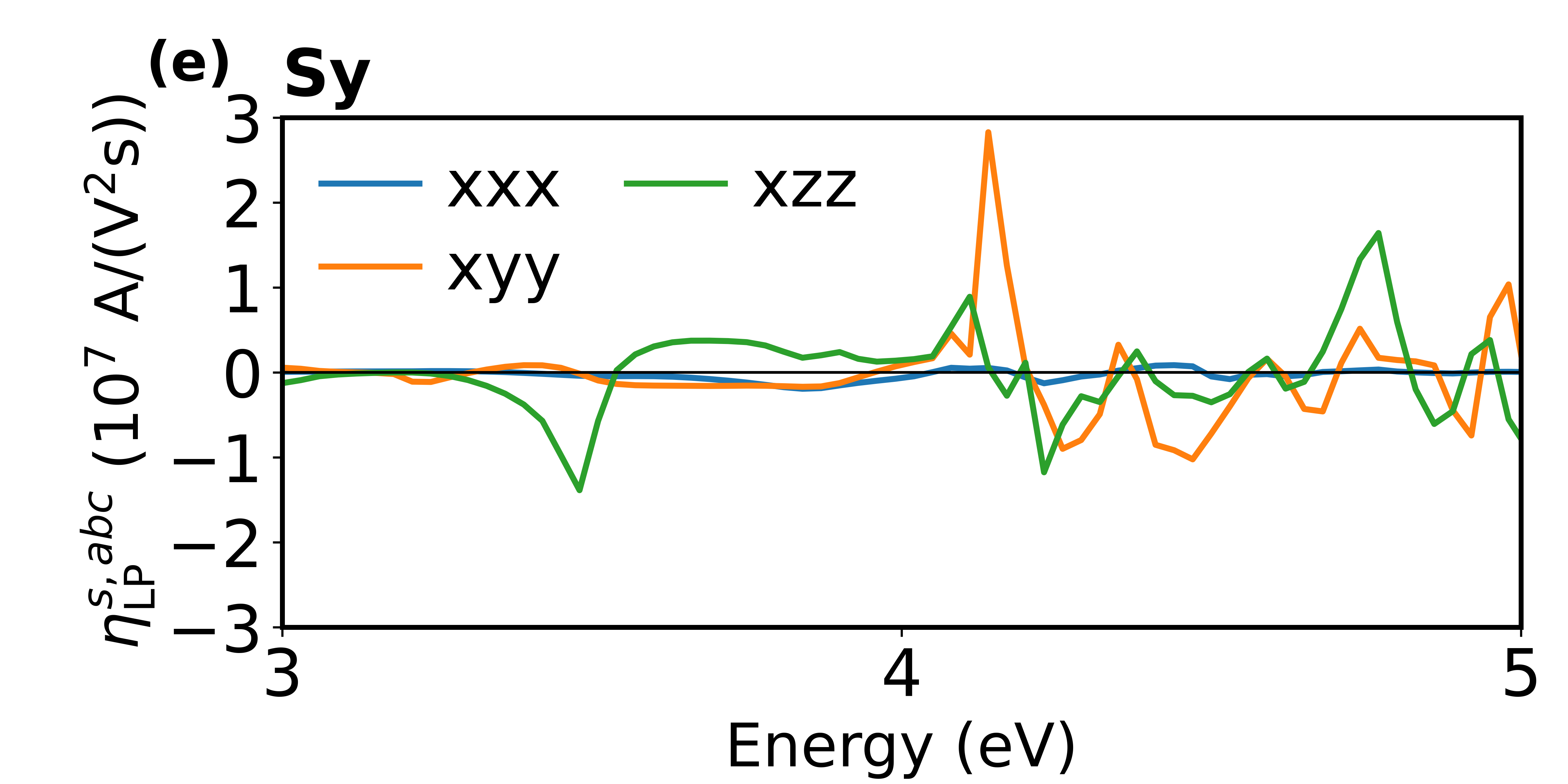}\hfill
  \includegraphics[width=0.32\textwidth]{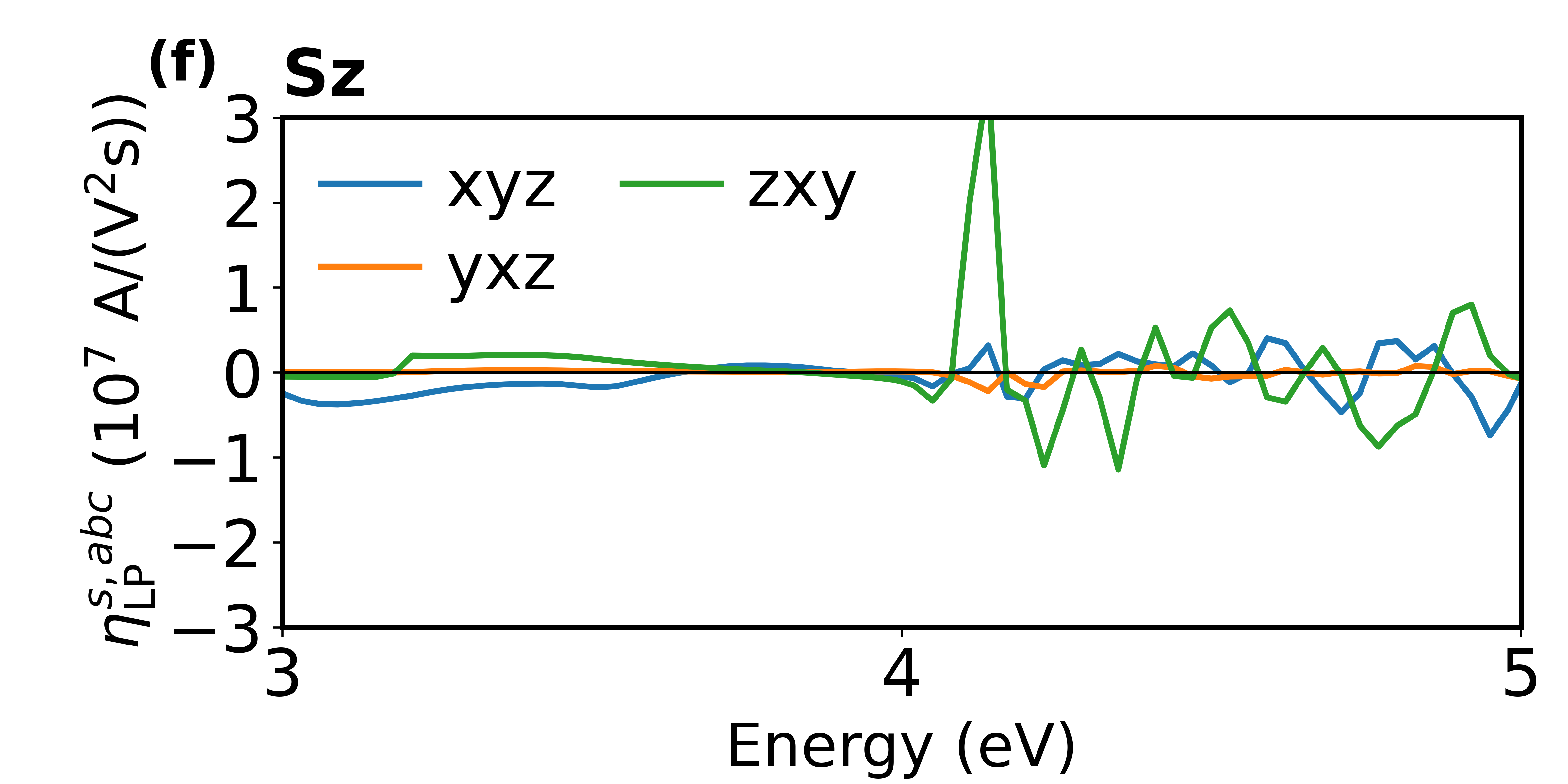}%
}
\caption{Main components of the LP spin photoconductivity for $\mathbf N\parallel y$, resolved by spin-polarization direction. Panels (a)--(c) show the shift photoconductivity, while panels (d)--(f) show the injection photoconductivity. The three columns correspond to the spin polarizations $S_x$, $S_y$, and $S_z$, respectively. For the injection photoconductivity, note the different vertical-axis scale in panel (d) compared with panels (e) and (f).} 
\label{spinphotocurrents}
\end{figure*}

\subsubsection{LP shift conductivity}

The symmetry-allowed elements of the spin shift conductivity for LP
light are listed in Table~\ref{tab:TRodd_spinphotoconductivities}(a).
%Since this tensor is odd under time reversal, the set of allowed elements depends on the orientation of the N\'eel vector. The color coding reflects the group--subgroup hierarchy introduced for the SML: purple elements are allowed in the collinear spin-group limit, orange elements emerge in the coplanar spin-group limit, and black elements require the full MSG. 
The first-principles calculations reproduce all the symmetry-allowed elements for the three N\'eel-vector
orientations. Figure~\ref{spinphotocurrents}(a)--(c) shows the
calculated spectra for the easy-axis configuration
$\mathbf N\parallel y$, while the results for
$\mathbf N\parallel x$ and $\mathbf N\parallel z$ are reported in the
SM.

We first consider the purple elements, whose spin polarization is
parallel to the N\'eel vector and which represent the altermagnetic contribution. These elements
remain finite in the collinear configuration when SOC is switched off in the DFT calculations
(see SM), confirming their nonrelativistic origin. They also provide
the largest spin shift response: for $\mathbf N\parallel y$, $\sigma_{\mathrm{LP}}^{y,yyy}$ and
$\sigma_{\mathrm{LP}}^{y,yzz}$ reach approximately
$8~\mu\text{A}/\text{V}^2$ and $6~\mu\text{A}/\text{V}^2$,
respectively, near $\hbar\omega=4.2$~eV
[Fig.~\ref{spinphotocurrents}(b)]. These peaks occur in the ultraviolet,
well above the fundamental absorption onset, and are consistent with
optical transitions from the strongly spin-polarized valence bands at
$-1.5\lesssim E\lesssim-1.2$~eV to the conduction band near $2.8$~eV.
For both tensor elements, the spin current flows along $y$, parallel to
both its spin polarization and the N\'eel vector.

We next consider the orange elements, whose spin polarization is
parallel to the weak-ferromagnetic magnetization: along $z$ for
$\mathbf N\parallel y$ and along $y$ for
$\mathbf N\parallel z$. Consistent with their classification in the coplanar
spin-group limit, these elements remain finite when SOC is
switched off in the DFT calculations while the canted magnetic configuration is retained, but
vanish when the canting and net magnetization are removed. Thus, although they do not explicitly require SOC
once the canted state has been established, their physical realization
relies on SOC to stabilize weak ferromagnetism. The largest of these
elements, $\sigma_{\mathrm{LP}}^{z,yzy}$, reaches approximately
$3~\mu\text{A}/\text{V}^2$ near $\hbar\omega=4.2$~eV
[blue curve in Fig.~\ref{spinphotocurrents}(c)]. Remarkably, despite the small
weak-ferromagnetic moment, this response remains comparable in
magnitude to the dominant nonrelativistic contributions, showing that
weak spin canting can generate sizable spin photocurrents.

Finally, the black elements require the full MSG and describe the
purely relativistic contribution. For $\mathbf N\parallel y$, their
spin polarization is along $x$, and they vanish when SOC is switched
off, confirming their relativistic origin. They are also the smallest
calculated elements: $\sigma_{\mathrm{LP}}^{x,xzz}$ reaches
approximately $0.6~\mu\text{A}/\text{V}^2$
[Fig.~\ref{spinphotocurrents}(a)], about one order of magnitude below
the largest nonrelativistic peaks.

The LP spin shift conductivity therefore reproduces the hierarchy
identified in the equilibrium spin texture. The nonrelativistic
altermagnetic contribution dominates, the contribution associated with
weak ferromagnetism is smaller but remains sizable, and the purely
SOC-induced contribution is substantially weaker.

%\begin{figure}[t!]
%\centering
%\includegraphics[width=0.5\textwidth]{Main_M_b_Spin_shift_Sx.png}\\[4pt]
%\includegraphics[width=0.5\textwidth]{Main_M_b_Spin_shift_Sy.png}\\[4pt]
%\includegraphics[width=0.5\textwidth]{Main_M_b_Spin_shift_Sz.png}
%\caption{Main components of the spin linear shift current for the Néel vector $\mathbf{N}\parallel$ $y$ and for the different spin projections S$_x$, S$_y$ and S$_z$.}
%\label{fig:spin_linear_shift_M_b}
%\end{figure}

\subsubsection{LP injection conductivity}

The symmetry-allowed elements of the LP spin injection conductivity are listed in Table~\ref{tab:TRodd_spinphotoconductivities}(d). Because this response is even under time reversal, the same elements are allowed for all three N\'eel-vector orientations, although their classification within the group--subgroup hierarchy depends on $\mathbf N$. Unlike the LP shift conductivity, the LP injection conductivity has no contribution in the collinear spin-group limit. It therefore separates into a coplanar sector (orange) and a purely SOC-induced sector requiring the full MSG (black). For $\mathbf N\parallel x$, weak ferromagnetism is forbidden and the magnetic state remains collinear, so all allowed elements belong to the latter sector.

For $\mathbf N\parallel y$ and $\mathbf N\parallel z$, the two sectors can be distinguished by their spin polarization. Injection conductivity  elements with spin component $S_x$ are allowed in the coplanar spin-group limit and can be associated with the subdominant $Q_y^{(1)}$ dipole of the SML, whereas those with spin components $S_y$ and $S_z$ require the full MSG. The first-principles calculations confirm this classification: when SOC is switched off while retaining the coplanar magnetic configuration, the elements with spin component $S_x$ remain finite, whereas the elements with spin component $S_y$ and $S_z$ vanish.

For the easy-axis configuration $\mathbf N\parallel y$, the calculated spectra are shown in Fig.~\ref{spinphotocurrents}(d)--(f). The response is dominated by the coplanar sector. In particular, $\eta_{\mathrm{LP}}^{x,yyy}$, describing a spin current flowing along $y$ with spin polarization along $x$ under $y$-polarized light, reaches approximately $10^{9}~\mathrm{A}/(\mathrm{V}^{2}\mathrm{s})$ near $\hbar\omega=4.2$~eV [blue curve in Fig.~\ref{spinphotocurrents}(d)]. This value is approximately 30 times larger than those of the largest purely SOC-induced elements, among which $\eta_{\mathrm{LP}}^{y,xyy}$ and $\eta_{\mathrm{LP}}^{z,zxy}$ reach peak values of approximately $3\times10^{7}~\mathrm{A}/(\mathrm{V}^{2}\mathrm{s})$ [Fig.~\ref{spinphotocurrents}(e),(f)].

Thus, whereas the LP shift conductivity is dominated by the nonrelativistic altermagnetic sector, the LP injection conductivity is dominated by the coplanar sector associated with the subdominant $Q_y^{(1)}$ dipole of the spin texture.

\subsubsection{CP shift and injection conductivities}

For CP light, the time-reversal parities of the shift and injection responses are interchanged relative to LP light. The time-reversal-even CP shift conductivity therefore follows the same group--subgroup hierarchy as the LP injection conductivity, whereas the time-reversal-odd CP injection conductivity follows that of the LP shift conductivity. Their symmetry-allowed elements are listed in Table~\ref{tab:TRodd_spinphotoconductivities}(c) and (b), respectively.

Accordingly, the CP shift conductivity has no contribution in the collinear spin-group limit. For $\mathbf N\parallel x$, all its allowed elements require the full MSG and are purely SOC-induced. For $\mathbf N\parallel y$ and $\mathbf N\parallel z$, elements with spin component $S_x$ are instead allowed in the coplanar spin-group limit and are associated with the $Q^{(1)}_y$ dipole of the spin texture, while the remaining elements require the full MSG. The calculated response is dominated by this coplanar sector: $\sigma_{\mathrm{CP}}^{x,zyz}$ reaches maximum absolute values of approximately $6~\mu\text{A}/\text{V}^2$ and $3~\mu\text{A}/\text{V}^2$ for $\mathbf N\parallel y$ and $\mathbf N\parallel z$, respectively. By comparison, the purely SOC-induced elements with spin components $S_y$ and $S_z$ reach only about $0.2~\mu\text{A}/\text{V}^2$.

The CP injection conductivity contains contributions from all three levels of the hierarchy. The collinear spin-group sector has spin polarization parallel to the N\'eel vector and represents the nonrelativistic altermagnetic contribution. For $\mathbf N\parallel y$ and $\mathbf N\parallel z$, the coplanar sector has spin polarization parallel to the weak-ferromagnetic moment, while the remaining elements are purely SOC-induced. The calculated response is dominated by the nonrelativistic sector: $\eta_{\mathrm{CP}}^{s,xxy}$ and $\eta_{\mathrm{CP}}^{s,zyz}$, with $s=x,y,z$ for $\mathbf N\parallel x,y,z$, respectively, reach peak absolute values of the order of $10^{8}~\mathrm{A}/(\mathrm{V}^{2}\mathrm{s})$. The coplanar and purely SOC-induced elements are approximately one order of magnitude smaller, with maximum absolute values of the order of $10^{7}~\mathrm{A}/(\mathrm{V}^{2}\mathrm{s})$. The complete CP spectra are reported in the SM.

Thus, the exchange of time-reversal parity between the shift and injection responses is reflected in their dominant contributions: the CP shift conductivity, like the LP injection conductivity, is dominated by the coplanar sector corresponding to the $Q^{(1)}_y$ dipole of the spin texture, whereas the CP injection conductivity, like the LP shift conductivity, is dominated by the nonrelativistic altermagnetic sector.

\begin{figure*}[t!]
\centering
\includegraphics[width=1.0\textwidth]{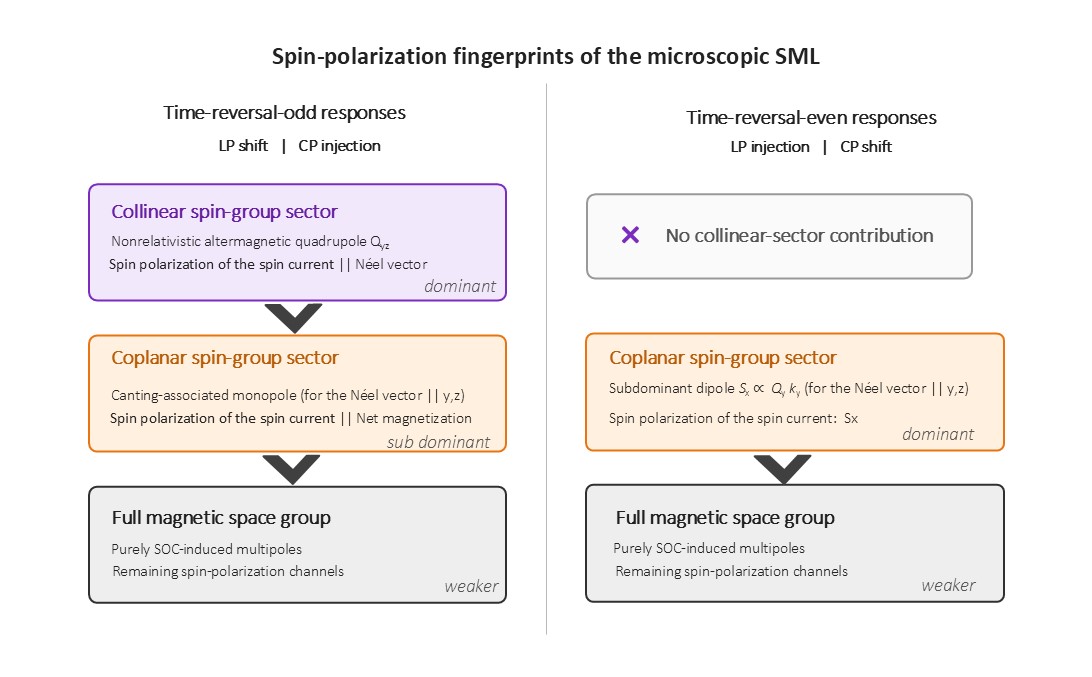}

\caption{Schematic relation between the microscopic contributions to SML and the spin-photogalvanic response. The time-reversal-odd LP shift and CP injection conductivities contain contributions from all three levels of the group--subgroup hierarchy. Spin currents with spin polarization parallel to the N\'eel vector probe the nonrelativistic altermagnetic quadrupole $Q^{(2)}_{yz}$; those polarized along the weak-ferromagnetic moment probe the canting-associated monopole $Q^{(0)}$; and the remaining spin-polarization channels probe purely SOC-induced multipoles. Their calculated magnitudes generally decrease in the same order. The time-reversal-even LP injection and CP shift conductivities contain no contribution from the collinear spin-group sector. For $\mathbf N\parallel y,z$, they are dominated by spin currents with spin component $S_x$, associated with the coplanar $Q^{(1)}_y$ dipole, whereas the remaining components require the full MSG and are purely SOC-induced. }
\label{figure_summary}
\end{figure*}

\section{Discussion and conclusions}

We have established a direct connection between the microscopic SML of the altermagnetic multiferroic BaCuF$_4$ and its nonlinear spin-photogalvanic response, as summarized in Fig. \ref{figure_summary}. By combining first-principles calculations with spin-group and MSG symmetry analysis, we have shown that the spin texture arises from the superposition of distinct spin multipoles with different microscopic origins.

The dominant nonrelativistic contribution is described by the quadrupole $Q^{(2)}_{yz}$, which produces a $d$-wave spin polarization parallel to the N\'eel vector. This contribution originates primarily from the rotation of the CuF$_6$ octahedra. Because this rotation is coupled to the polar distortion, the nonrelativistic SML reverses upon ferroelectric switching, realizing the altermagnetoelectric effect.

SOC introduces additional contributions whose form depends on the orientation of the N\'eel vector. For $\mathbf N\parallel x$, the magnetic state remains collinear, and the relativistic spin texture contains Rashba-like dipoles and additional quadrupolar components. For $\mathbf N\parallel y$ and $\mathbf N\parallel z$, SOC also induces spin canting and weak ferromagnetism, allowing further momentum-dependent multipoles. Importantly, some of these contributions, such as the dipole $Q_y^{(1)}$, are allowed already by the coplanar spin group. They do not explicitly require SOC once the canted configuration has been established, although SOC is required to stabilize that configuration. The momentum dependence of a spin texture alone is therefore insufficient to establish its microscopic origin, which instead follows from the hierarchy between spin-group and MSG symmetries.

This hierarchy is directly reflected in the spin-photogalvanic response. For the time-reversal-odd LP shift and CP injection conductivities, spin currents polarized along the N\'eel vector probe the nonrelativistic altermagnetic sector. In the weakly ferromagnetic configurations, currents polarized along the net magnetization identify the coplanar sector, while the remaining spin-polarization channels probe purely SOC-induced contributions. Their calculated magnitudes follow the same hierarchy: the nonrelativistic contributions are generally the largest, those associated with spin canting remain sizable despite the small weak-ferromagnetic moment, and the purely SOC-induced contributions are typically weaker. The time-reversal-even LP injection and CP shift conductivities provide complementary information and are dominated by the coplanar contribution associated with the $Q^{(1)}_y$ dipole of the spin texture.

Experimentally, the spin photogalvanic response could be detected by converting the optically generated spin current into a charge signal. In a BaCuF$_4$/heavy-metal heterostructure, for example, a Pt, W, or Ta layer could convert the injected spin current into a transverse voltage through the inverse spin Hall effect. Rotating the N\'eel vector, reversing the ferroelectric polarization, or varying the light polarization would allow spin currents polarized along the N\'eel vector, along the weak-ferromagnetic moment, and perpendicular to both to be distinguished. The measured voltage could thus provide an indirect but symmetry-resolved probe of the different microscopic contributions to the spin texture.

The group--subgroup framework developed here is not restricted to BaCuF$_4$. It can be applied more generally to noncentrosymmetric altermagnets and multiferroics in which nonrelativistic altermagnetic SML coexists with Rashba-like, Weyl-like, or persistent relativistic spin textures and with deviations from collinear magnetic order. Our results therefore provide a general route for connecting the microscopic origin of complex spin textures to experimentally accessible nonlinear spin-transport responses.

\begin{acknowledgments}
The authors thank J. Ibañez-Azpiroz and J. Sivianes for useful discussions. This research was supported by the Foundation for Polish Science project “MagTop” no. FENG.02.01-IP.05-0028/23 co-financed by the European Union from the funds of Priority 2 of the European Funds for a Smart Economy Program 2021–2027 (FENG). C. A. was supported by the Polish National Agency for Academic Exchange (NAWA) under the Bekker Programme, grant no. BPN/BEK/2025/1/00244/DEC/1.
We further acknowledge access to the computing facilities of the Interdisciplinary Center of Modeling at the University of Warsaw, Grant g91-1418, g91-1419, g96-1808, g96-1809 and g103-2540 for the availability of high-performance computing resources and support. We acknowledge HPC resources and support provided by CINECA under the ISCRA IsB28 HEXTIM, IsCc2 SFERA,  IsCc9 BRIMS, IsCc9 IDMVW and IsCd7 CSPA projects. We acknowledge the access to the computing facilities of the Poznan Supercomputing and Networking Center, Grants No. pl0267-01, pl0365-01 and pl0471-01.
\end{acknowledgments}

\appendix
\section{Tensor transformation properties in Jahn notation}

\subsection{Spin multipoles}\label{app.spin_mult}
The spin polarization transforms as an axial vector that is odd under time reversal, with Jahn symbol $aeV$, whereas the crystal momentum transforms as a time-reversal-odd polar vector. It follows that the multipolar coefficients $Q^{(n)}_{i\alpha\beta\gamma\ldots}$ are characterized by the Jahn symbol $a^{n+1}eV[V^n]$. Accordingly, the zeroth-order (monopole) term transforms as the net magnetization, while the first-order (dipole) and second-order (quadrupole) terms are characterized by $eV^2$ and $aeV[V^2]$, respectively.

In this notation, each $V$ represents a vector index, with the total number of $V$s specifying the tensor rank. Square brackets $[]$ denote symmetrization over the enclosed indices, whereas curly brackets $\{\}$ denote antisymmetrization. The symbols $a$ and $e$ encode additional transformation properties: $a$ indicates odd parity under time reversal, while $e$ denotes axial character.

These transformation properties determine the components of each spin multipole directly from the relevant spin group or magnetic space group, thereby providing a systematic framework for distinguishing nonrelativistic and relativistic contributions to the spin-momentum locking.

\begin{table}[h!]
\centering
\caption{Jahn symbols for the second-order spin photoconductivities.}
\begin{tabular}{c|c|c}
\hline
\hline
conductivity & LP & CP \\
\hline
Shift & $aeVV[V^2]$ & $eVV\{V^2\}$ \\
Injection & $eVV[V^2]$ & $aeVV\{V^2\}$ \\
\hline
\hline
\end{tabular}
\label{tab:Jahn_photo}
\end{table}

\subsection{Second-order spin photoconductivity}\label{app.spin_cond}

The Jahn symbols of the second-order spin-photoconductivity tensors can be constructed directly from the transformation properties of their indices. %We denote the spin photoconductivity by $\sigma^{s}_{abc}$, where $a$ specifies the spatial direction of the spin current, $s$ its spin-polarization direction, and $b$ and $c$ the polarization directions of the two optical fields. 
The current direction is polar, whereas the spin polarization is axial, giving the factor $eVV$ in the Jahn notation: the two $V$ factors correspond to the current and spin-polarization indices, respectively, while $e$ accounts for the axial character associated with the latter.

The two optical-field indices provide the remaining tensorial factor. For LP light, they enter symmetrically under interchange of their indices and therefore contribute $[V^2]$. For CP light, they enter antisymmetrically and contribute $\{V^2\}$. The tensorial structures of the LP and CP spin-photoconductivity tensors are thus $eVV[V^2]$ and $eVV\{V^2\}$, respectively.

The spin LP shift and CP injection conductivities are odd under time reversal and therefore carry the additional factor $a$, whereas the spin LP injection and CP shift conductivities are time-reversal even and do not. Together with the axial character and index symmetries discussed above, these time-reversal properties yield the Jahn symbols summarized in Table~\ref{tab:Jahn_photo}.

%The factor $a$ is determined by the time-reversal parity of the conductivity. The spin LP shift and CP injection conductivities are odd under time reversal and therefore acquire an additional factor $a$, whereas the spin LP injection and CP shift conductivities are time-reversal even and do not. Combining the axial character, tensor-index symmetries, and time-reversal parity gives the transformation properties summarized in Table~\ref{tab:Jahn_photo}.

As for the spin multipoles, these transformation properties determine the symmetry-allowed spin-photoconductivity components. Comparing the constraints imposed by the nonrelativistic spin group and the corresponding magnetic space group then allows these components to be distinguished according to their nonrelativistic or relativistic origin.

\bibliography{references}

\clearpage
\onecolumngrid

\begin{center}
\textbf{\large Supplemental Material}
\end{center}
\vspace{0.5cm}

\setcounter{section}{0}
\setcounter{subsection}{0}
\setcounter{figure}{0}
\setcounter{table}{0}
\setcounter{equation}{0}
\renewcommand{\thesection}{S\arabic{section}}
\renewcommand{\thefigure}{S\arabic{figure}}
\renewcommand{\thetable}{S\arabic{table}}
\renewcommand{\theequation}{S\arabic{equation}}

\makeatletter
\let\@sectioncntformat\@seccntformat
\def\@hangfrom@section#1#2#3{\@hangfrom{#1#2}\MakeTextUppercase{#3}}
\def\thesubsection{\Alph{subsection}}
\makeatother

\section{Density of states}

\begin{figure}[h]
\centering
\includegraphics[width=6.6cm,angle=270]{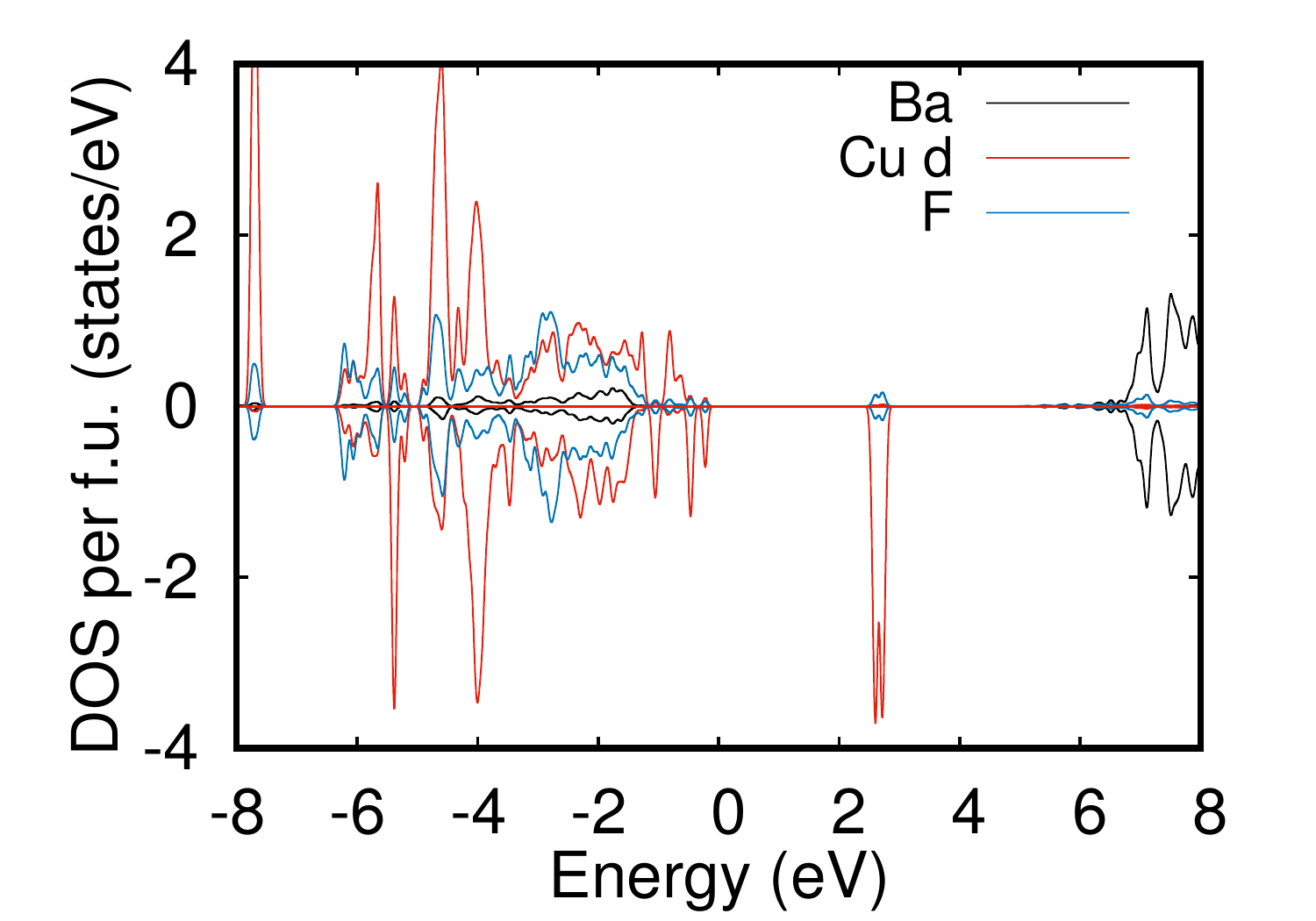}
\caption{Atom-projected DOS of BaCuF$_4$ in the $Cmc2_1$ ground state. Positive and negative values correspond to the spin-up and spin-down channels, respectively, and the energy is referred to the Fermi level. The DOS is projected onto Ba, Cu $d$-orbitals and F states.} %(b) Cu $d$-orbital-resolved DOS, decomposed into the $d_{xy}$, $d_{yz}$, $d_{z^2}$, $d_{xz}$ and $d_{x^2-y^2}$ components.}
\label{DOS} %figure S1
\end{figure}

Figure~\ref{DOS} shows the atom-projected density of states (DOS) of BaCuF$_4$ in the $Cmc2_1$ phase. The system is insulating, as also observed in the band structure discussed in the main text. The valence states are primarily derived from Cu $d$ orbitals hybridized with F $p$-states, while Ba states lie well above the Fermi level and contribute mainly to the high-energy conduction region. The inequivalence of the spin-up and spin-down channels on the Cu sites reflects the local magnetic moments of the Cu$^{2+}$ ($d^9$) ions. The elongation of the CuF$_6$ octahedra lifts the $e_g$ degeneracy, as discussed in Sec. III-A of the main text. This results in a minority-spin gap of about 2.7~eV.
%To assess the role of spin–orbit coupling (SOC), Fig.~\ref{fig:soc_nosoc} compares the band structures near the Fermi level calculated with and without SOC. SOC leaves the band gap essentially unchanged.

\newpage

\section{Wannierization procedure}
\begin{figure}[h!]
\centering
\includegraphics[width=6.6cm,angle=270]{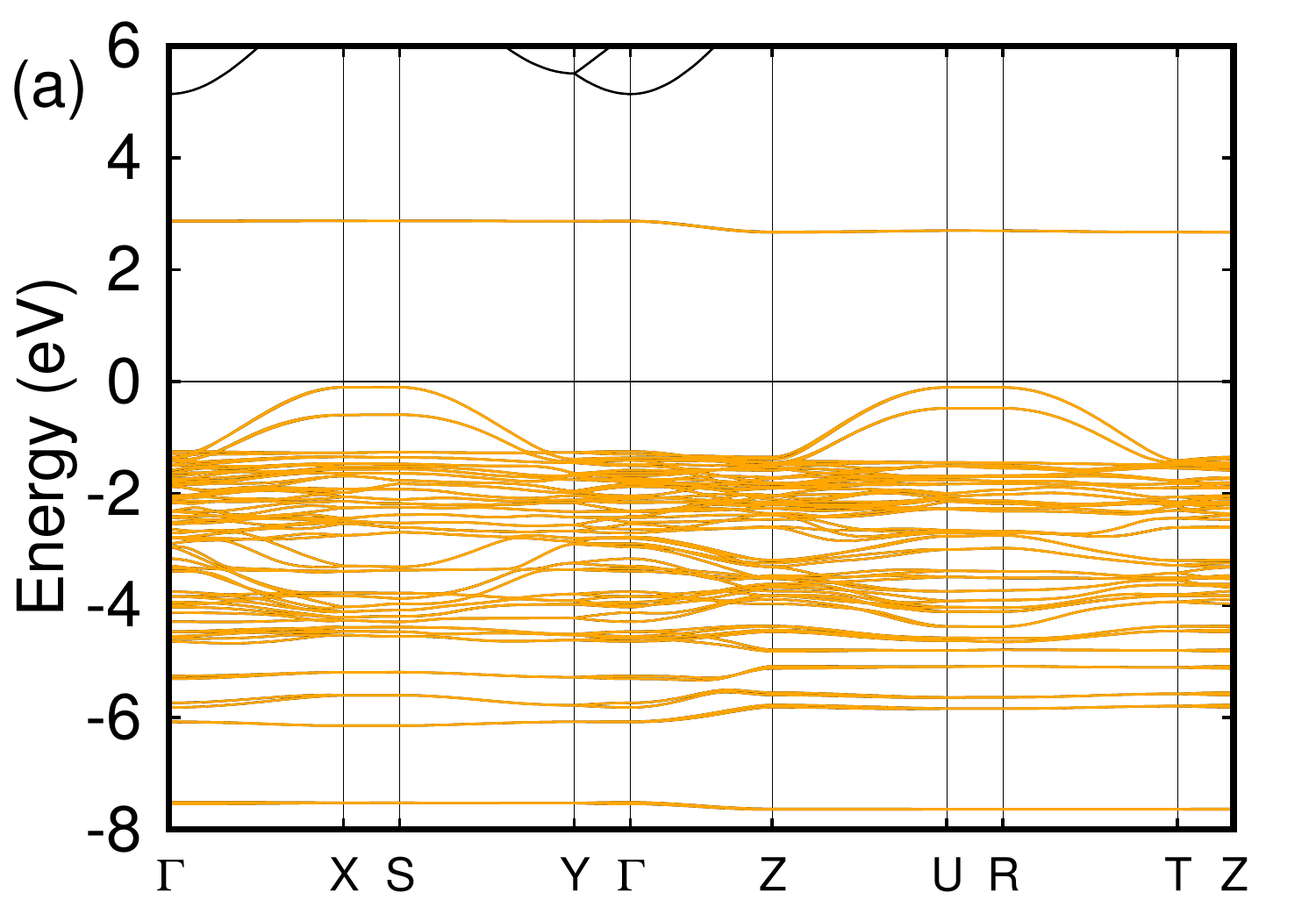}
\includegraphics[width=6.6cm,angle=270]{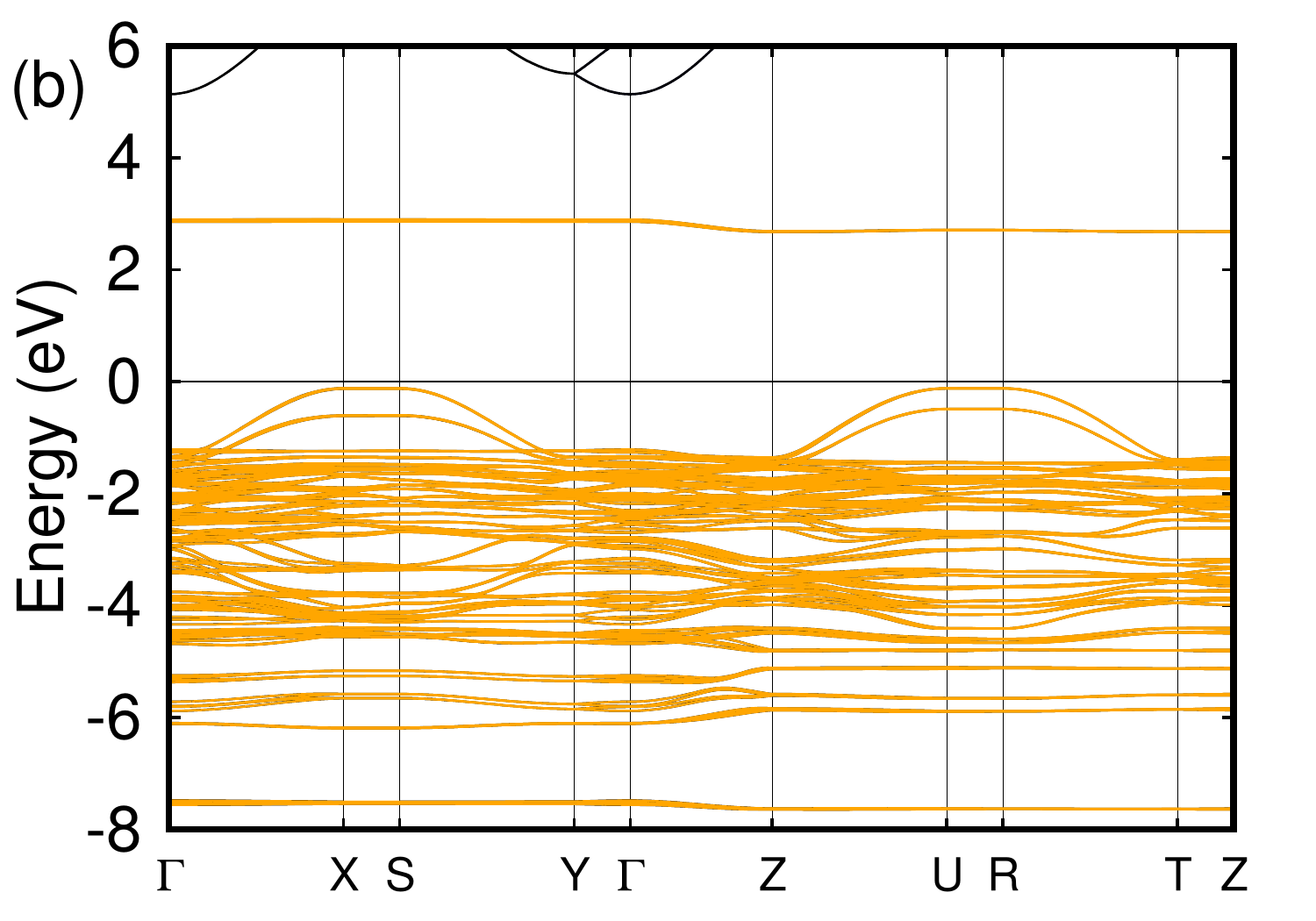}
\caption{DFT band structure (in black) and the Wannier-interpolated band structures (in orange) (a) without SOC and (b) with SOC. In all plots, the Fermi energy is set at 0 eV.}\label{Wannierized_bands}
\end{figure}

We evaluated the second-order photocurrent responses following the methodology reported in Refs.~\cite{Azpiroz18,Lihm22,Puente23}. 
This approach is based on the interpolation of the DFT band structure via maximally localized Wannier functions (MLWFs)~\cite{Marzari97,Mostofi08}, as implemented in the \textsc{Wannier90} package~\cite{Pizzi2020}. The Cu $d$ and F $p$ orbitals were taken into account, resulting in a total of 136 spinor bands in the presence of SOC. The wannierized bands are reported in Fig. \ref{Wannierized_bands} along the orthorhombic BZ path $\Gamma$-X-S-Y-$\Gamma$-Z-U-R-T-Z, both without and with the inclusion of SOC. The DFT bands are well reproduced in a wide energy range around the Fermi level, namely from -8 to 4 eV. As we can see from the two panels, SOC leaves the band gap essentially unchanged.

 %S3

%\begin{figure}
  %  \centering
 %   \includegraphics[width=6.6cm,angle=270]{DFT_BaCuF4_SOC_NOSOC.eps}
  %  \caption{DFT band structure of BaCuF$_4$ near the Fermi level along the high-symmetry path, computed without (solid) and with (dashed) SOC.}
  %  \label{fig:soc_nosoc}
%\end{figure}% S2

%The Density Functional Theory (DFT) calculations were performed using the projector-augmented wave (PAW) method, as implemented in the \textsc{VASP} code~\cite{Kresse93,Kresse96,Kresse96b}. 
%The exchange-correlation interactions were described within the generalized gradient approximation (GGA) of Perdew–Burke–Ernzerhof for solids (PBEsol)~\cite{Perdew08,Perdew96}.

\newpage

%We carried out structural optimization of the ground state $Cmc2_1$ structure and hypothetical nonpolar reference $Cmcm$ structure \cite{DANCE_ACuF4, Claude_2006_BaMF4} by relaxing both the ionic positions and lattice parameters until the Hellmann–Feynman forces on all atoms were reduced below 0.005 eV/\AA. Structural optimization has been performed keeping the A-AFM \cite{Garcia18} magnetic order of the Cu magnetic moments, which is observed at high temperature\cite{DANCE_ACuF4}. Symmetry-adapted mode (SAM) analysis has been performed using ISODISTORT\cite{Isodistort,Isodistort1} software.

\begin{comment}
\begin{figure}
\centering
%\includegraphics[width=6.6cm,angle=270]{BaCuF4_Neely_Sx_xaxis_OLD.eps}
%\includegraphics[width=6.6cm,angle=270]{BaCuF4_Neely_Sy_xaxis_OLD.eps}
%\includegraphics[width=6.6cm,angle=270]{BaCuF4_Neely_Sz_xaxis_OLD.eps}
%\caption{Spin-resolved band structure for the N\'eel vector along the y-axis and along the k$_x$-axis. The spin-momentum locking is Q$_x$, Q$_x$ and Q$_{0}$ for S$_x$, S$_y$ and S$_z$, respectively, as reported in Table \ref{tab:RSML_xaxis}.}\label{spinresolved_neely_xaxis}
\includegraphics[width=6.6cm,angle=270]{BaCuF4_Neely_Sy_xaxis.pdf}
\includegraphics[width=6.6cm,angle=270]{BaCuF4_Neely_Sz_xaxis.pdf}
\caption{Spin-resolved band structure for the N\'eel vector along the y-axis, shown along the k$_x$-direction. The spin-momentum locking is Q$_x$ and Q$_{0}$ for S$_y$ and S$_z$, respectively, as reported in Table \ref{tab:RSML_xaxis}.}\label{spinresolved_neely_xaxis}
\end{figure}% TableI figure S4
\end{comment}

\clearpage

\section{Spin-momentum locking along high-symmetry lines of the BZ}

Using DFT calculations and the multipole-extraction method described in Ref.~[\onlinecite{gong2026symmetryprotectednodalplanesaccidental}], we calculate the magnetic multipoles in the reciprocal space. The results, summarized in Table~\ref{tab:RSML}, are consistent with the symmetry analysis presented in the main text.

The Rashba terms $Q_y$ and $Q_x$, associated with the spin components $S_x$ and $S_y$, respectively, are present for all three orientations of the N\'eel vector. The $Q_{yz}$ term describes the nonrelativistic spin–momentum locking and appears along the diagonal of the table. Upon excluding the Rashba terms, the matrix is symmetric about its diagonal. The $Q_0$ term represents weak ferromagnetism, while the remaining terms describe quadrupolar contributions.

Below, we examine how the relativistic spin–momentum locking obtained from DFT simplifies along high-symmetry lines and for high-symmetry configurations.

\begin{table}[h!]
\centering
\resizebox{0.49\textwidth}{!}{%
\begin{tabular}{|c|c|c|c|}
\hline
 & \multicolumn{3}{|c|}{Spin components} \\
\hline
 & $S_x$ & $S_y$ & $S_z$ \\
\hline
$\mathbf{N} \parallel x$ & \textcolor{purple}{$Q_{yz}$},$Q_y$ & $Q_{xz}$,$Q_x$ & $Q_{xy}$ \\
\hline
$\mathbf{N} \parallel y$ & $Q_{xz}$,\textcolor{orange}{$Q_y$} & \textcolor{purple}{$Q_{yz}$},$Q_x$ & \textcolor{orange}{$Q_{0}$} \\
\hline
$\mathbf{N} \parallel z$ & $Q_{xy}$,\textcolor{orange}{$Q_y$} & \textcolor{orange}{$Q_{0}$},$Q_x$ & \textcolor{purple}{$Q_{yz}$} \\
\hline
\end{tabular}
}
\caption{Magnetization directions $N \parallel x,y,z$ for BaCuF$_4$ and corresponding spin-momentum locking for the spin components $S_x$, $S_y$ and $S_z$. The dominant component has spin-momentum locking as the quadrupole $Q_{yz}$, while the subdominant components have the d-wave $Q_{xz}$, $Q_{xy}$ and the s-wave $Q_{0}$ depending on the N\'eel vector orientation as reported in the Table. As in the main text, purple-coloured terms are those that would survive for collinear spin groups, while orange-coloured ones are those allowed by coplanar spin groups.}
\label{tab:RSML}
\end{table}

% Table SI -- Figure S2 and S3
Table~\ref{tab:RSML_xaxis} summarizes the relativistic spin--momentum locking along the $k_x$ axis.

\begin{table}[h!]
\centering
\resizebox{0.49\textwidth}{!}{%
\begin{tabular}{|c|c|c|c|}
\hline
 & \multicolumn{3}{|c|}{Spin components for k$_y$=k$_z$=0} \\
\hline
 & $S_x$ & $S_y$ & $S_z$ \\
\hline
$\mathbf{N} \parallel x$ & 0 & $Q_x$ & 0 \\
\hline
$\mathbf{N} \parallel y$ &
0 & $Q_x$ & $Q_{0}$ \\
\hline
$\mathbf{N} \parallel z$ & 0 & $Q_{0}$,$Q_x$ & 0\\
\hline
\end{tabular}
}
\caption{Magnetization directions $N \parallel x,y,z$ for BaCuF$_4$ and corresponding spin-momentum locking for the spin components $S_x$, $S_y$ and $S_z$ and with $k_y$=$k_z$=0.}
\label{tab:RSML_xaxis}
\end{table} % table S1

\begin{figure}[h!]
\centering
\includegraphics[width=6.6cm,angle=270]{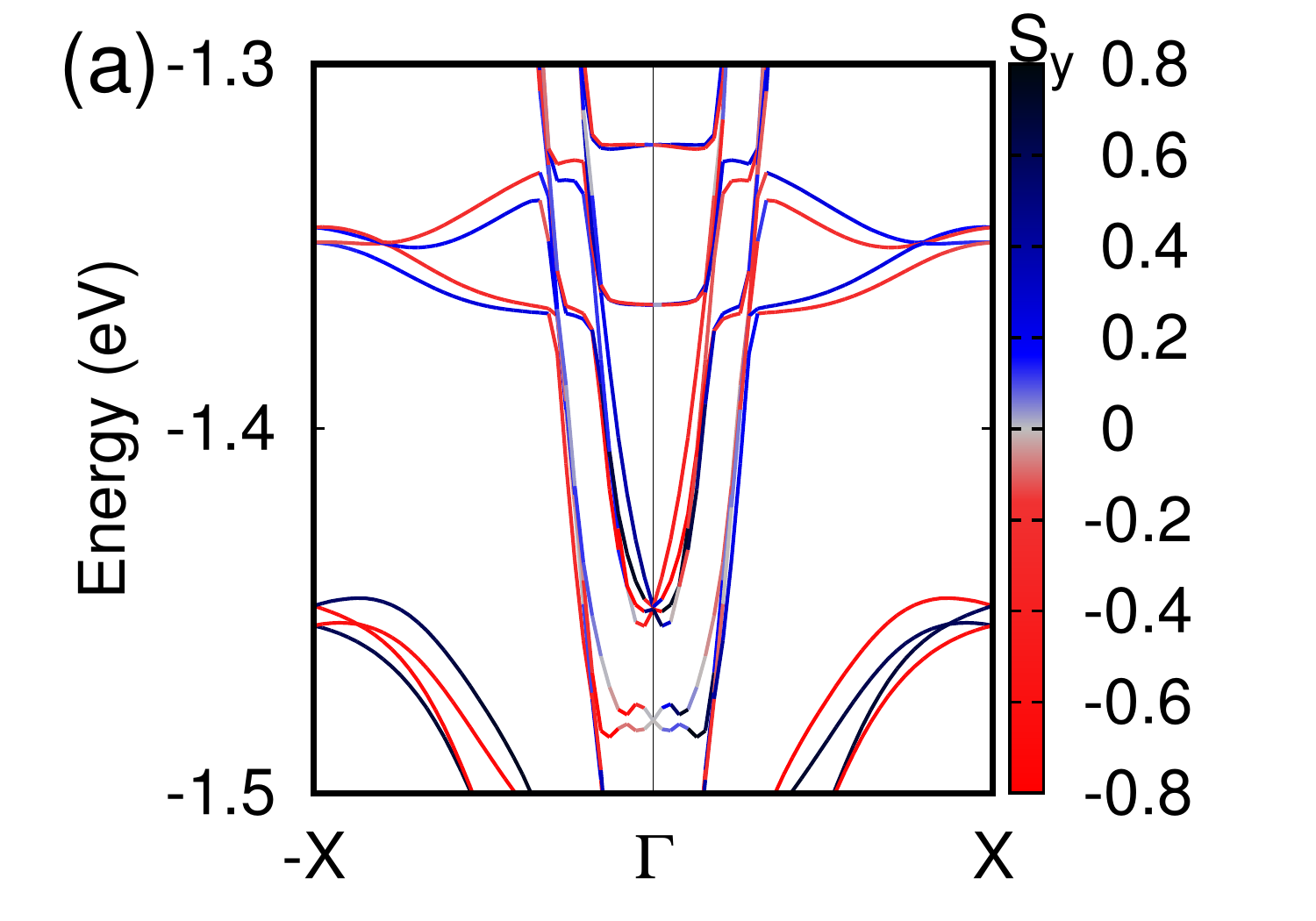}
\includegraphics[width=6.6cm,angle=270]{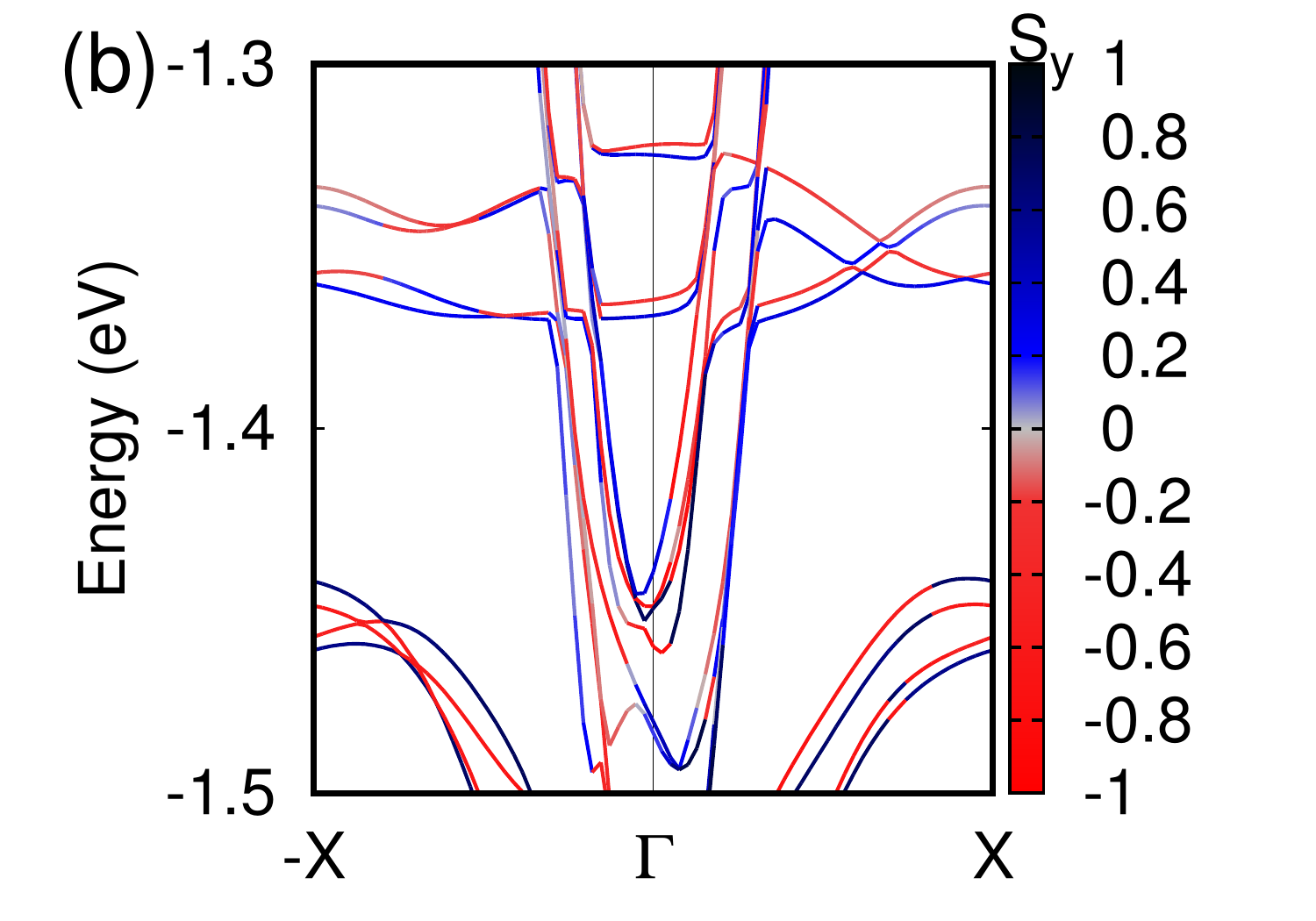}
\caption{Spin-resolved band structure for the N\'eel vector along the x- and z-axis, shown along the $k_x$-direction. The spin-momentum locking is $Q_x$ and $Q_{0}$+$Q_x$ for $S_y$, when the N\'eel vector is oriented along the x- and z-axis, respectively, as reported in Table \ref{tab:RSML_xaxis}.}\label{spinresolved_neelxz_xaxis} 
\end{figure}

For $\mathbf N\parallel x$, the $S_y$ component exhibits $Q_x$ symmetry [Fig.~\ref{spinresolved_neelxz_xaxis}(a)], whereas for $\mathbf N\parallel z$, it contains both $Q_0$ and $Q_x$ contributions [Fig.~\ref{spinresolved_neelxz_xaxis}(b)].

% TableI Figure S5

\begin{figure}[h!]
\centering
%\includegraphics[width=6.6cm,angle=270]{BaCuF4_Neely_Sx_xaxis_OLD.eps}
%\includegraphics[width=6.6cm,angle=270]{BaCuF4_Neely_Sy_xaxis_OLD.eps}
%\includegraphics[width=6.6cm,angle=270]{BaCuF4_Neely_Sz_xaxis_OLD.eps}
%\caption{Spin-resolved band structure for the N\'eel vector along the y-axis and along the k$_x$-axis. The spin-momentum locking is Q$_x$, Q$_x$ and Q$_{0}$ for S$_x$, S$_y$ and S$_z$, respectively, as reported in Table \ref{tab:RSML_xaxis}.}\label{spinresolved_neely_xaxis}
\includegraphics[width=6.6cm,angle=270]{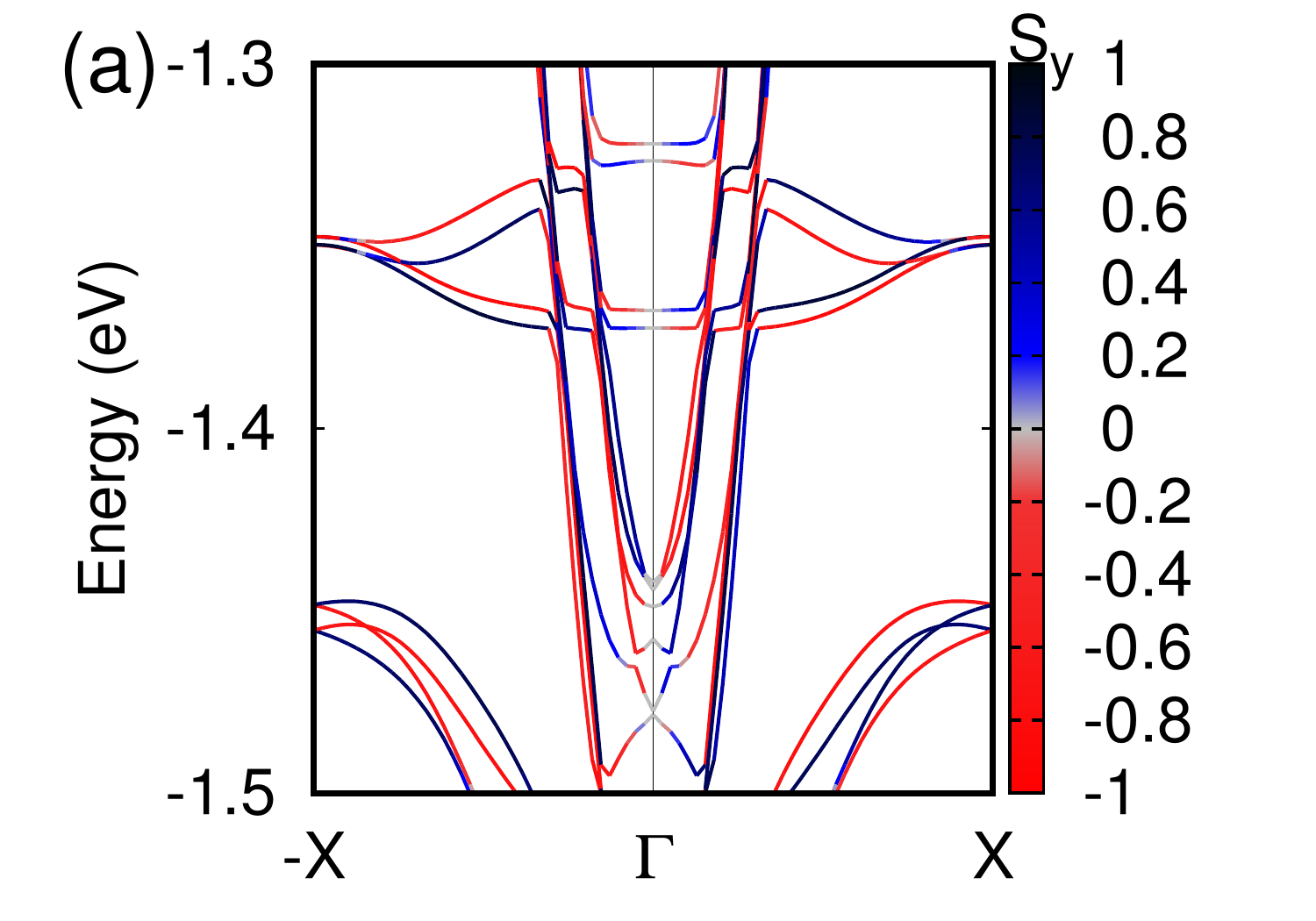}
\includegraphics[width=6.6cm,angle=270]{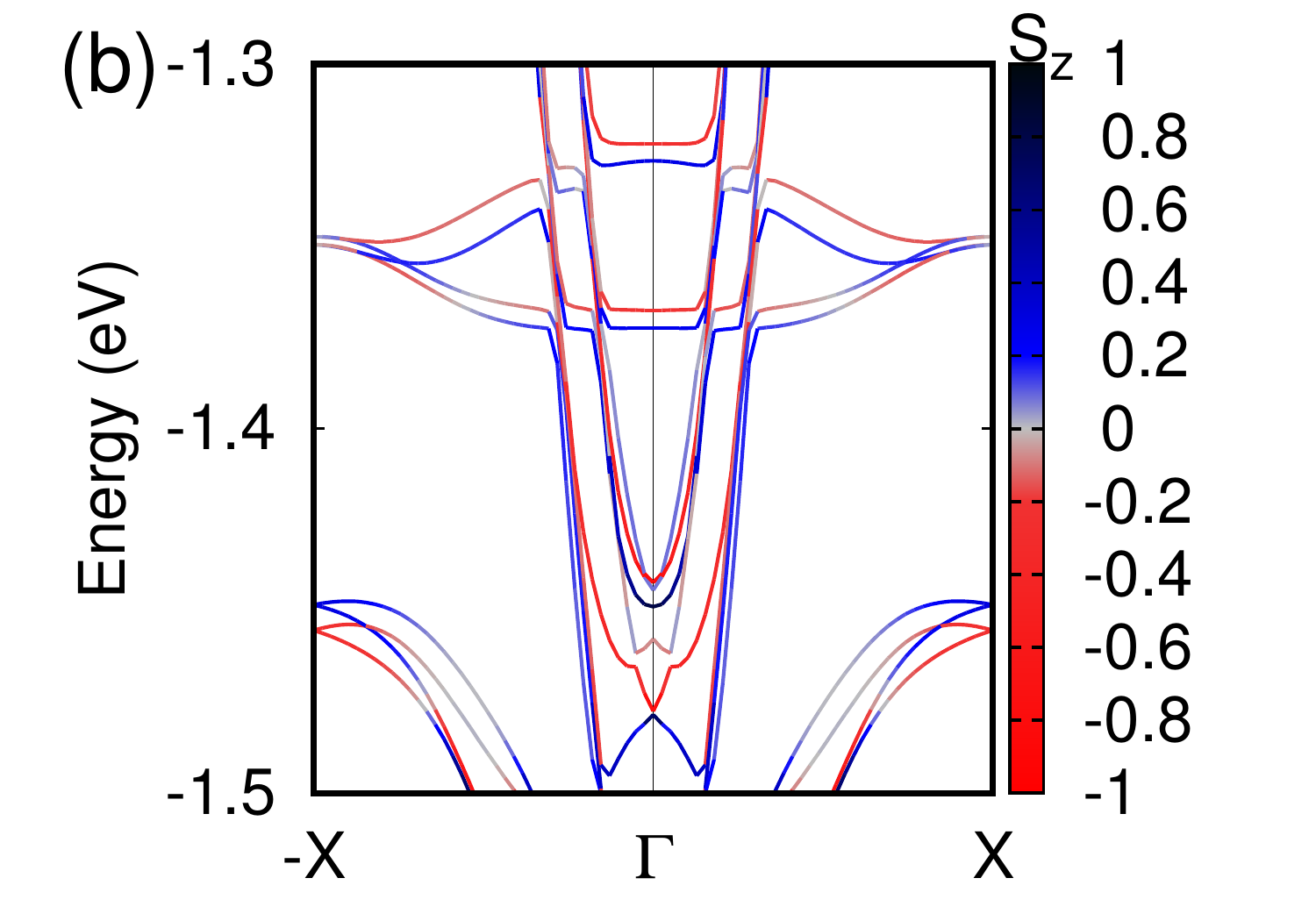}
\caption{Spin-resolved band structure for the N\'eel vector along the y-axis, shown along the $k_x$-direction. The spin-momentum locking is $Q_x$ and $Q_{0}$ for $S_y$ and $S_z$, respectively, as reported in Table \ref{tab:RSML_xaxis}.}\label{spinresolved_neely_xaxis}
\end{figure}

For $\mathbf N\parallel y$, the $S_y$ and $S_z$ components exhibit $Q_x$ and $Q_0$ symmetries, respectively [Fig.~\ref{spinresolved_neely_xaxis}(a), (b)].

% TableI figure S4

\clearpage

%With the N\'eel vector along the x-axis, the system is a pure altermagnet with Rashba; therefore, at the Gamma point the system exhibits double degeneracy as reported in Figure~\ref{spinresolved_neelxz_xaxis}(a). With the N\'eel vector along the y-axis, the system is a weak ferromagnet with magnetization parallel to the ferroelectric polarization; therefore, there is a splitting of the Rashba bands at $\Gamma$ \cite{Bihlmayer2022}.
%With the N\'eel vector along the z-axis, there is a shift in the Rashba splitting in Figure~\ref{spinresolved_neelxz_xaxis}(b), which is due to the weak ferromagnetism orthogonal to the ferroelectric polarization\cite{Bihlmayer2022}.
%In this compound, both the Rashba shift and the Rashba splitting at $\Gamma$ are clearly visible due to the small magnetic moment. In the case of a ferromagnet with a larger magnetic moment and a correspondingly larger splitting at $\Gamma$, it would be difficult to follow the evolution of the band structure.

%%%%%%%%%%%%%%%%%%%%%%%%%%%%%%%%%%%%%%%%%%%%%%%%%

% Table II -- Figure S4 and S5
Table~\ref{tab:RSML_yaxis} summarizes the relativistic spin--momentum locking along the $k_y$ axis.

\begin{table}[h!]
\centering
\resizebox{0.49\textwidth}{!}{%
\begin{tabular}{|c|c|c|c|}
\hline
 & \multicolumn{3}{|c|}{Spin components for k$_x$=k$_z$=0} \\
\hline
 & $S_x$ & $S_y$ & $S_z$ \\
\hline
$\mathbf{N} \parallel x$ & $Q_y$ & 0 & 0 \\
\hline
$\mathbf{N} \parallel y$ & $Q_y$ & 0 & $Q_{0}$ \\
\hline
$\mathbf{N} \parallel z$ & $Q_y$ & $Q_{0}$ & 
0\\
\hline
\end{tabular}
}
\caption{Magnetization directions $N \parallel x,y,z$ for BaCuF$_4$ and corresponding spin-momentum locking for the spin components $S_x$, $S_y$ and $S_z$ and with $k_x$=$k_z$=0.}
\label{tab:RSML_yaxis}
\end{table} % table SII

\begin{figure}[h!]
\centering
\includegraphics[width=6.6cm,angle=270]{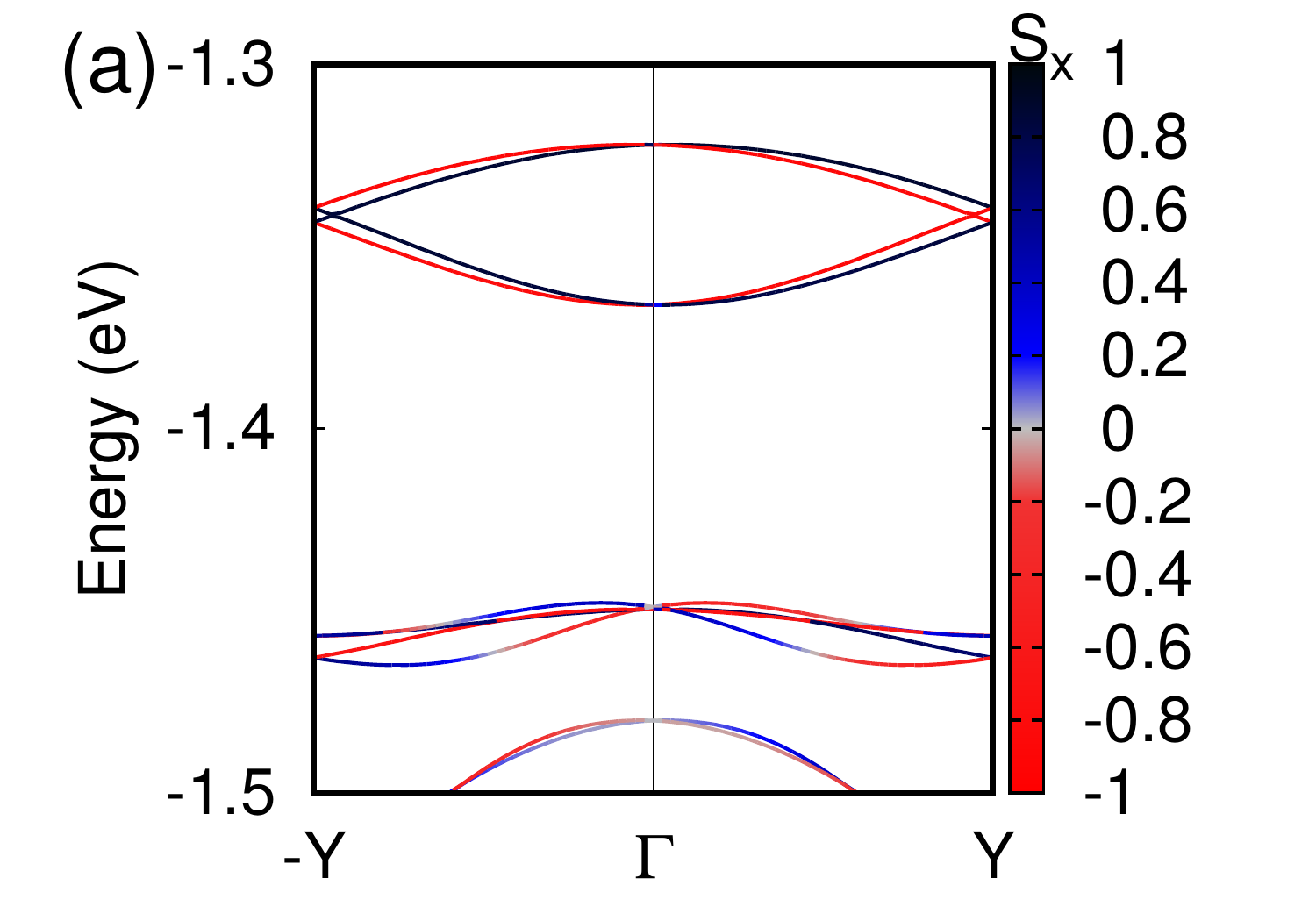}
\includegraphics[width=6.6cm,angle=270]{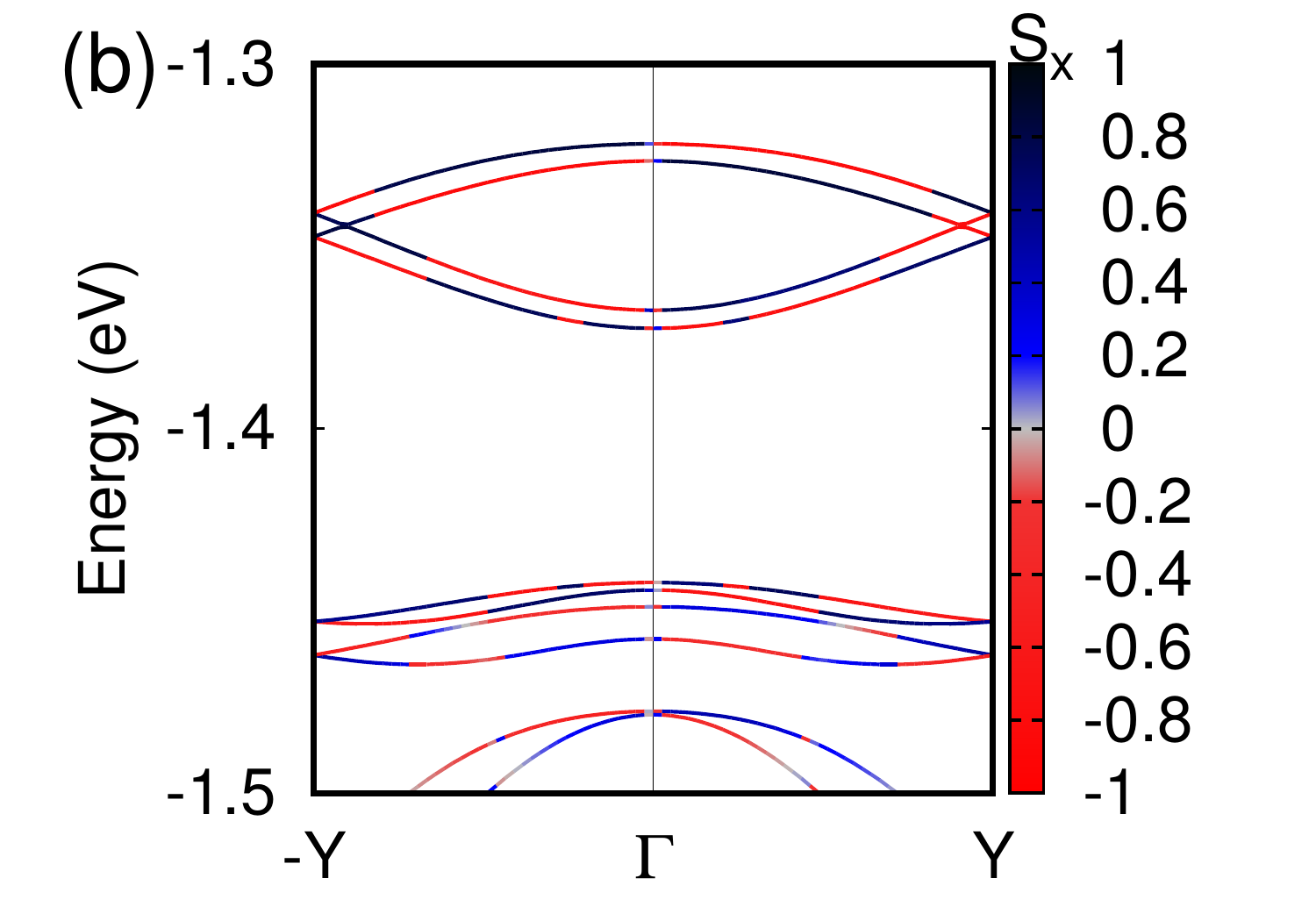}
\caption{Spin-resolved band structure for the N\'eel vector along the x- and y-axis, shown along the k$_y$-direction. For both case, the spin-momentum locking is Q$_y$ as reported in Table \ref{tab:RSML_yaxis}.}\label{spinresolved_neelxy_yaxis}
\end{figure}

For $\mathbf N\parallel x$ and $\mathbf N\parallel y$, the spin--momentum locking exhibits $Q_y$ symmetry [Fig.~\ref{spinresolved_neelxy_yaxis}(a), (b)].

% TableII Figure S6

\begin{figure}[h!]
\centering
\includegraphics[width=6.6cm,angle=270]{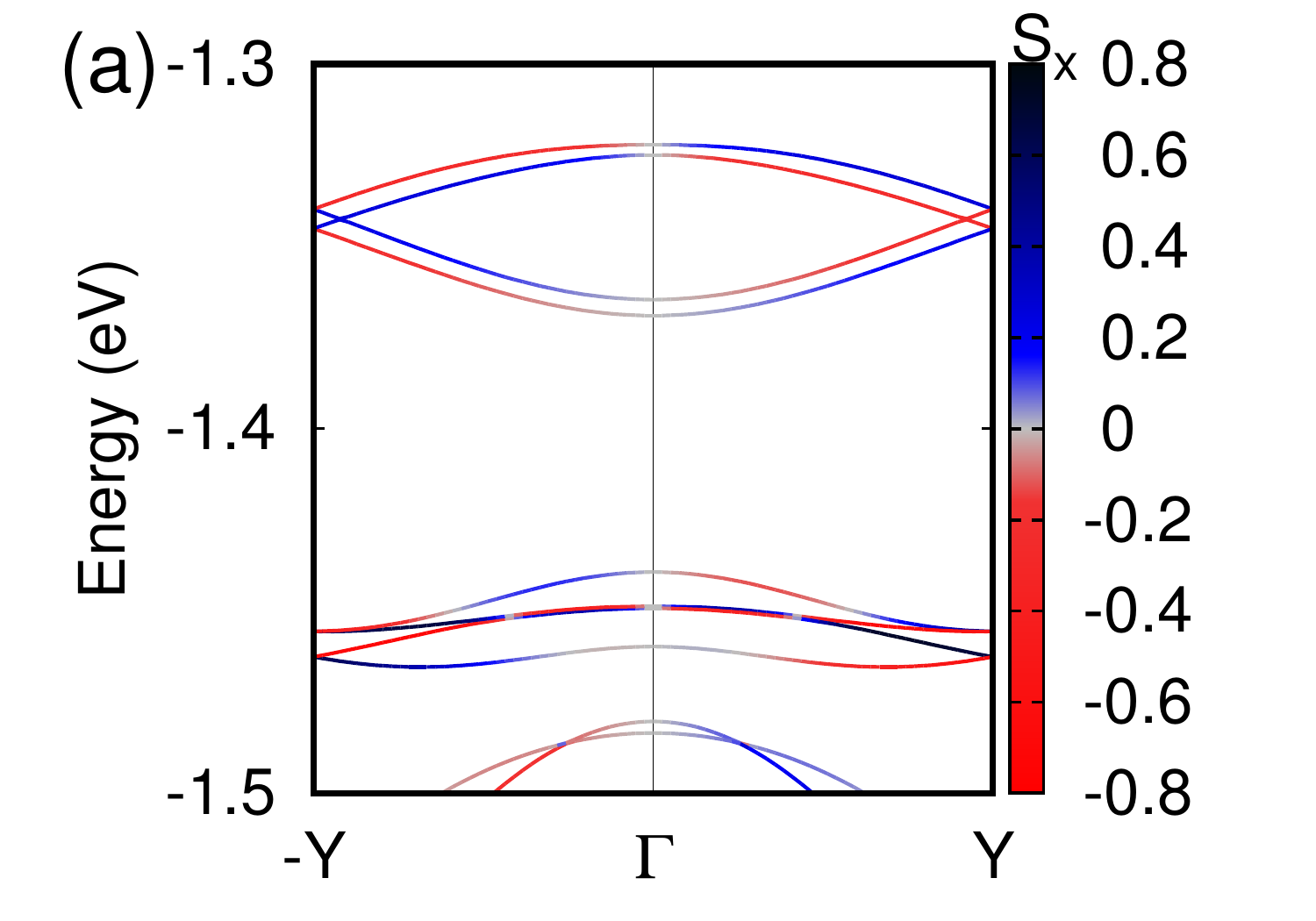}
\includegraphics[width=6.6cm,angle=270]{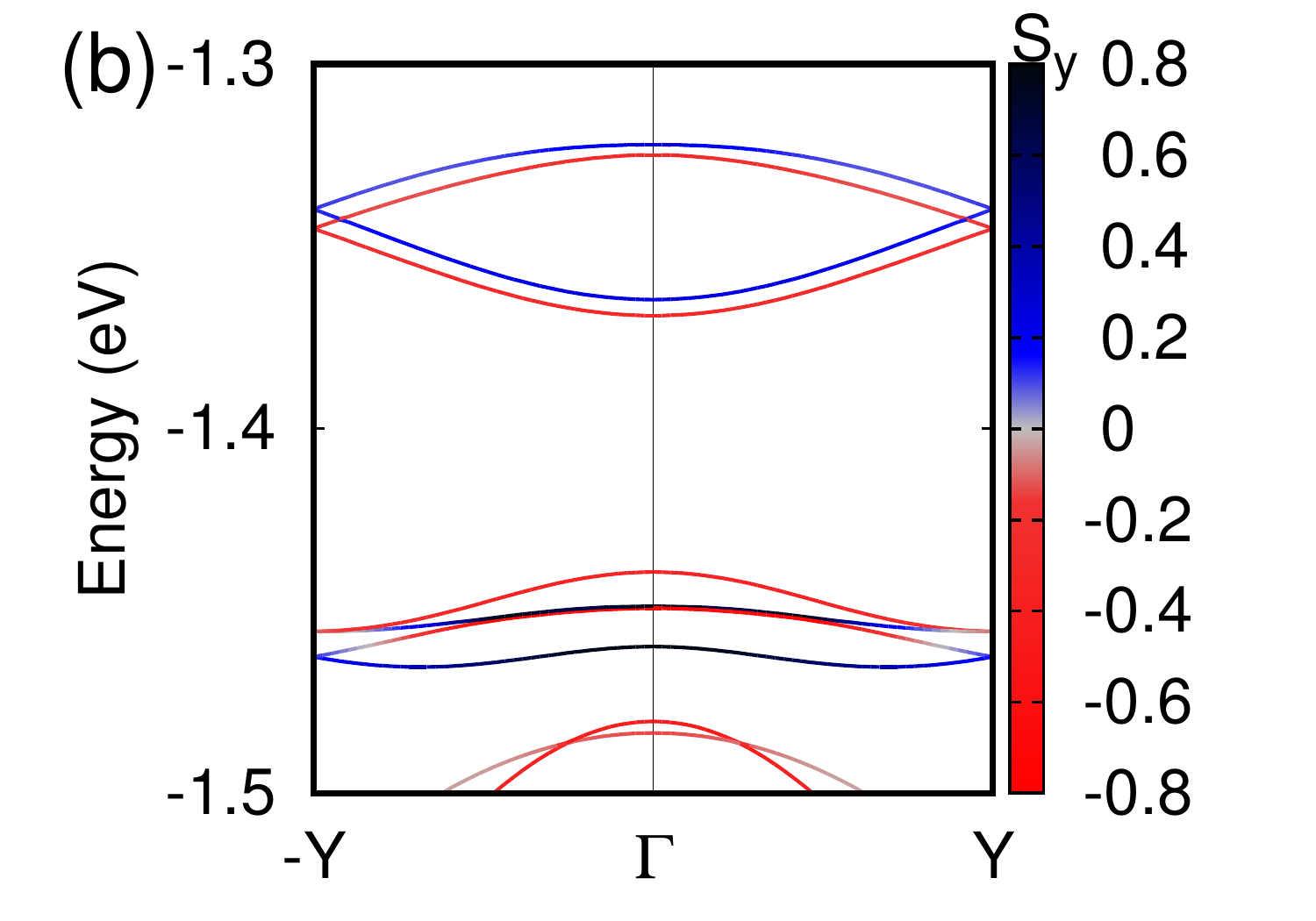}
%\includegraphics[width=6.6cm,angle=270]{BaCuF4_Neelz_Sz_yaxis.eps}
%\caption {Spin-resolved band structure along the k$_y$-axis. The spin-momentum locking is Q$_y$, Q$_{0}$, and Q$_y$ for S$_x$, S$_y$ and S$_z$, respectively, N\'eel vector along the z-axis, as reported in Table \ref{tab:RSML_yaxis}.}
\caption {Spin-resolved band structure along the z-axis, shown along the $k_y$-direction. The spin-momentum locking is $Q_y$ and $Q_{0}$ for $S_x$ and $S_y$, as reported in Table \ref{tab:RSML_yaxis}.}
\label{spinresolved_neelz_yaxis} 
\end{figure}

For $\mathbf N\parallel z$, the $S_x$ and $S_y$ components exhibit $Q_y$ and $Q_0$ symmetries, respectively [Fig.~\ref{spinresolved_neelz_yaxis}(a), (b)].

% TableII Figure S7

%%%%%%%%%%%%%%%%%%%%%%%%%%%%%%%%%%%%%%%%%%%%%%%%%

%%%%%%%%%%%%%%%%%%%%%%%%%%%%%%%%%%%%%%%%%%%%%%%%%%%%%%
\clearpage

\begin{figure}[h!]
\centering
\includegraphics[width=6.6cm,angle=270]{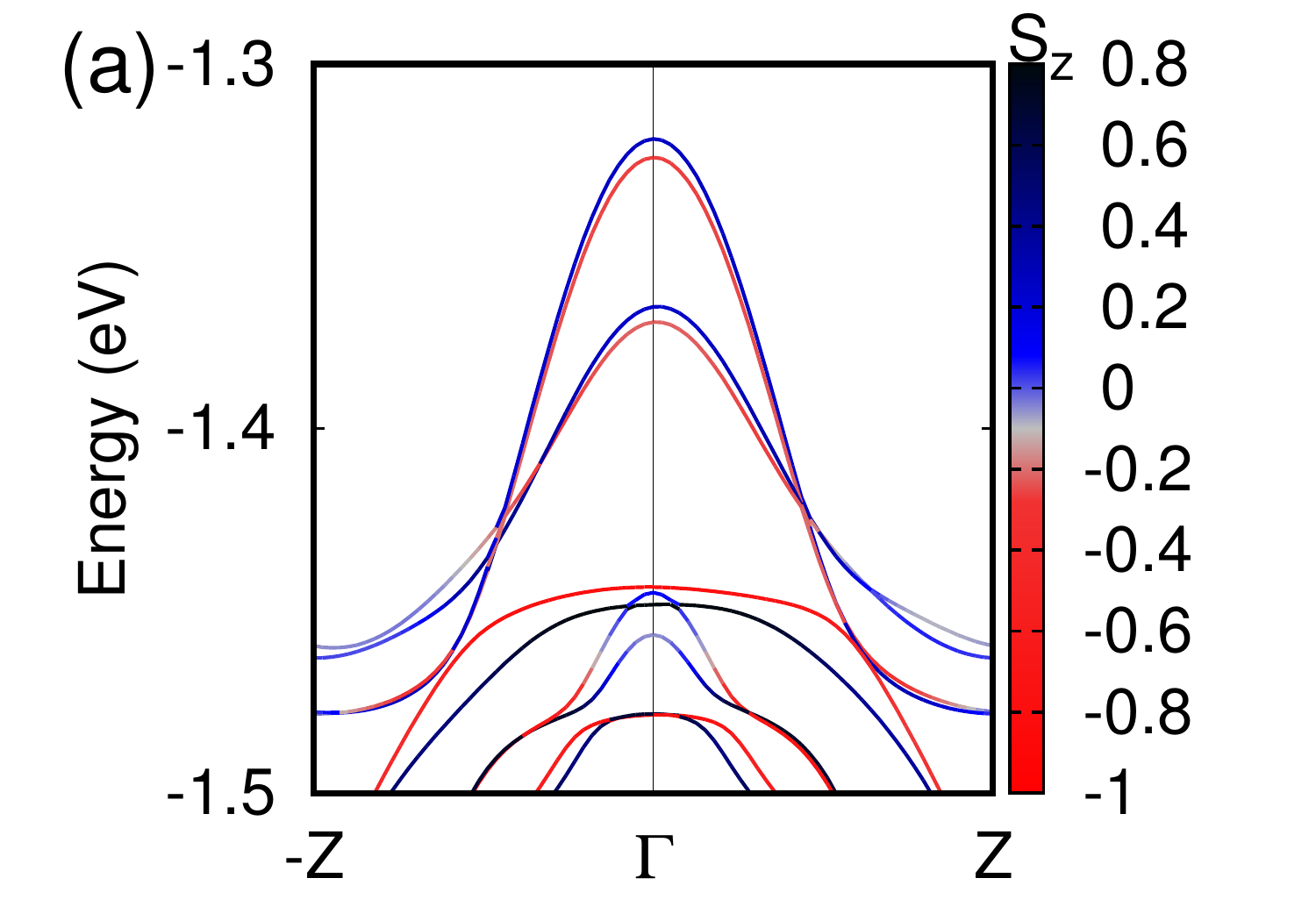}
\includegraphics[width=6.6cm,angle=270]{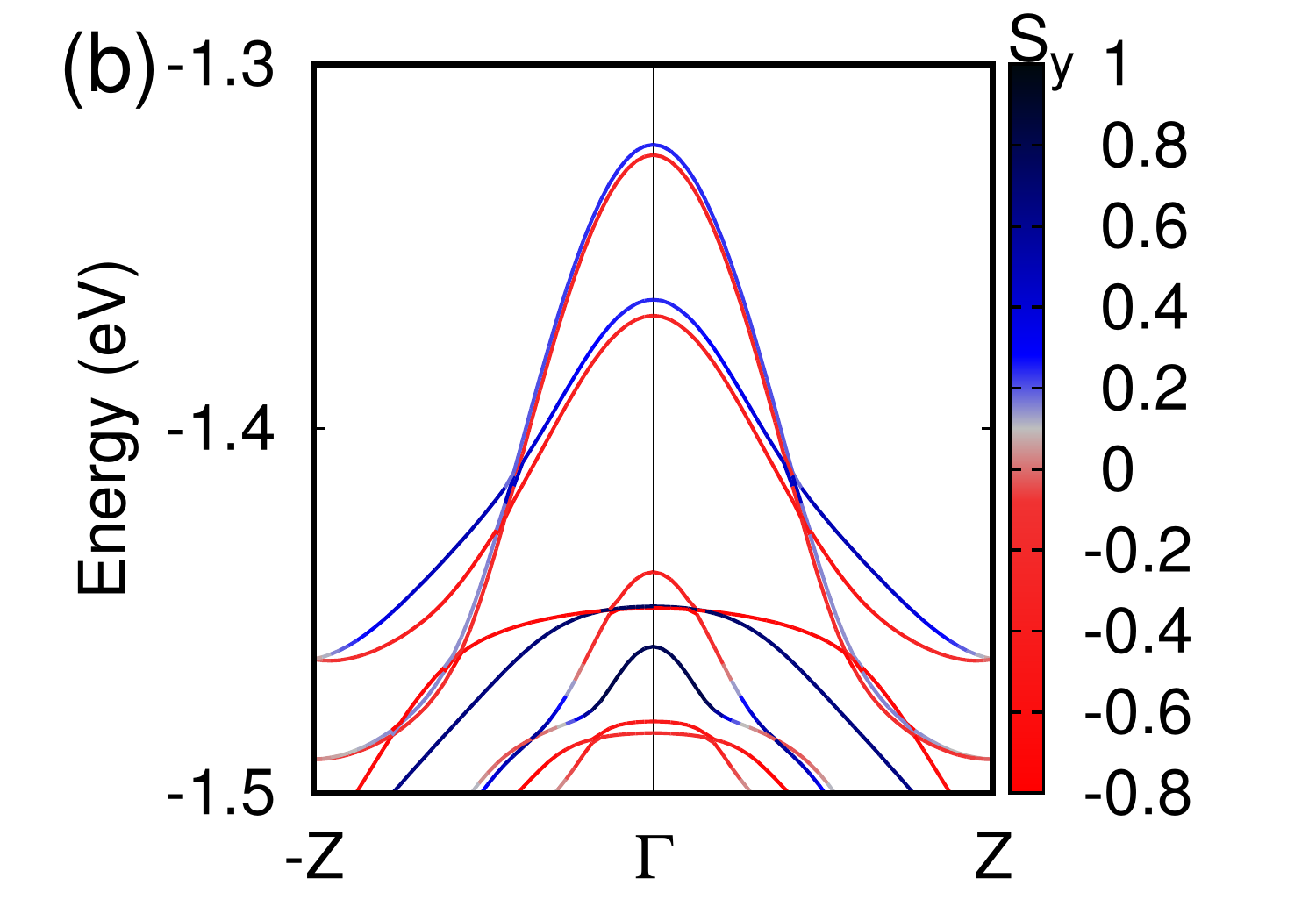}
\caption{Spin-resolved band structure for the N\'eel vector along the y- and z-axis, shown along the $k_z$-direction. For both case, the spin-momentum locking is $Q_0$ as reported in Table \ref{tab:RSML_zaxis}.}\label{spinresolved_neelx_zaxis} % TableIII Figure S8
\end{figure}

Table~\ref{tab:RSML_zaxis} summarizes the spin--momentum locking along the $k_z$ axis, where only $Q_0$ components are present [Fig.~\ref{spinresolved_neelx_zaxis}].

\begin{table}[h!]
\centering
\resizebox{0.49\textwidth}{!}{%
\begin{tabular}{|c|c|c|c|}
\hline
 & \multicolumn{3}{|c|}{Spin components for k$_x$=k$_y$=0} \\
\hline
 & $S_x$ & $S_y$ & $S_z$ \\
\hline
$\mathbf{N} \parallel x$ & 0 & 0 & 0 \\
\hline
$\mathbf{N} \parallel y$ & 0 & 0 & $Q_{0}$ \\
\hline
$\mathbf{N} \parallel z$ & 0 & $Q_{0}$ & 0\\
\hline
\end{tabular}
}
\caption{Magnetization directions $N \parallel x,y,z$ for BaCuF$_4$ and corresponding spin-momentum locking for the spin components $S_x$, $S_y$ and $S_z$ and with $k_x$=0 and $k_y$=0.}
\label{tab:RSML_zaxis}
\end{table} % Table SIII

Along $\Gamma$--$Z$, the dominant spin component vanishes, and the band dispersion is slightly asymmetric about $\Gamma$, as seen in Fig.~\ref{spinresolved_neelx_zaxis}(a), (b).

%%%%%%%%%%%%%%%%%%%%%%%%%%

%The dipole $Q_x$ terms in Fig.~\ref{spinresolved_neely_xaxis}(a) and the $Q_y$ term in Fig.~~\ref{spinresolved_neelz_yaxis}(c) are not induced by an electric field along the $z$ axis. These terms are smaller since they exhibit low spin polarization and appear only in the presence of weak ferromagnetism. Since they are dipolar terms, they do not seem to influence the spin photocurrents. Further investigation is required to clarify their origin; therefore, they are neglected in the subsequent analysis.

%%%%%%%%%%%%%%%%%%%%%%%%%%

\begin{figure}[h!]
\centering
\includegraphics[width=6.6cm,angle=270]{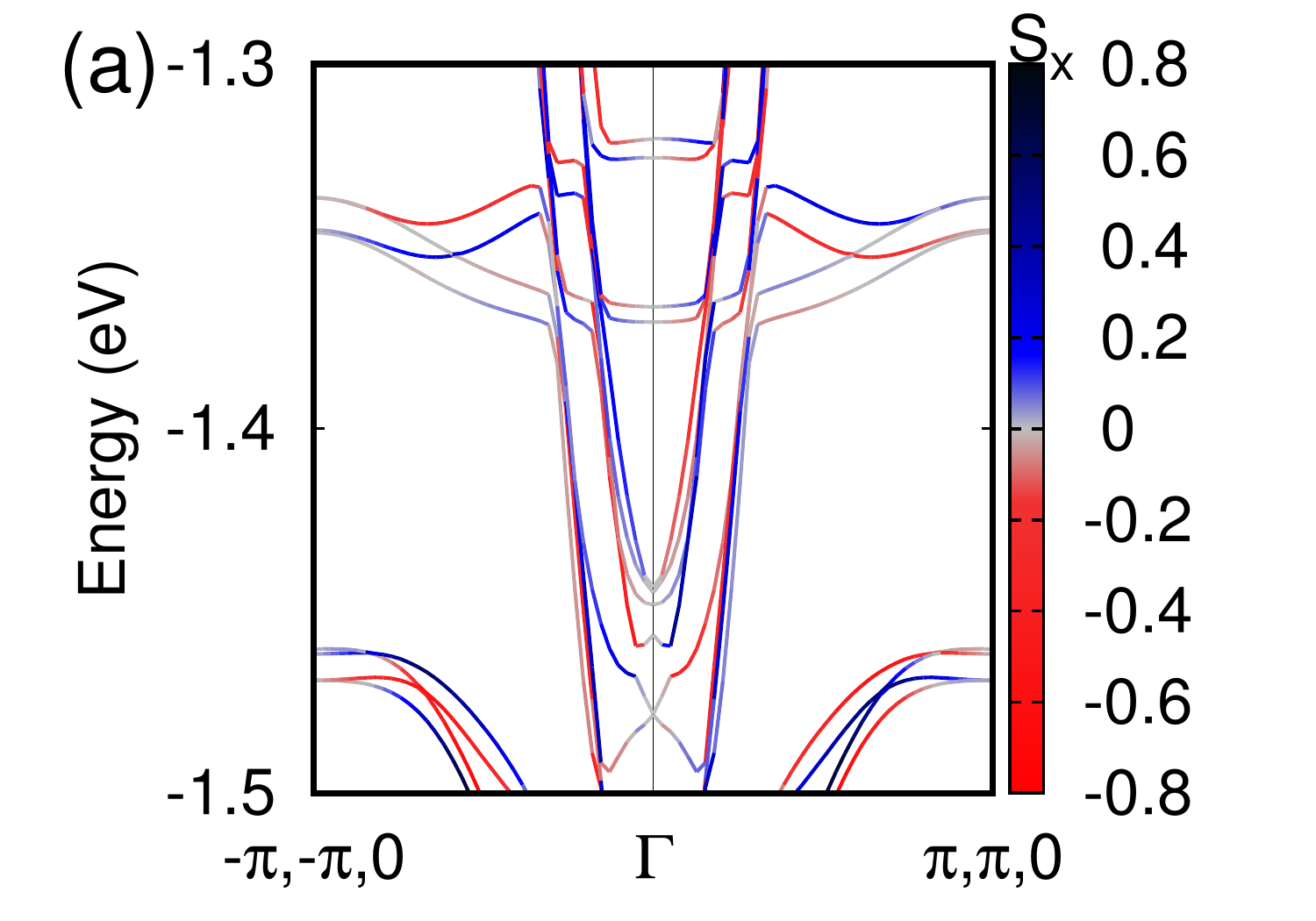}
\includegraphics[width=6.6cm,angle=270]{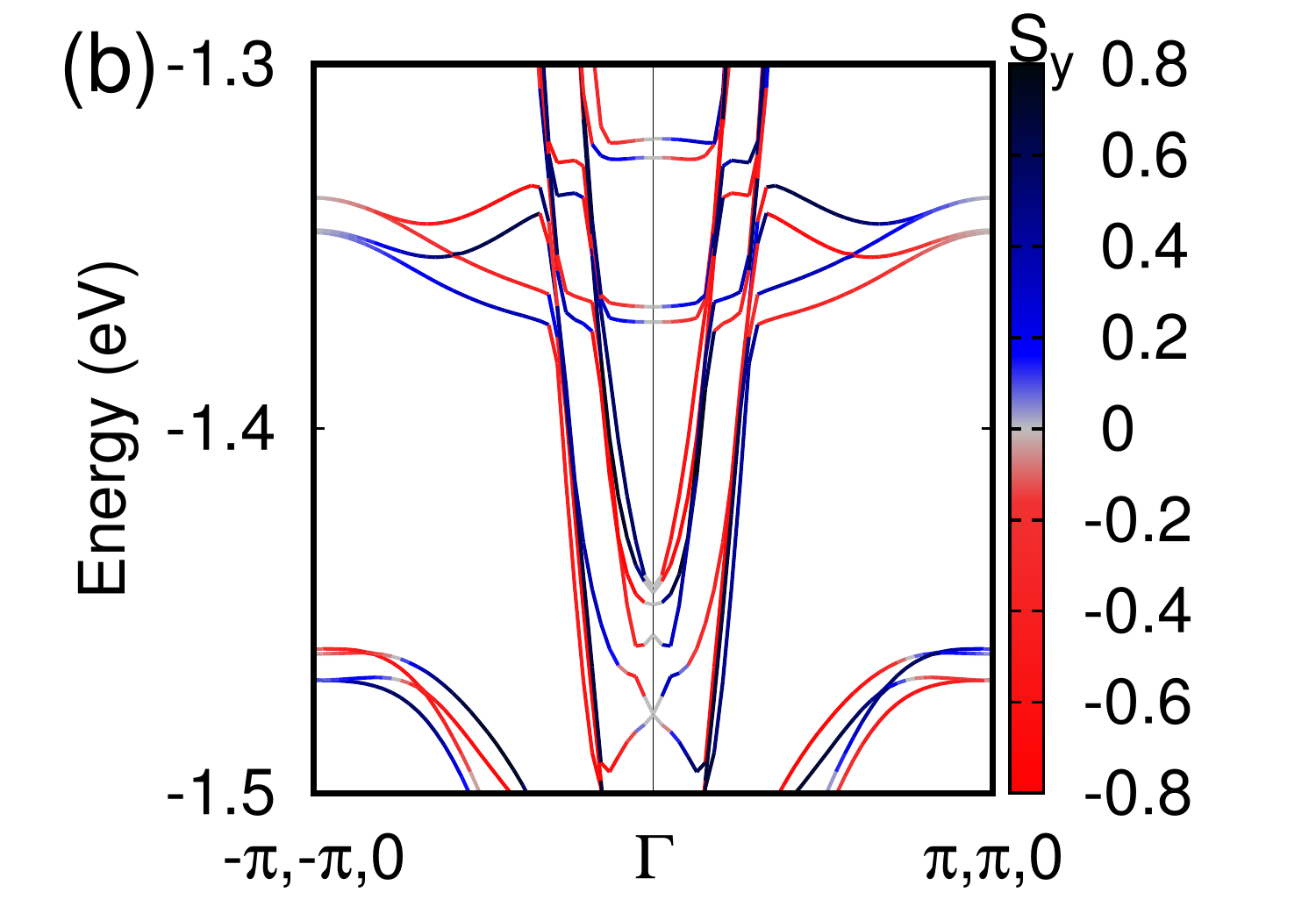}
\includegraphics[width=6.6cm,angle=270]{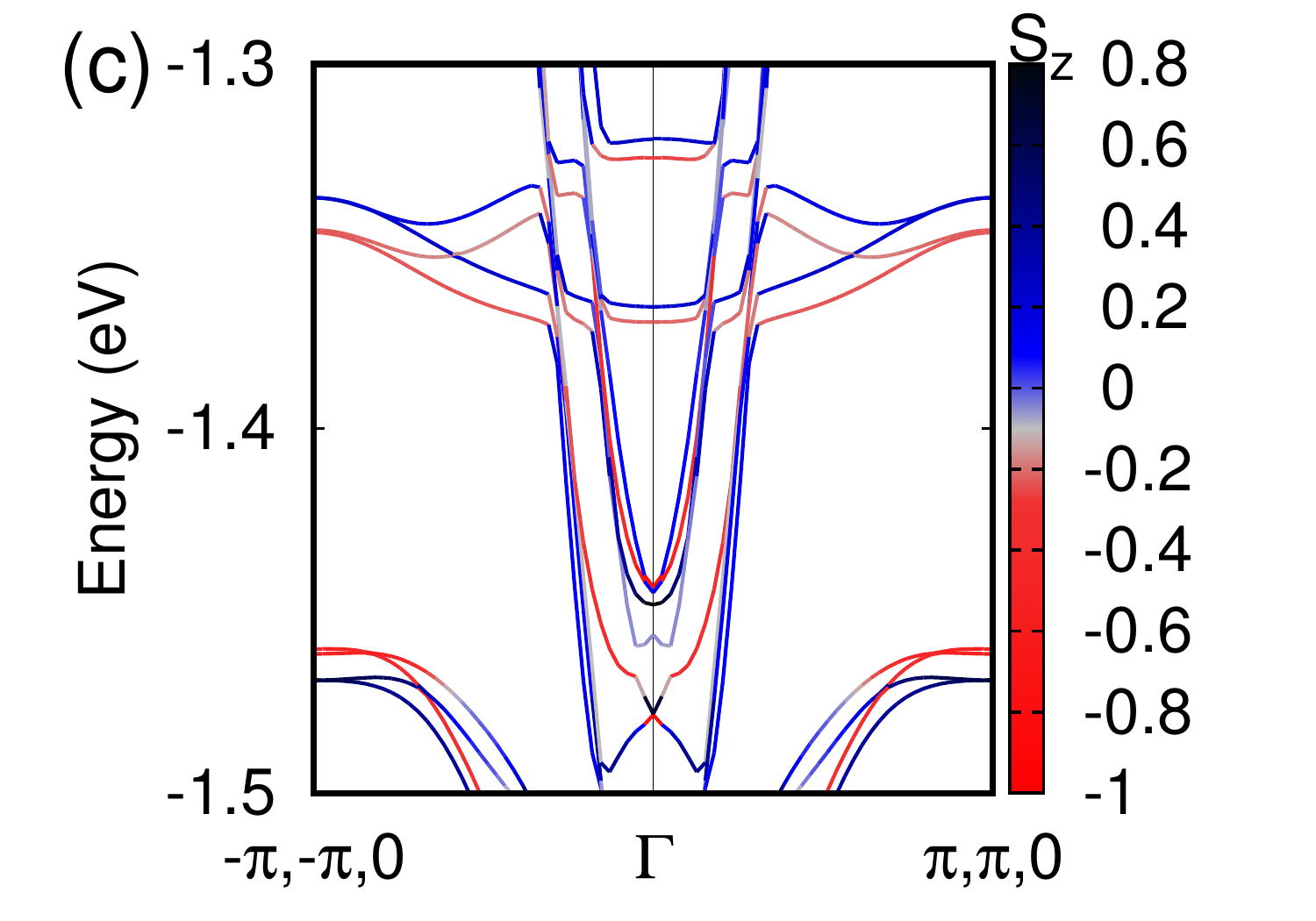}
\caption{Spin-resolved band structure for the N\'eel vector along the y-axis, shown along the diagonal for which $k_x$=$k_y$ and $k_z$=0. The spin-momentum locking is $Q_x$, $Q_x$ and $Q_{0}$ for $S_x$, $S_y$ and $S_z$, respectively.}\label{spinresolved_neely_kz0_diag1}
\end{figure}

Finally, we report in Fig. \ref{spinresolved_neely_kz0_diag1} the case of the spin momentum locking along the diagonal (110) of the k-space; therefore, we have $k_x$=$k_y$ and $k_z$=0. Other directions were tested but not plotted. As a result of all tested directions, we obtained the quadrupole terms of the main text.

 % Figure S9

\clearpage

\section{Spin-resolved Fermi surface}

\begin{figure}[h]
    \centering
    \includegraphics[width=6.5cm]{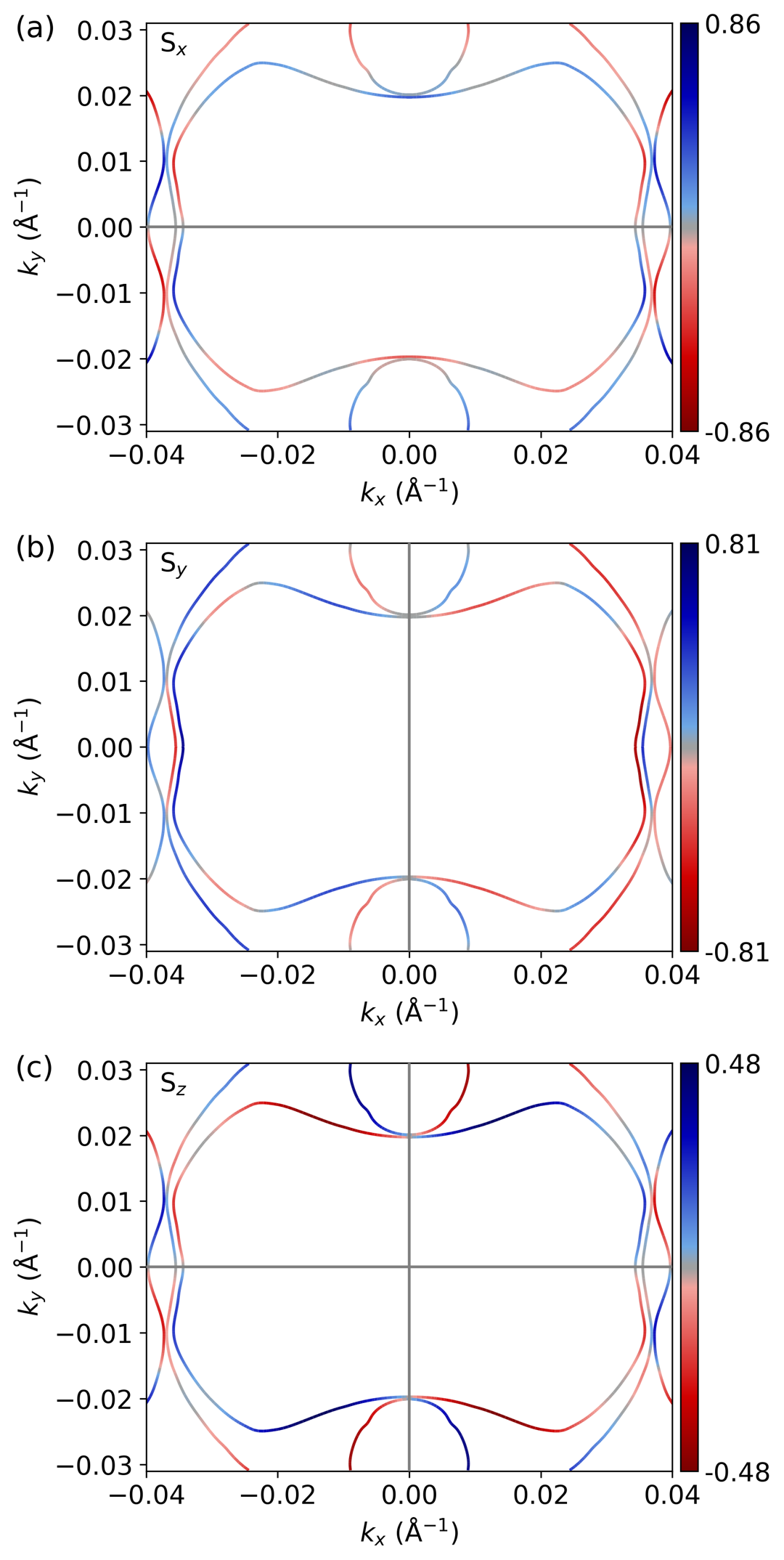}
    \caption{Spin-resolved constant-energy contours near the centre of the Brillouin zone of BaCuF$_4$ for $\mathbf N\parallel x$, in the $k_z=0$ plane at $3.2$~eV below the Fermi level. Panels (a), (b), and (c) show the $S_x$, $S_y$, and $S_z$ components, respectively. Solid black lines mark the intersections of the symmetry-protected nodal planes with the $k_z=0$ plane. Red and blue indicate opposite spin polarizations, with the colour scales giving the spectral weight of each spin component.}
    \label{fig:fermi2dkz0}
\end{figure}

For BaCuF$_4$ with $\mathbf N\parallel x$, the $S_x$, $S_y$, and $S_z$ spin components exhibit $Q_y$, $Q_x$, and $Q_{xy}$ symmetries, respectively, in the $k_z=0$ plane (Table~\ref{tab:RSML}). This is confirmed by the constant-energy contours at $3.2$~eV below the Fermi level, shown in Fig.~\ref{fig:fermi2dkz0}. The colour scales indicate that $S_x$ has the largest spectral weight, while $S_y$ and $S_z$ have comparable magnitudes. The $S_y$ component exhibits Rashba-like spin--momentum locking, whereas $S_z$ represents a subdominant relativistic contribution.

\clearpage
\section{Spin photogalvanic effect without SOC}

%In order to calculate the charge and spin photocurrents without SOC, we use the definitions adopted in Refs. \cite{Azpiroz18,Puente23}, where the two channels are treated separately.
%The charge photocurrents obtained without including SOC are presented in Fig.~\ref{NOSOC_charge_photocurrents}, where in panels (a) and (b) the linear shift current and the circular injection current are reported. We evaluate the contributions from the two spin channels separately and then compute their total response as $j^{\uparrow} + j^{\downarrow}$, with $j^{\uparrow}$ and $j^{\downarrow}$ denoting the respective currents from the spin-up and spin-down channels. 

%\begin{figure}[t!]
%\centering
%\includegraphics[width=5.5cm,angle=270]{NOSOC_Charge_shift_BaCuF4_1.eps}
%\includegraphics[width=5.5cm,angle=270]{NOSOC_Charge_circular_injection_BaCuF4_1.eps}
%\caption{Allowed components of the (a) charge linear shift current and (b) charge circular injection current without SOC.}
%\label{NOSOC_charge_photocurrents}
%\end{figure} 

To calculate the spin photoconductivities without SOC, we follow Refs.~\cite{Azpiroz18,Puente23} and treat the two spin channels separately. Each spin photoconductivity is then obtained as the difference between the corresponding spin-up and spin-down conductivities, consistent with defining the spin photocurrent as $j^{\uparrow}-j^{\downarrow}$.

\begin{figure}[h]
\centering
\includegraphics[width=5.5cm,angle=270]{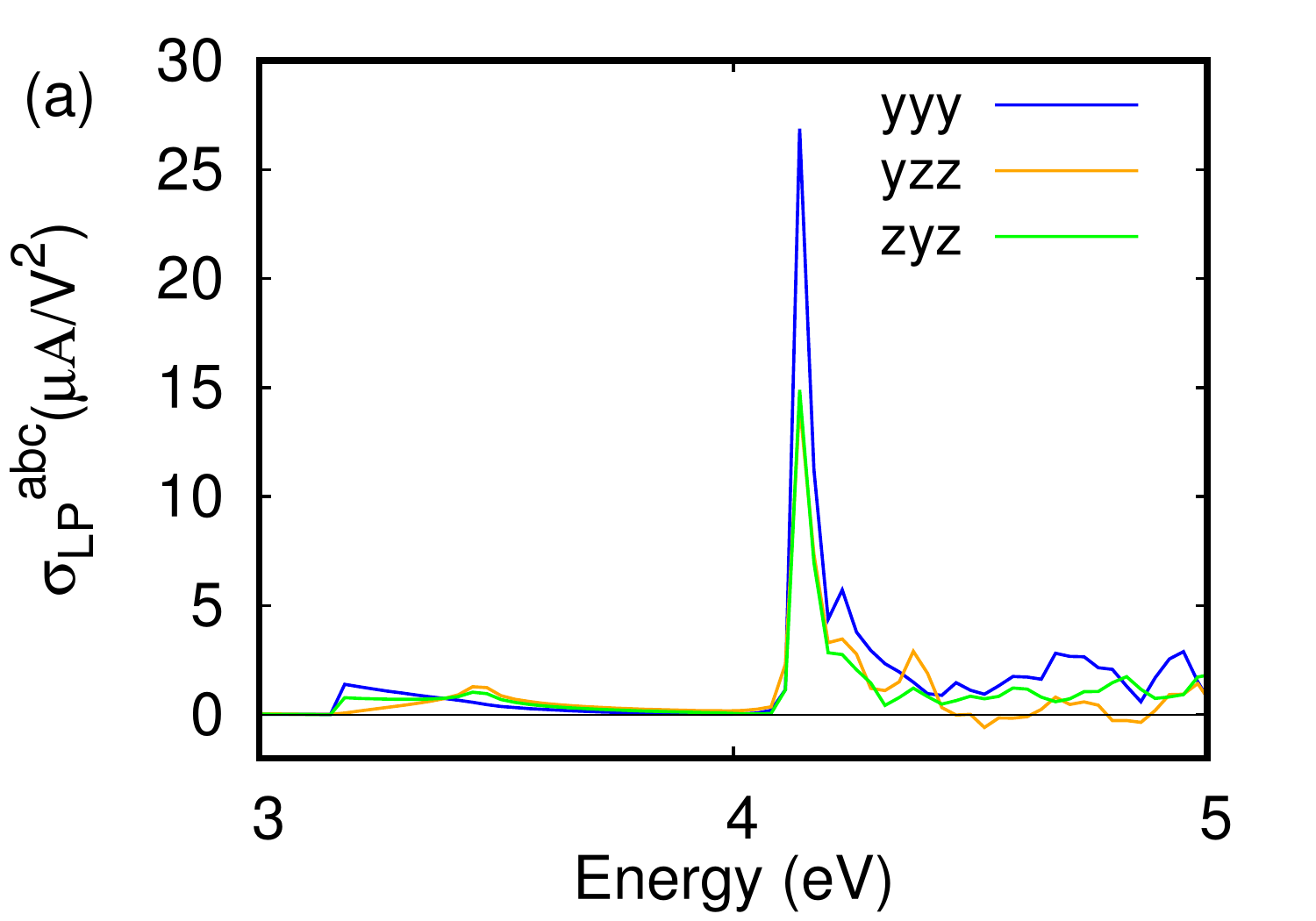}
\includegraphics[width=5.5cm,angle=270]{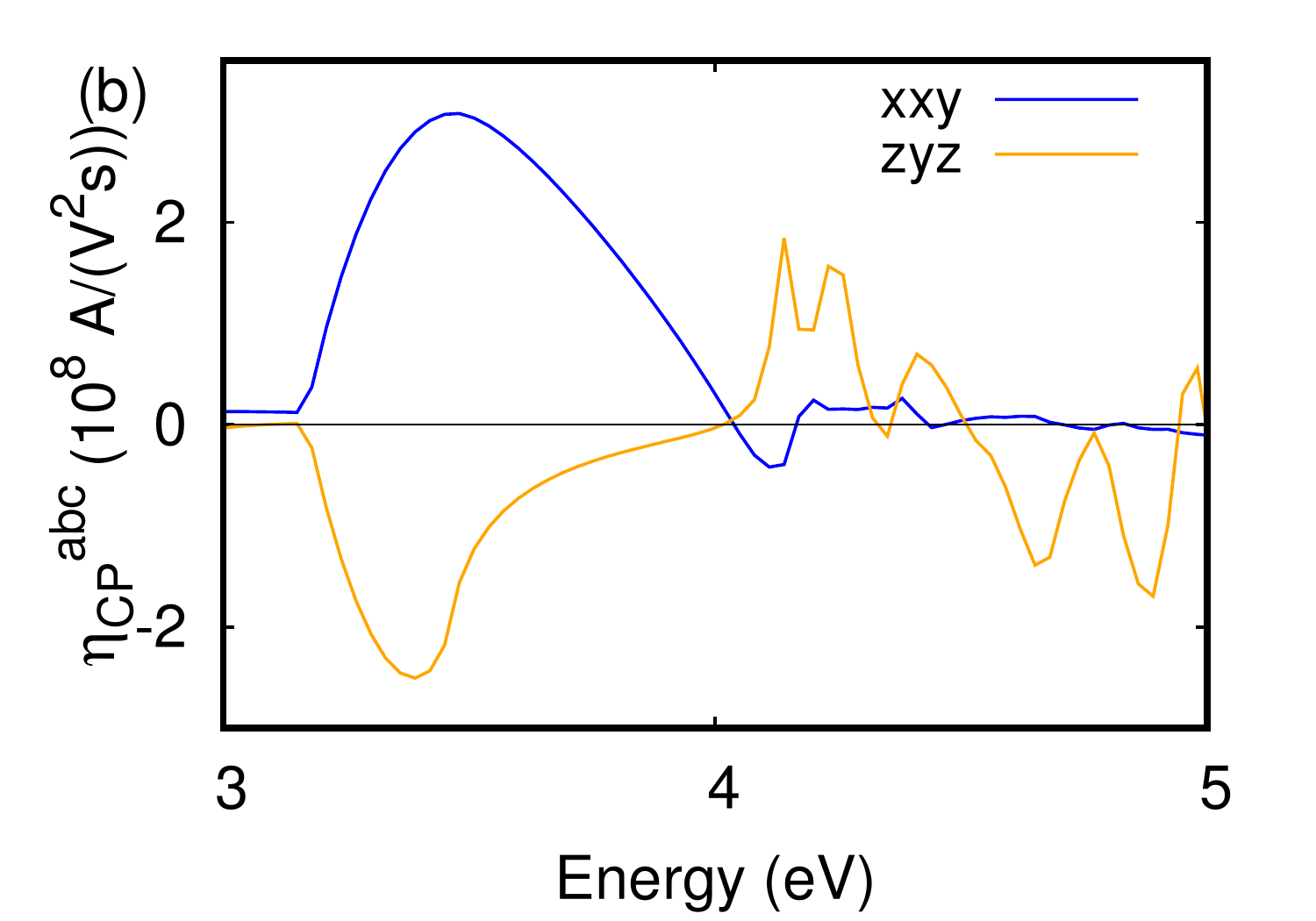}
\caption{Allowed components of the (a) LP spin shift photoconductivity and (b) CP spin injection photoconductivity without SOC.}
\label{NOSOC_spin_photocurrents}
\end{figure}

The LP spin shift and CP spin injection conductivities, shown in Fig.~\ref{NOSOC_spin_photocurrents}(a), (b), respectively, are the only nonzero contributions in the absence of SOC. The allowed tensor elements coincide with those permitted in the relativistic case for the spin parallel to the Néel vector. The dominant elements are $\sigma_{\mathrm{LP}}^{yyy}$, $\sigma_{\mathrm{LP}}^{yzz}$, and $\sigma_{\mathrm{LP}}^{zyz}$ for the LP shift conductivity, and $\eta_{\mathrm{CP}}^{xxy}$ and $\eta_{\mathrm{CP}}^{zyz}$ for the CP injection conductivity.

\clearpage

\section{Spin photoconductivities with SOC}

Here, we show the main elements of all the spin conductivities for the Néel vector orientations $\mathbf{N}\parallel x$, $\mathbf{N}\parallel y$, and $\mathbf{N}\parallel z$.

\subsection{LP shift conductivity}

\begin{figure*}[h]
\centering
\parbox{\textwidth}{%
  \centering
  \includegraphics[width=0.32\textwidth]{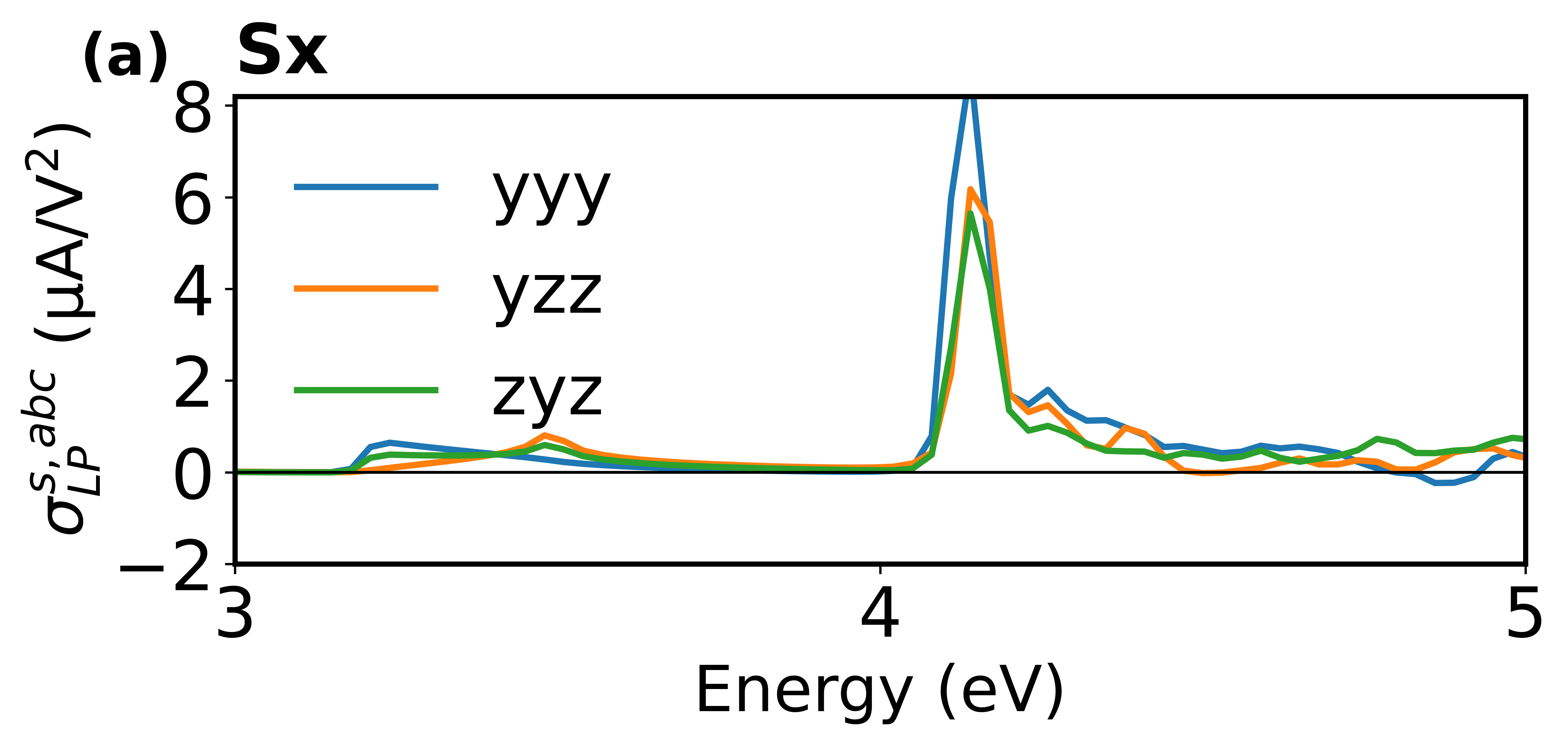}\hfill
  \includegraphics[width=0.32\textwidth]{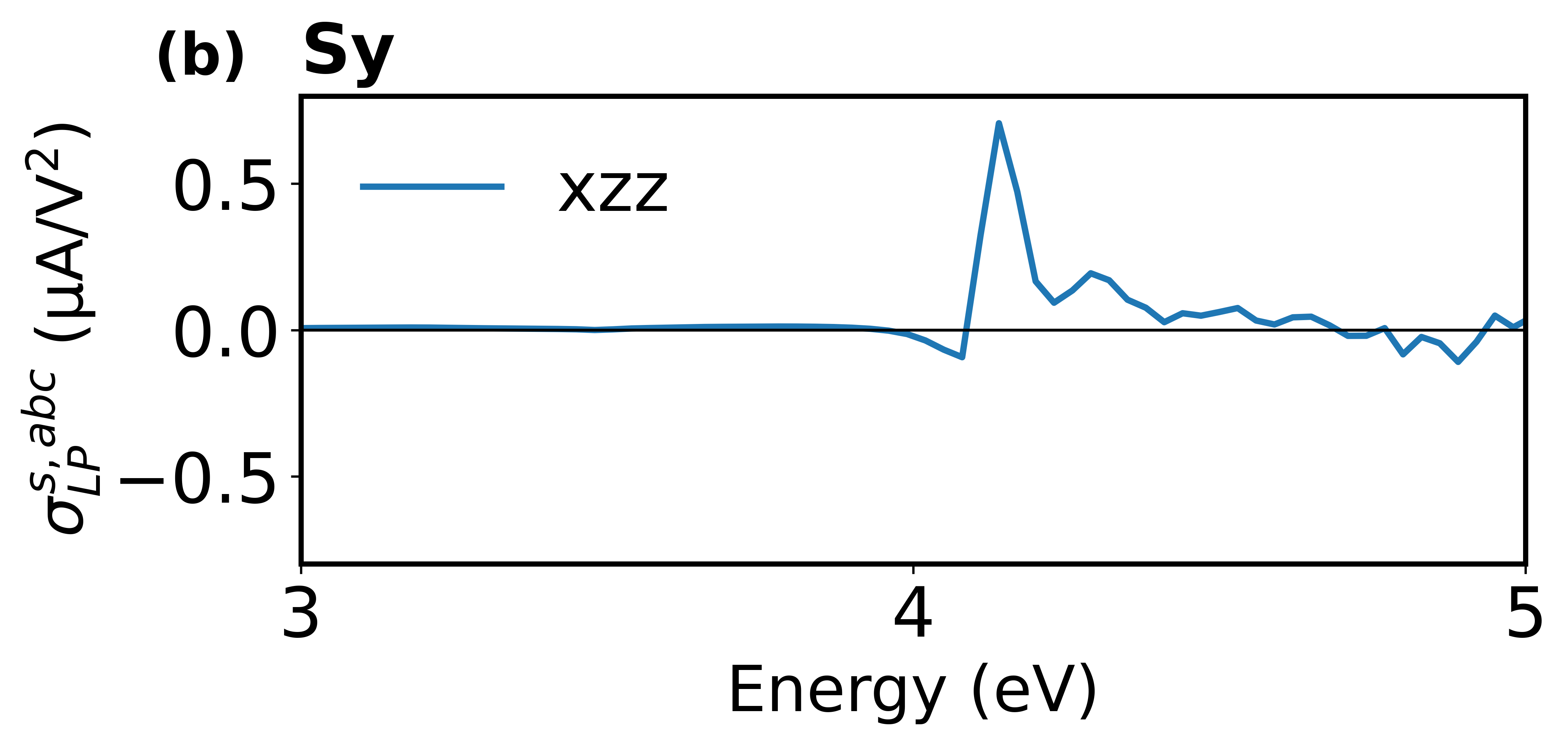}\hfill
  \includegraphics[width=0.32\textwidth]{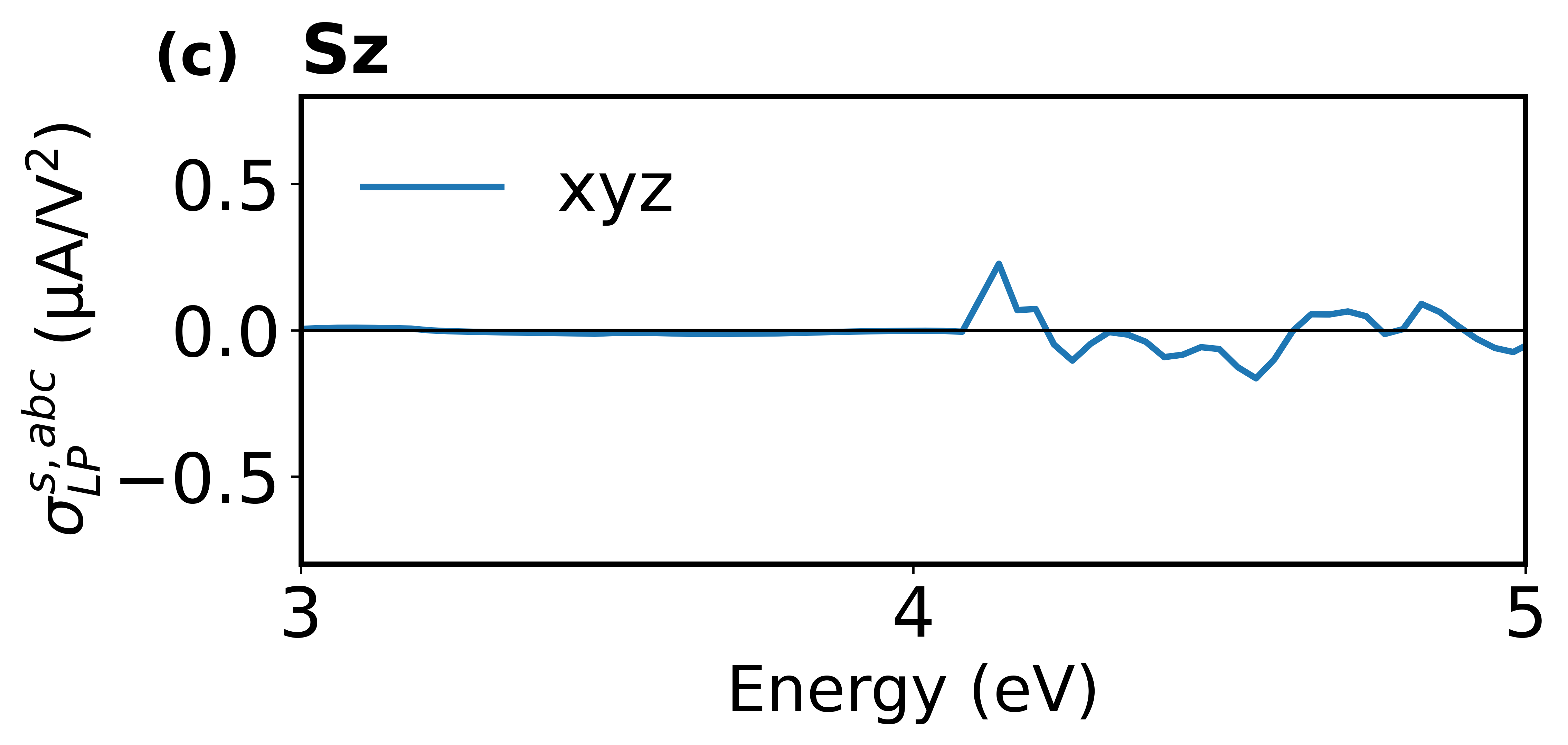}%
}
\vspace{1mm}
\parbox{\textwidth}{%
  \centering
  \includegraphics[width=0.32\textwidth]{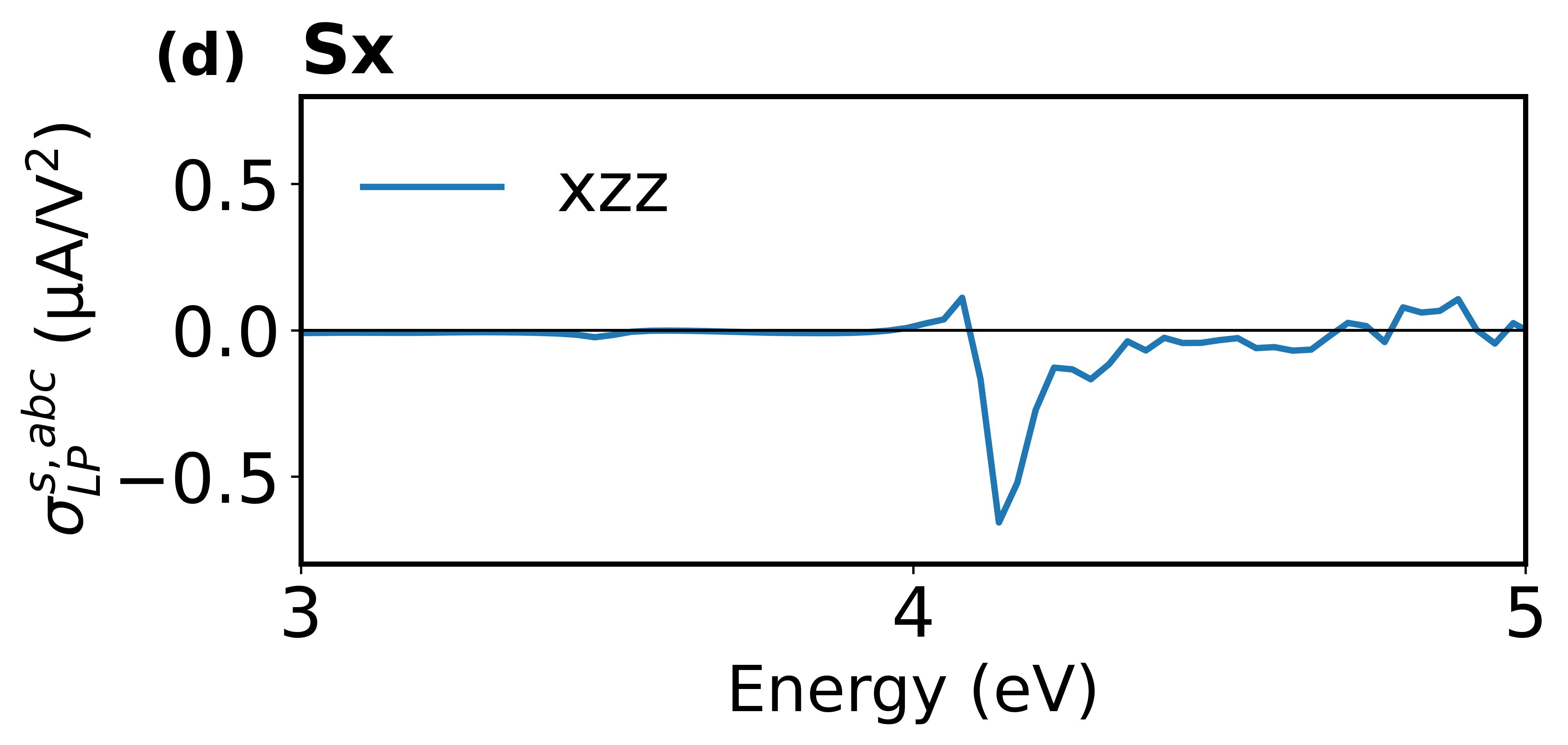}\hfill
  \includegraphics[width=0.32\textwidth]{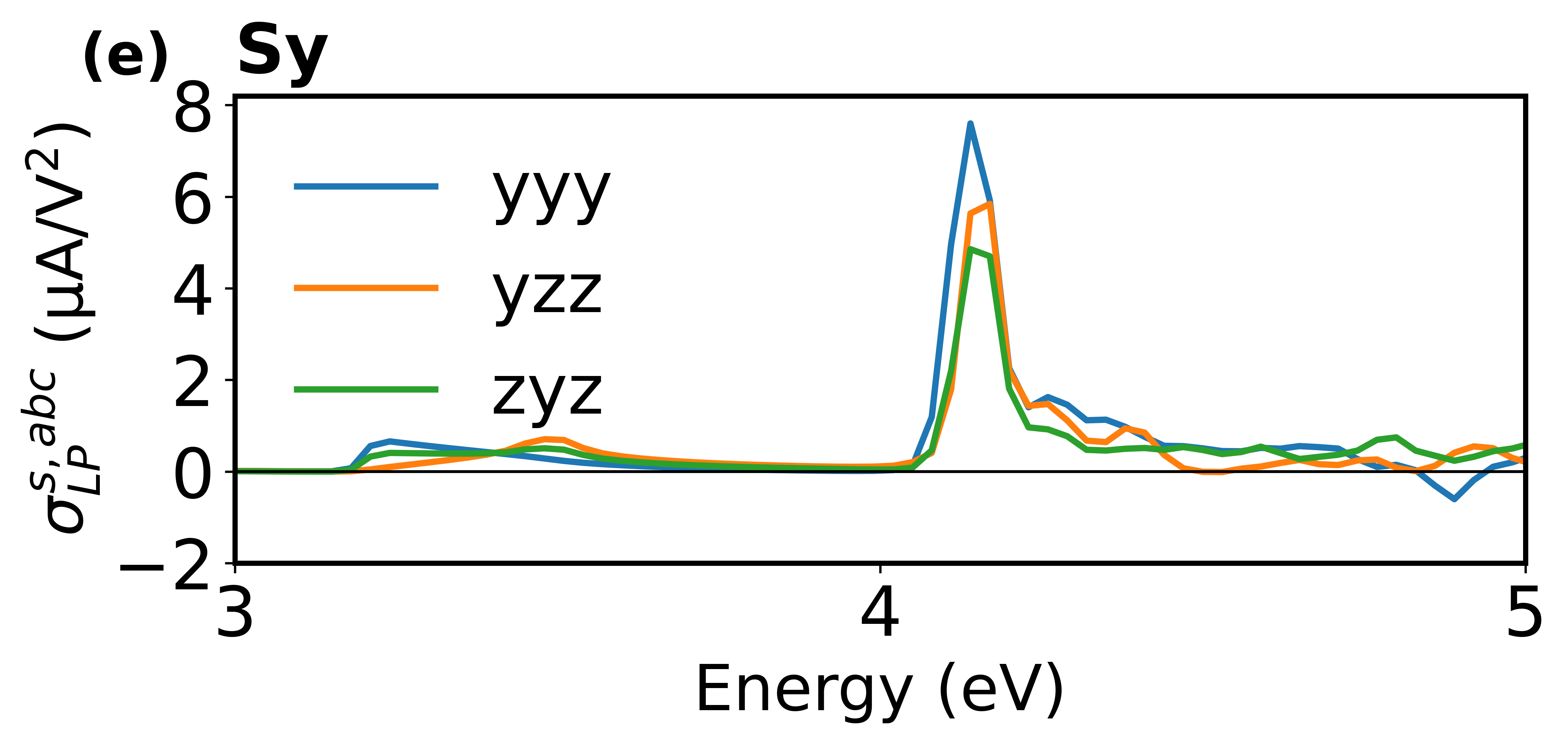}\hfill
  \includegraphics[width=0.32\textwidth]{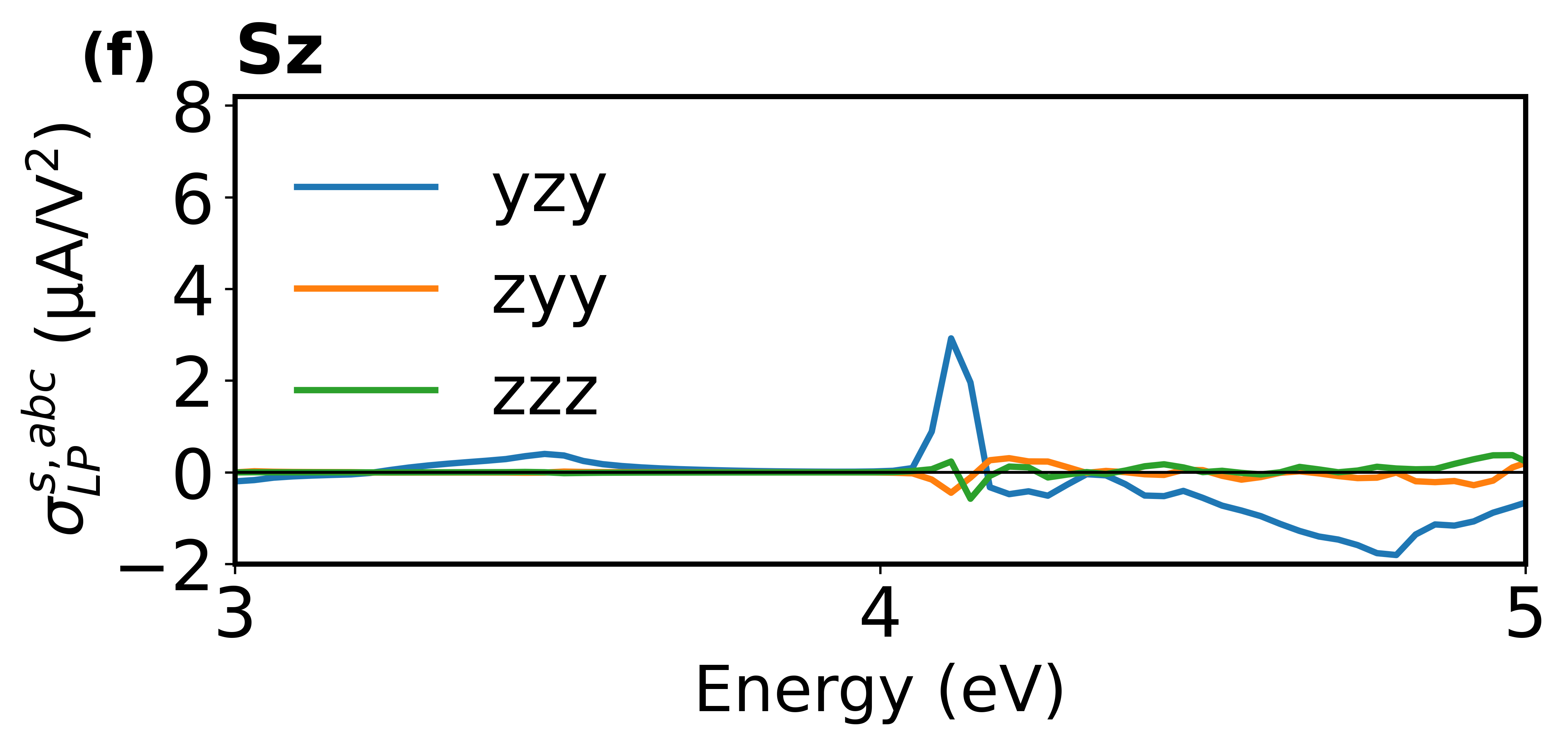}%
}
\vspace{1mm}
\parbox{\textwidth}{%
  \centering
  \includegraphics[width=0.32\textwidth]{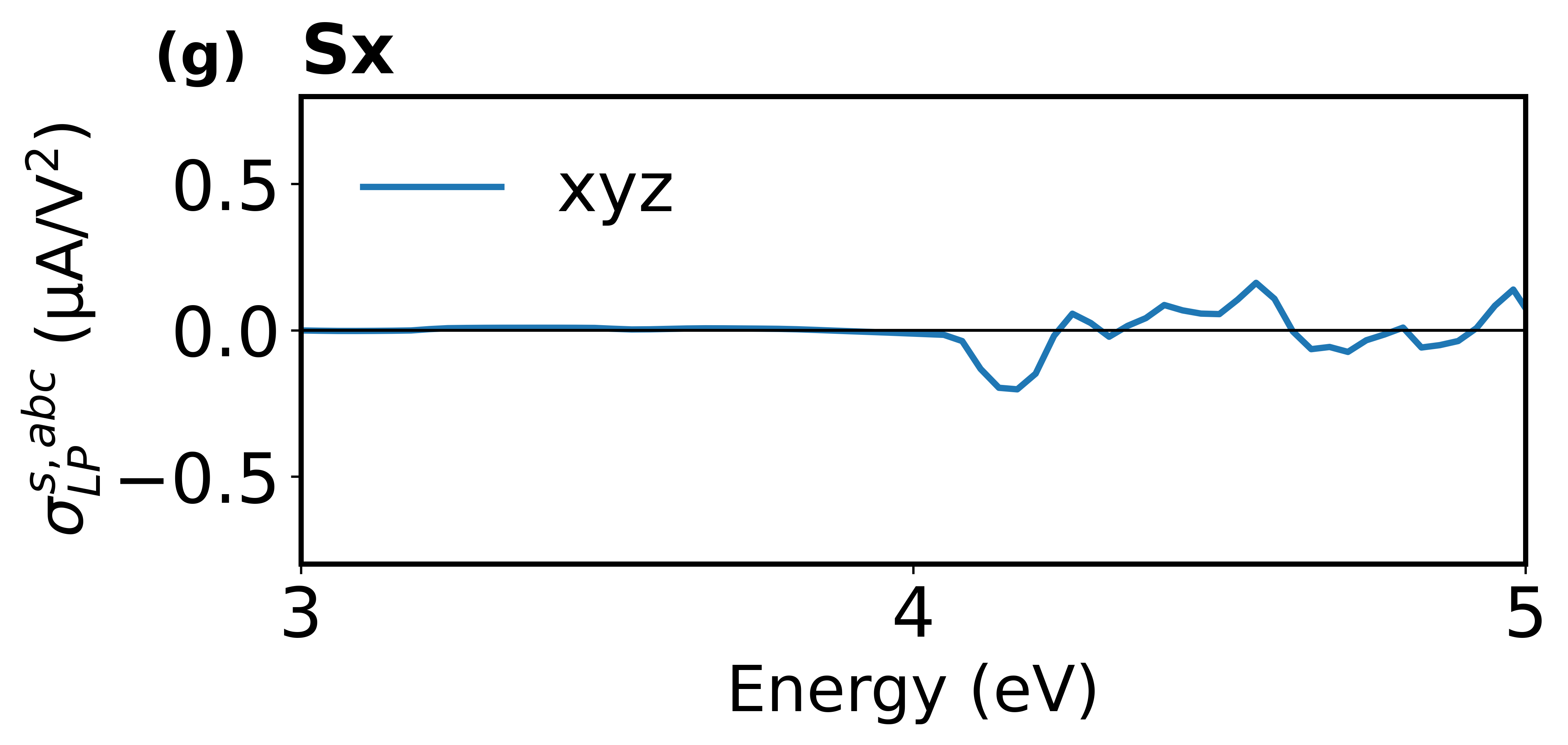}\hfill
  \includegraphics[width=0.32\textwidth]{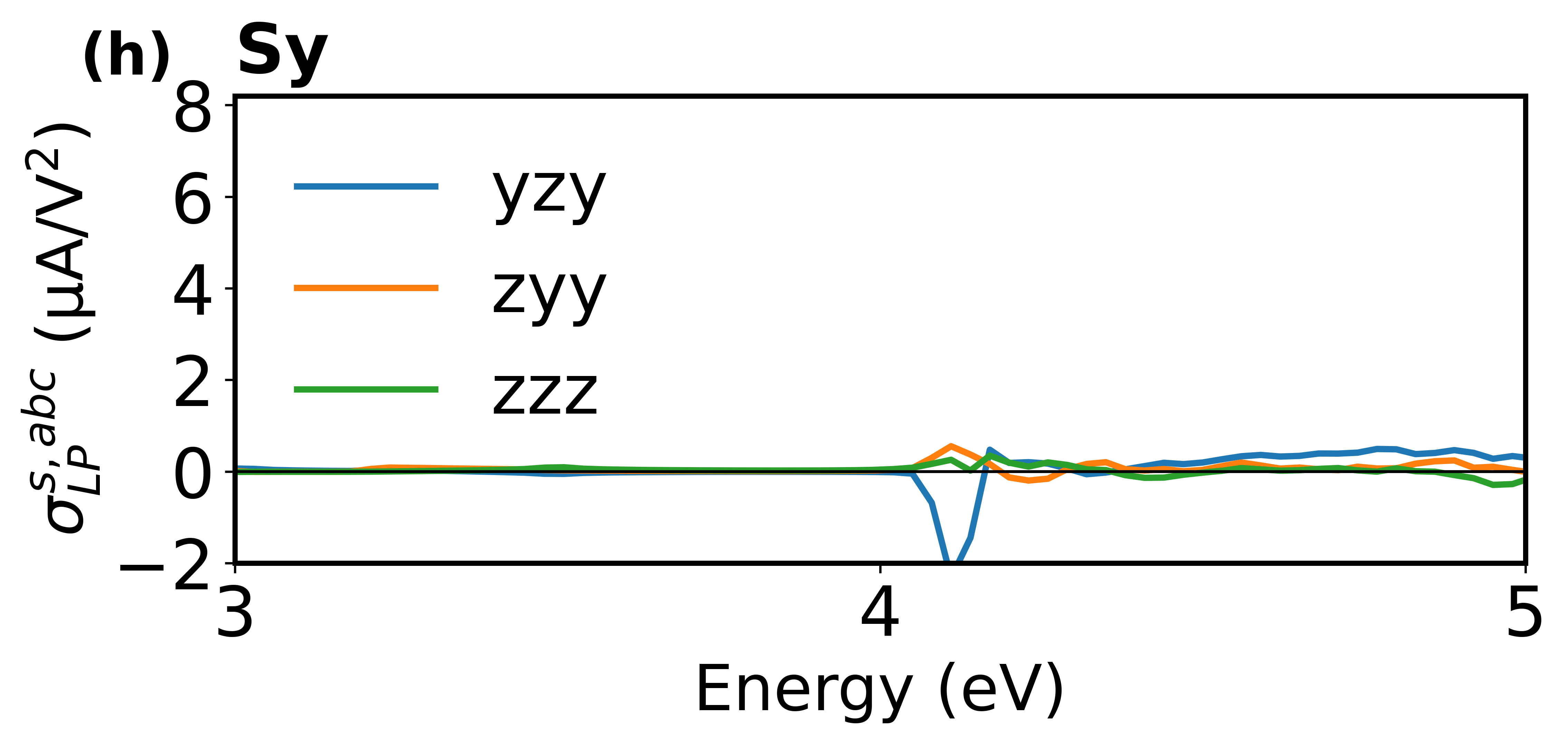}\hfill
  \includegraphics[width=0.32\textwidth]{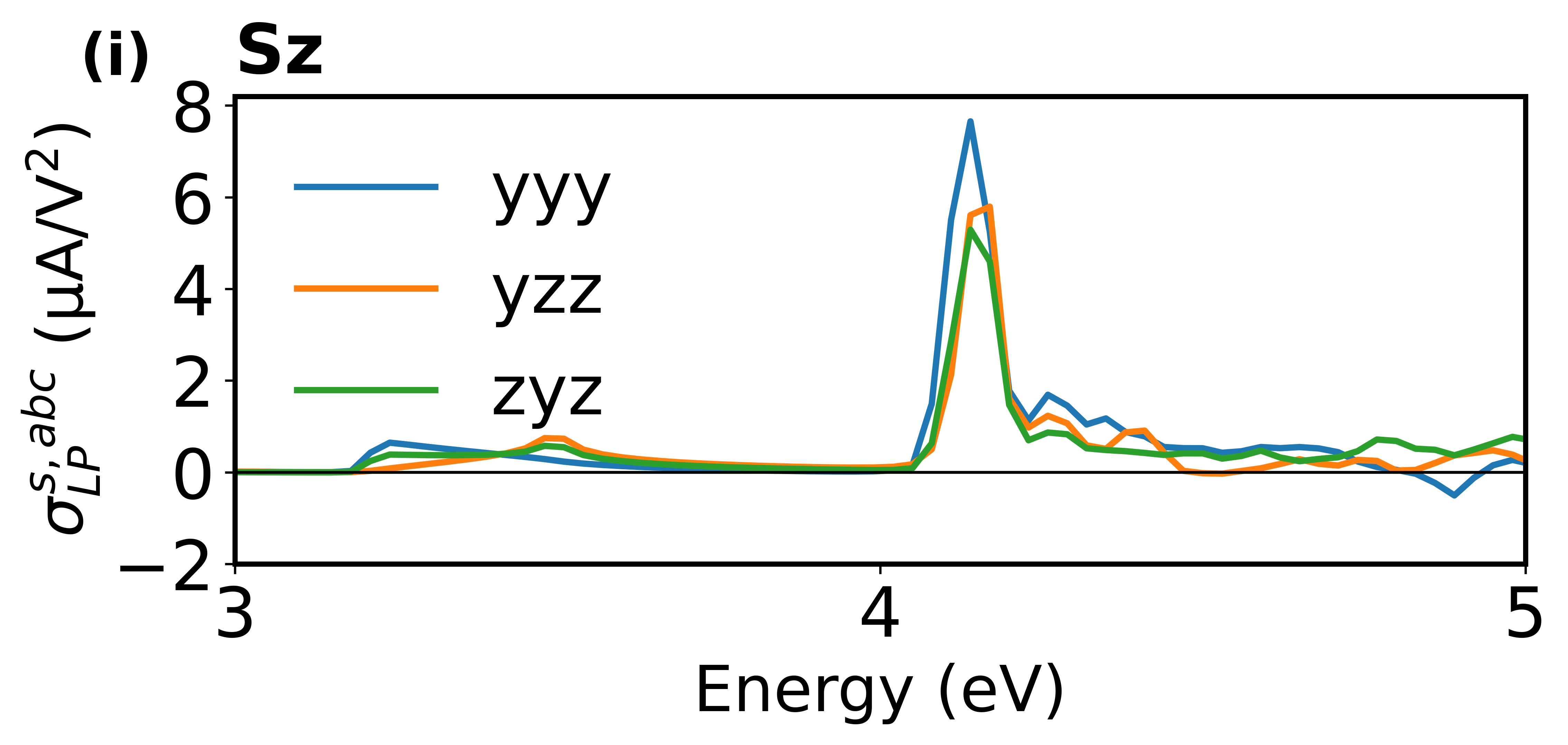}%
}
\caption{Main components of the LP spin shift photoconductivity for (a–c) $\mathbf N\parallel x$, (d–f) $\mathbf N\parallel y$, and (g–i) $\mathbf N\parallel z$, resolved into spin components $S_x$, $S_y$, and $S_z$. Components of comparable magnitude are plotted using the same $y$-axis range, while different scales are used for components of different orders of magnitude.}
\label{fig:spin_shift_9panel}
\end{figure*}

In Fig. \ref{fig:spin_shift_9panel} we report the main elements of the LP spin shift conductivity for the Néel vector orientations $\mathbf{N}\parallel x$ (a-c), $\mathbf{N}\parallel y$ (d-f), and $\mathbf{N}\parallel z$ (g-i) and for the spin components $S_x$, $S_y$ and $S_z$.
This response is odd under time reversal, and it shows different components depending on the Néel vector orientations. It shows nonrelativistic contributions and elements related to weak ferromagnetism and to relativistic effects.
The hierarchy is the same for all the cases: the dominant contributions correspond to the spin along the Néel vector, the intermediate ones arise from spin canting, and the smallest are SOC-driven.
By looking at Fig. \ref{fig:spin_shift_9panel} (a), (e), (i), we can see that the elements along the Néel vector direction are inherited from the nonrelativistic case, and the main contributions are $\sigma_{\mathrm{LP}}^{s,yyy}$, $\sigma_{\mathrm{LP}}^{s,yzz}$ and $\sigma_{\mathrm{LP}}^{s,zyz}$, with $s=x,y,z$ for $\mathbf N\parallel x,y,z$, reaching approximately
$8~\mu\text{A}/\text{V}^2$, $6~\mu\text{A}/\text{V}^2$ and $5~\mu\text{A}/\text{V}^2$ near $\hbar\omega=4.2$~eV,
respectively. 
The elements presented in Figs. \ref{fig:spin_shift_9panel}(f) and (h) originate from the weak ferromagnetism and are symmetry-allowed by coplanar spin groups, as shown in the symmetry analysis. 
The largest peaks for these contributions are reached by $\sigma_{\mathrm{LP}}^{z,yzy}$ and $\sigma_{\mathrm{LP}}^{y,yzy}$, for $\mathbf{N}\parallel$ $y$ and $\mathbf{N}\parallel$ $z$, respectively, of $3~\mu\text{A}/\text{V}^2$ near $\hbar\omega=4.2$~eV.
They persist even in the absence of SOC when the weakly ferromagnetic configuration is enforced, indicating that SOC is only required to stabilize the noncollinear magnetic structure. When the weak-ferromagnetic component is switched off, these contributions do not survive. The remaining elements are activated by including SOC and are the smallest; they are the ones showed in Figs. \ref{fig:spin_shift_9panel} (b), (c), (d) and (g). Here we show only the largest contributions, which reach $0.2$-$0.5~\mu\text{A}/\text{V}^2$, around one order of magnitude smaller than the dominant nonrelativistic elements.

\clearpage

\subsection{LP injection conductivity}

\begin{figure*}[h!]
\centering
\parbox{\textwidth}{%
  \centering
  \includegraphics[width=0.32\textwidth]{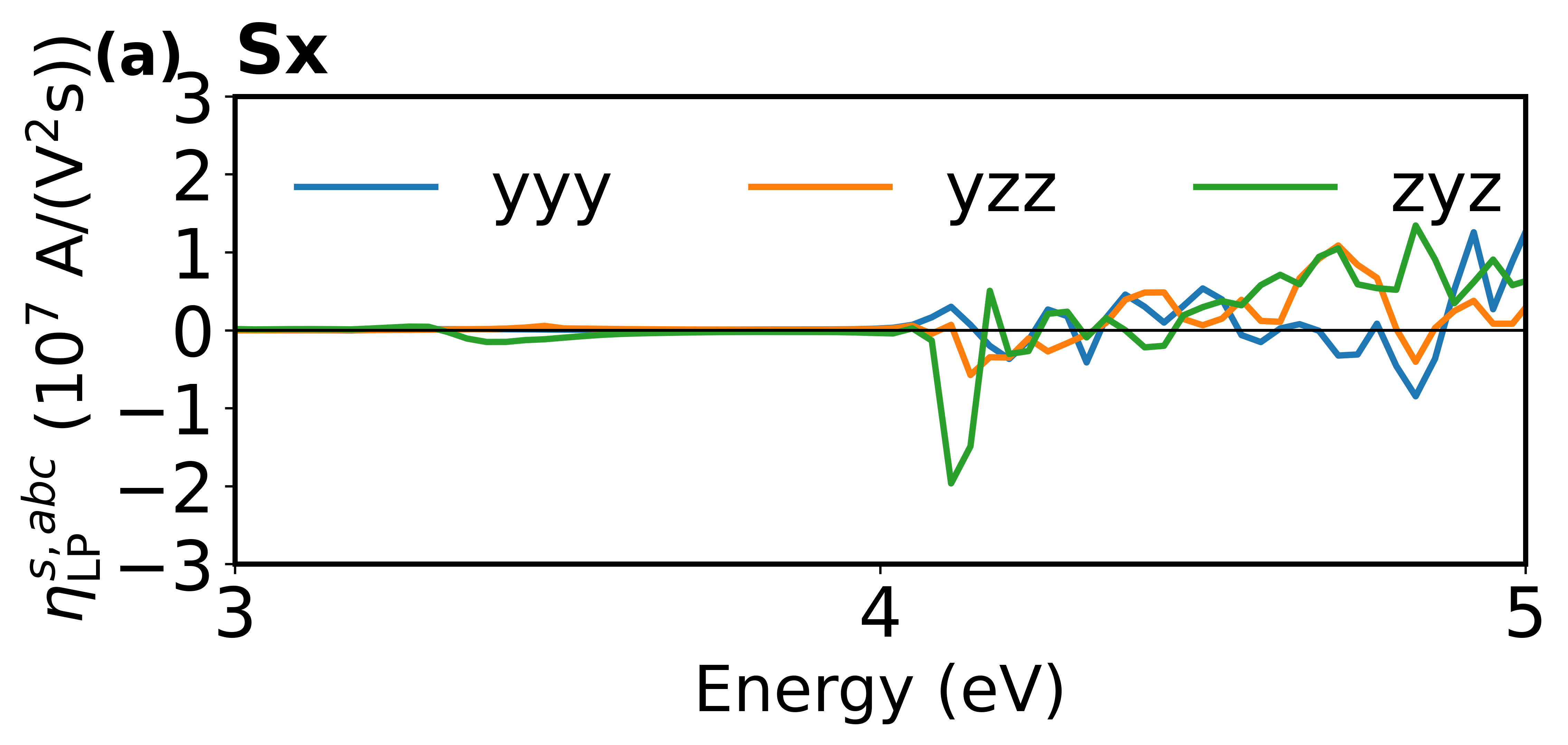}\hfill
  \includegraphics[width=0.32\textwidth]{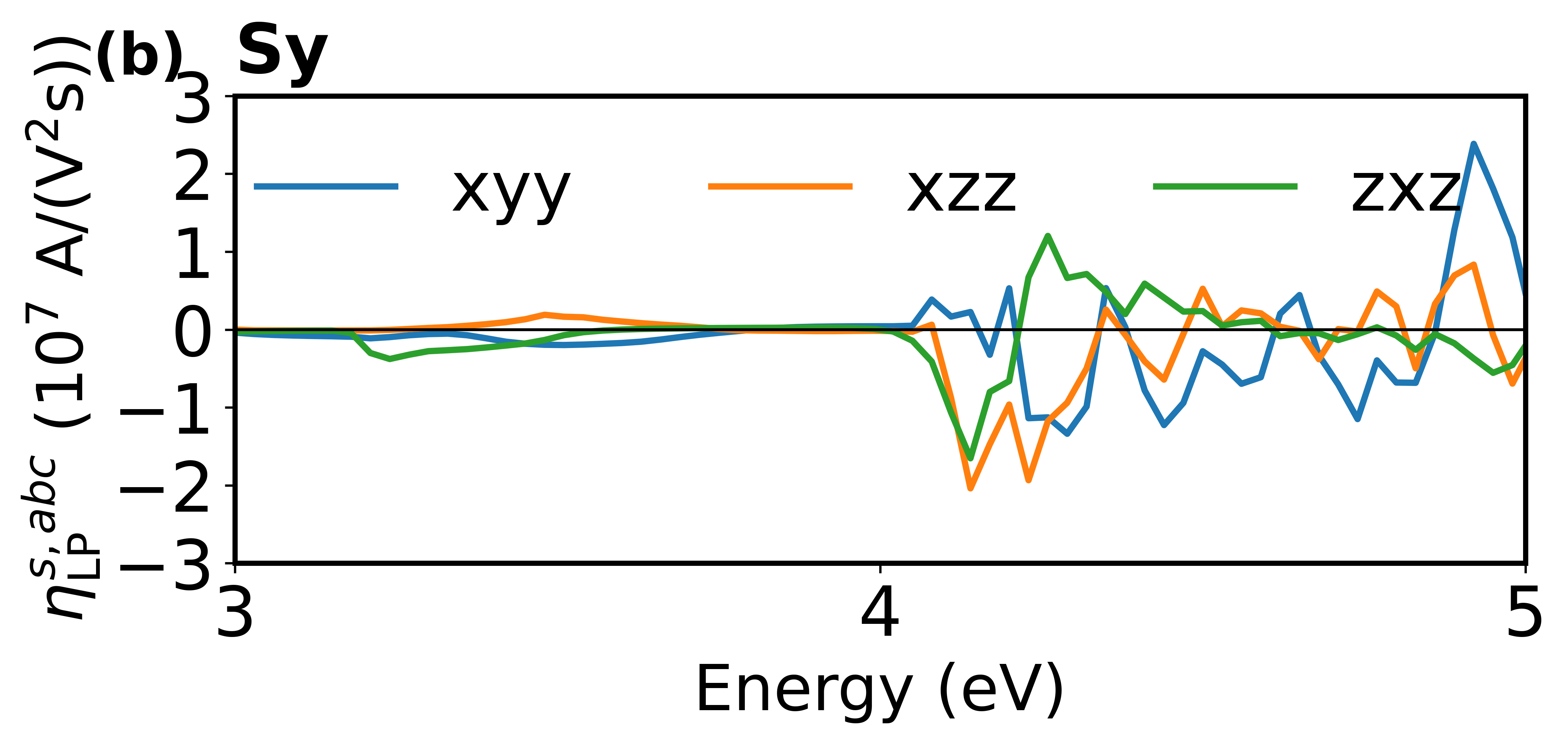}\hfill
  \includegraphics[width=0.32\textwidth]{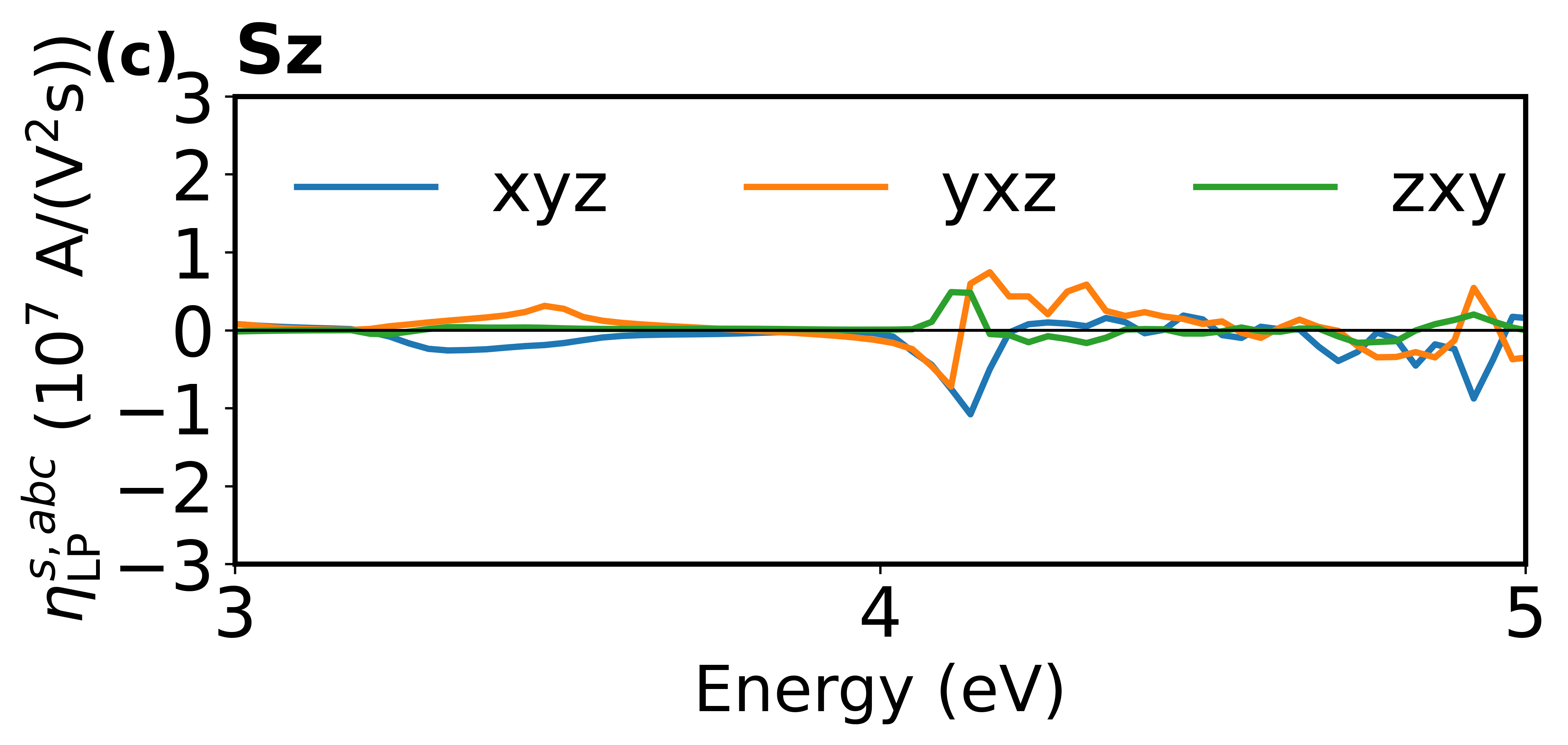}%
}
\vspace{1mm}
\parbox{\textwidth}{%
  \centering
  \includegraphics[width=0.32\textwidth]{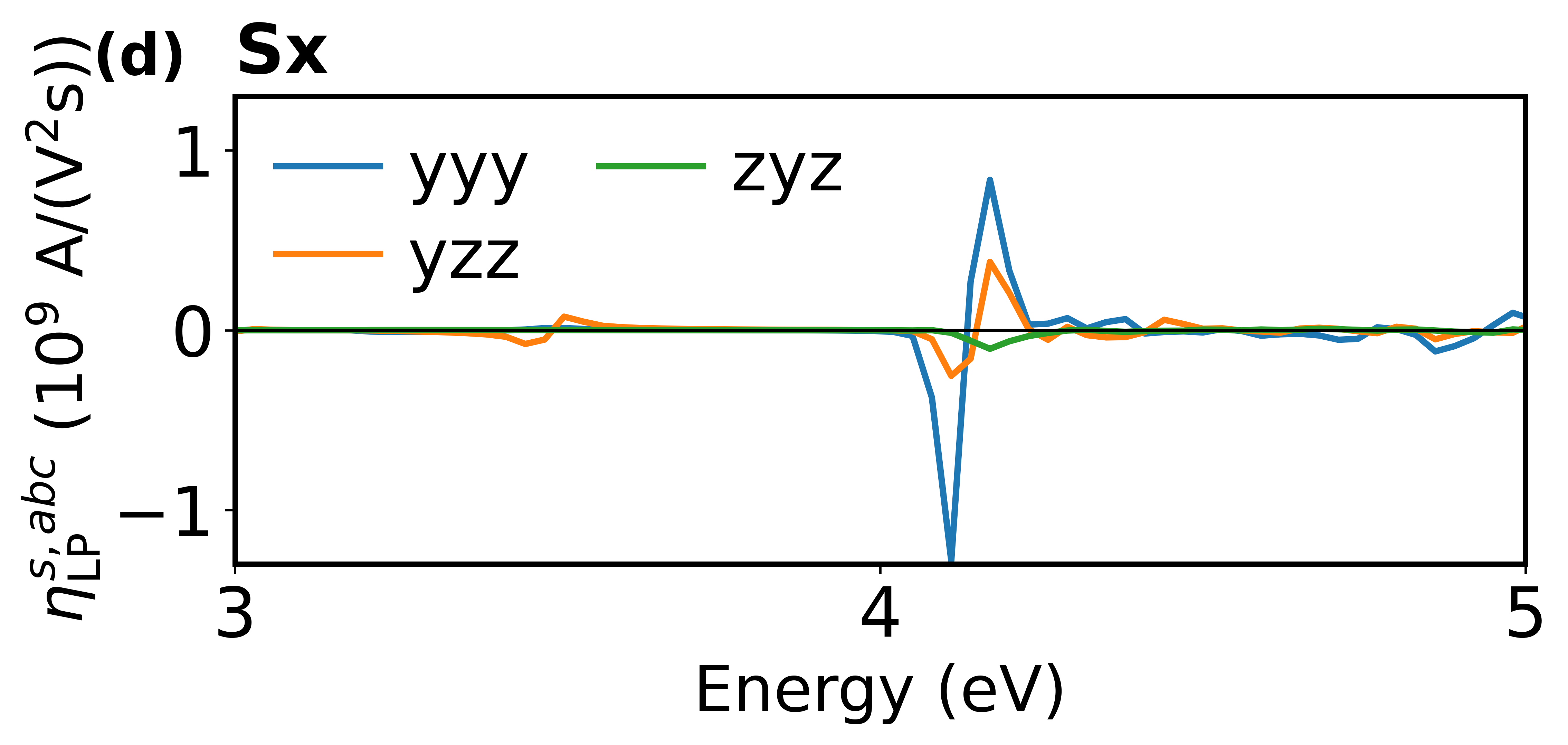}\hfill
  \includegraphics[width=0.32\textwidth]{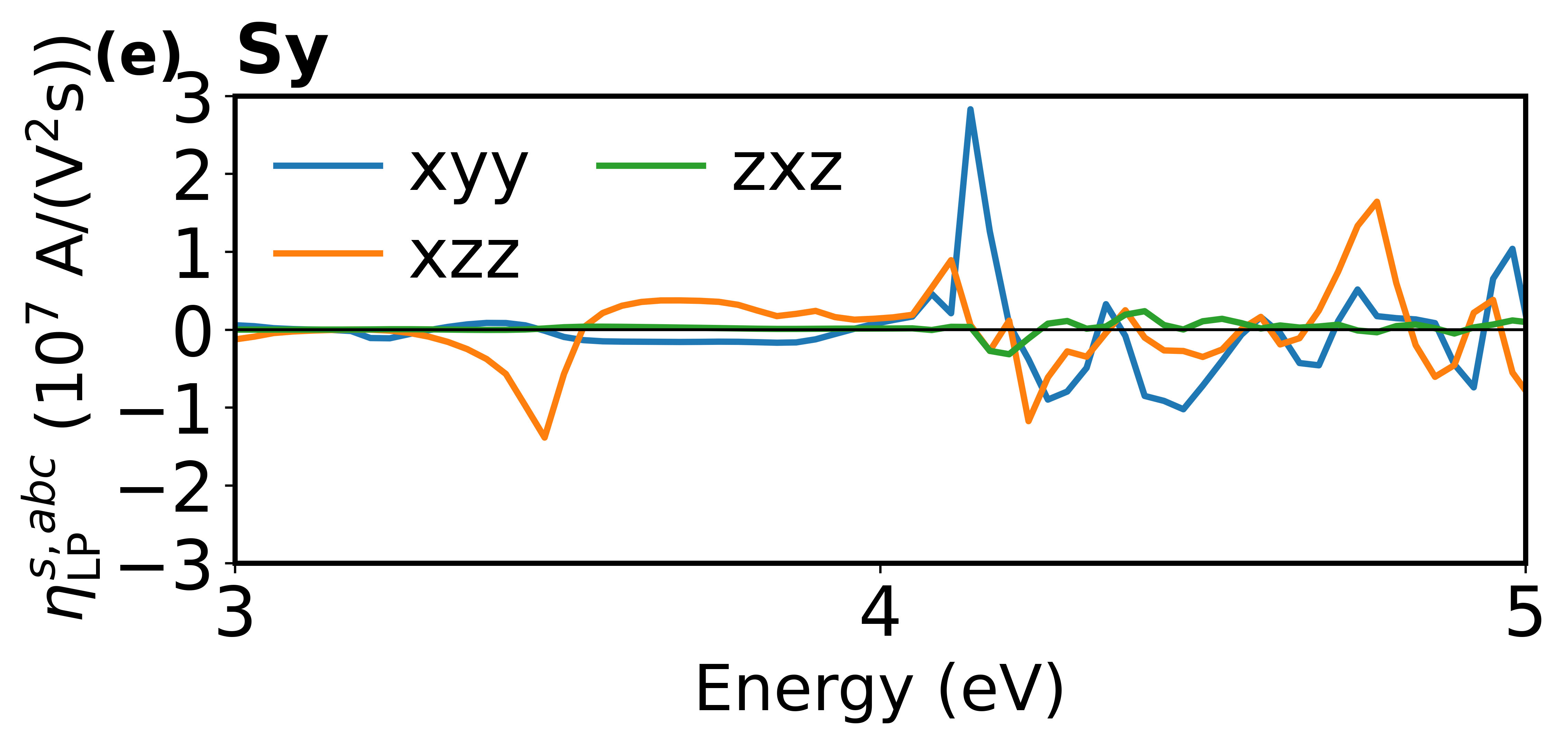}\hfill
  \includegraphics[width=0.32\textwidth]{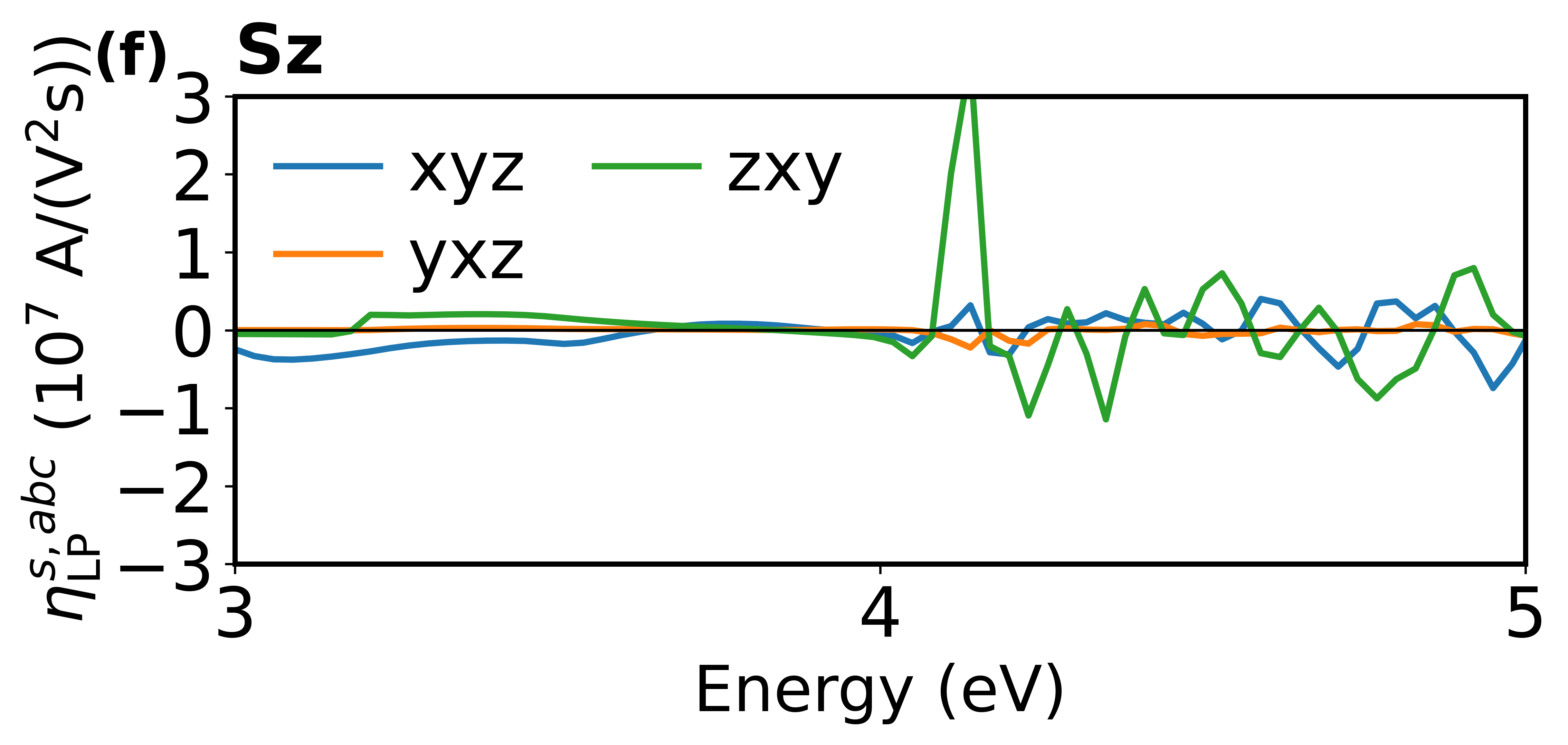}%
}
\vspace{1mm}
\parbox{\textwidth}{%
  \centering
  \includegraphics[width=0.32\textwidth]{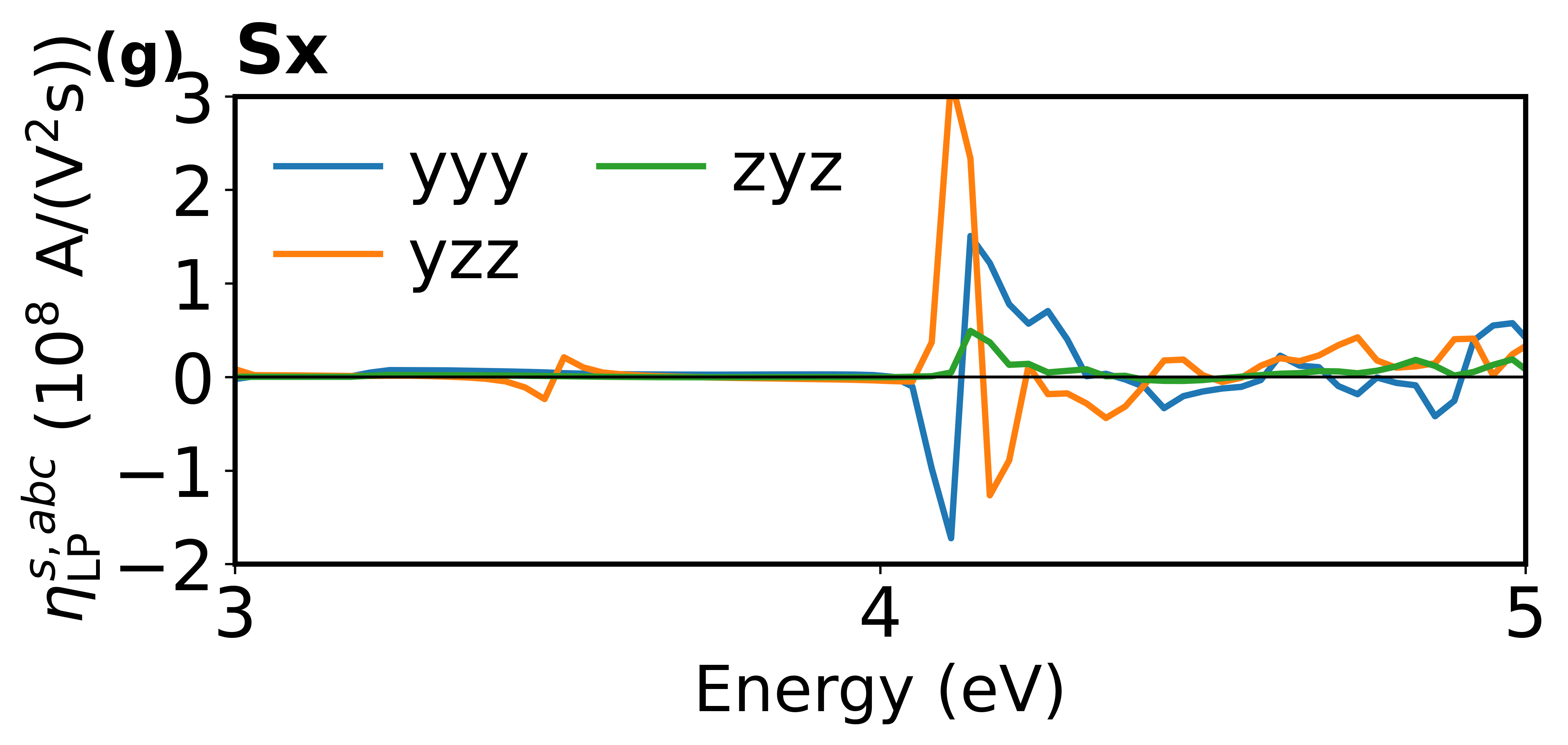}\hfill
  \includegraphics[width=0.32\textwidth]{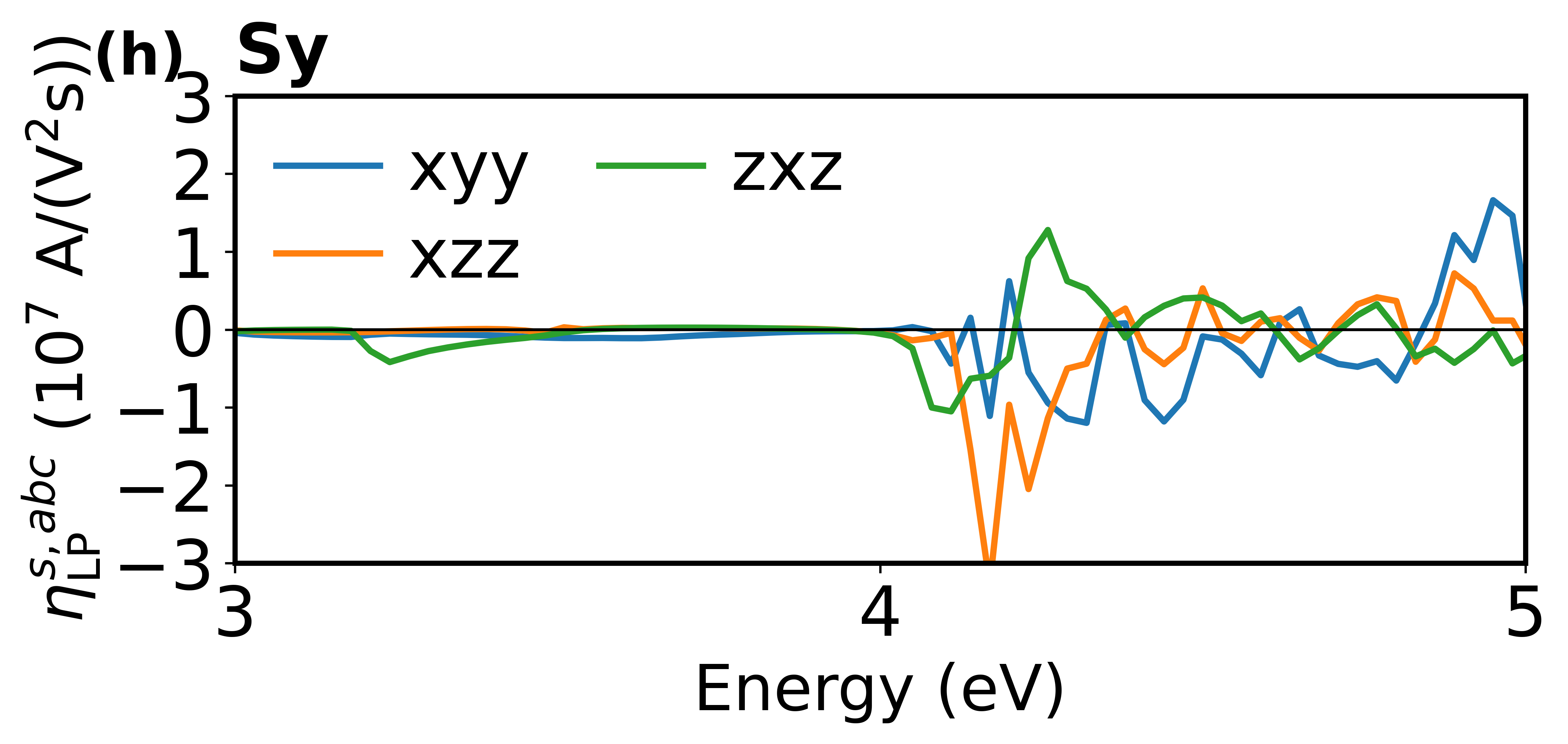}\hfill
  \includegraphics[width=0.32\textwidth]{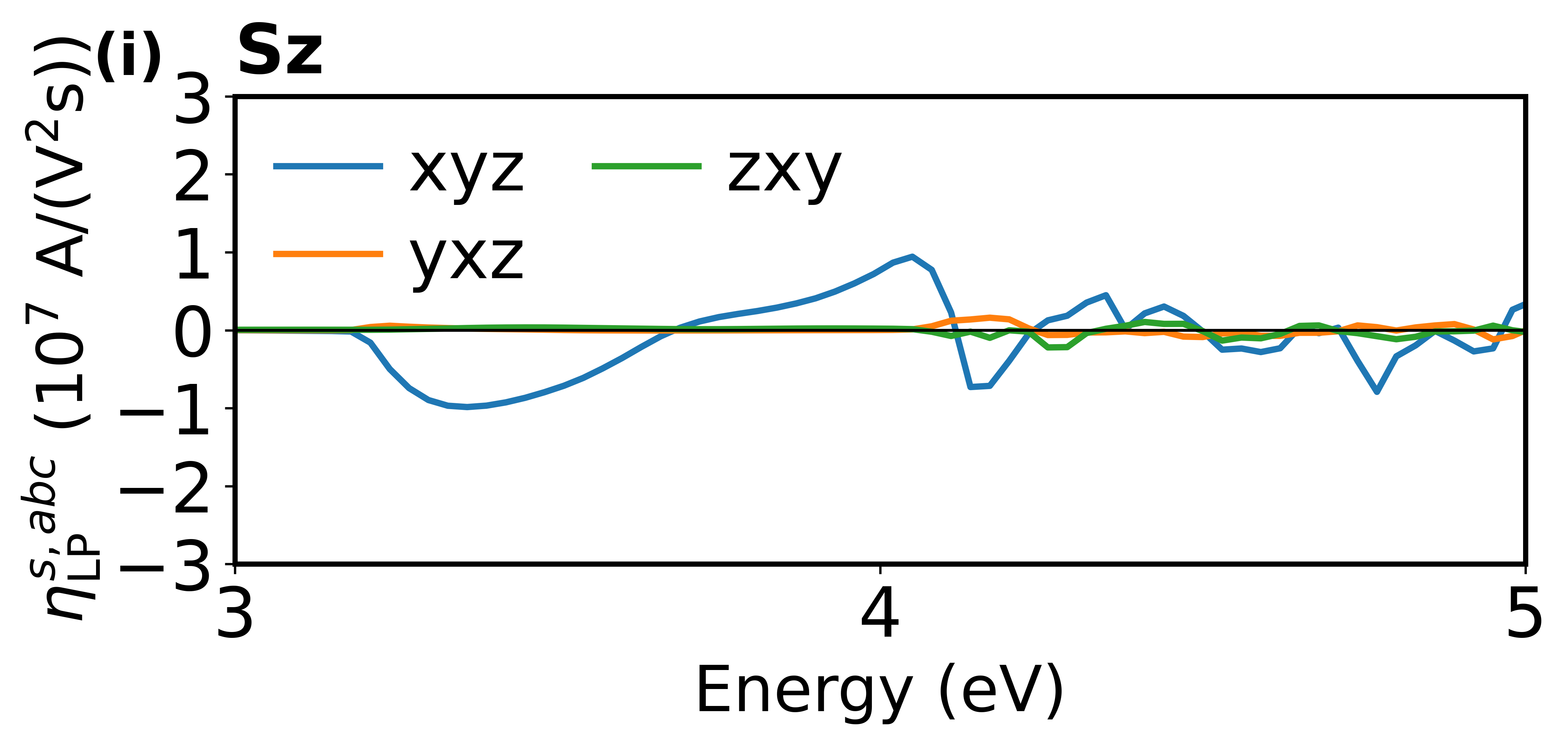}%
}
\caption{Main components of the LP spin injection photoconductivity for (a–c) $\mathbf N\parallel x$, (d–f) $\mathbf N\parallel y$, and (g–i) $\mathbf N\parallel z$, resolved into spin components $S_x$, $S_y$, and $S_z$. Components of comparable magnitude are plotted using the same $y$-axis range, while different scales are used for components of different orders of magnitude.}
\label{fig:spin_linear_injection_9panel}
\end{figure*}

In Fig. \ref{fig:spin_linear_injection_9panel} we report the main elements of the LP spin injection conductivity for the Néel vector orientations $\mathbf{N}\parallel x$ (a-c), $\mathbf{N}\parallel y$ (d-f), and $\mathbf{N}\parallel z$ (g-i) and for the spin components $S_x$, $S_y$ and $S_z$.
This response is even under time reversal, the same elements are allowed for all three Néel vector orientations, although their classification within the group--subgroup hierarchy depends on $\mathbf N$. Unlike the LP shift conductivity, the LP injection conductivity has no contribution in the collinear spin-group limit. It therefore separates into a coplanar sector and a purely SOC-induced sector.
The elements related to the coplanar sector are the dominant ones, namely those for $\mathbf{N}\parallel$ $y$ and  $\mathbf{N}\parallel$ $z$ with spin $S_x$.
$\eta_{\mathrm{LP}}^{x,yyy}$ reaches approximately $10^{9}~\mathrm{A}/(\mathrm{V}^{2}\mathrm{s})$ near $\hbar\omega=4.2$~eV for $\mathbf{N}\parallel$ $y$ (see Fig. \ref{fig:spin_linear_injection_9panel} (d)), while $\eta_{\mathrm{LP}}^{x,yzz}$ and $\eta_{\mathrm{LP}}^{x,yyy}$ reach approximately $10^{8}~\mathrm{A}/(\mathrm{V}^{2}\mathrm{s})$ near $\hbar\omega=4.2$~eV for $\mathbf{N}\parallel$ $z$ (see Fig. \ref{fig:spin_linear_injection_9panel} (g)).
The remaining contributions, namely those purely relativistic, are much smaller than these ones, and are of the order of $10^{7}~\mathrm{A}/(\mathrm{V}^{2}\mathrm{s})$, as shown in the other panels of Fig. \ref{fig:spin_linear_injection_9panel}.

\clearpage

\subsection{CP shift conductivity}

\begin{figure*}[h!]
\centering
\parbox{\textwidth}{%
  \centering
  \includegraphics[width=0.32\textwidth]{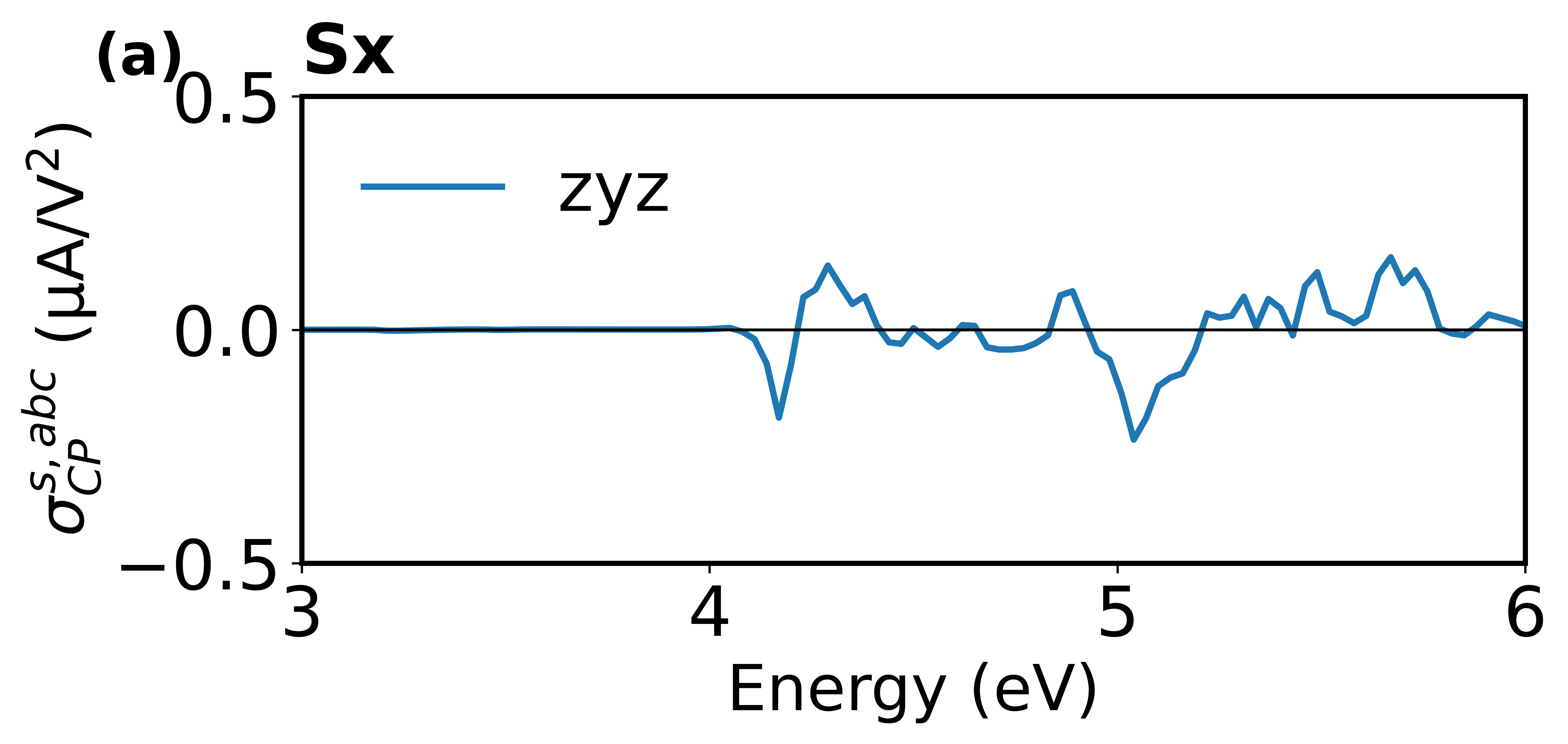}\hfill
  \includegraphics[width=0.32\textwidth]{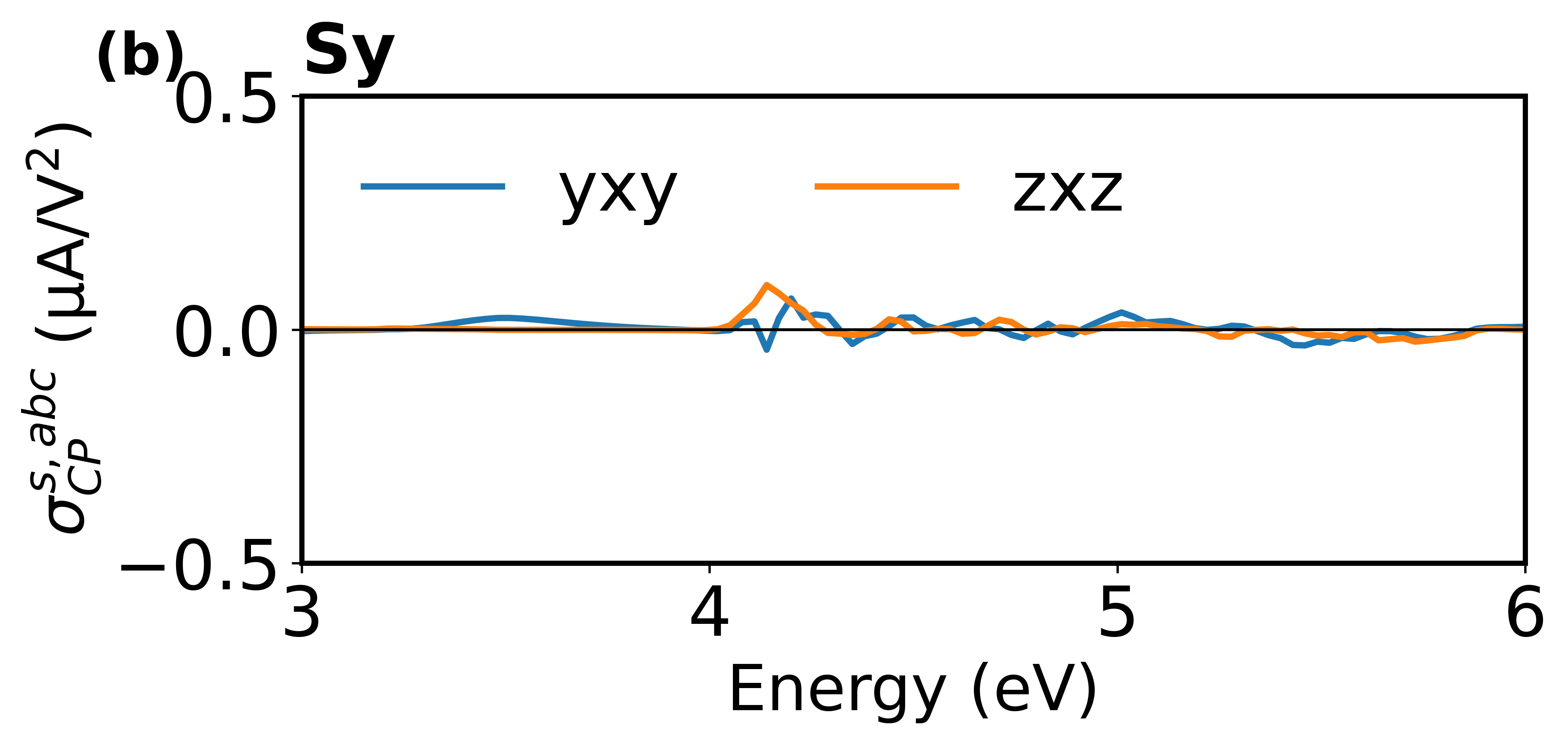}\hfill
  \includegraphics[width=0.32\textwidth]{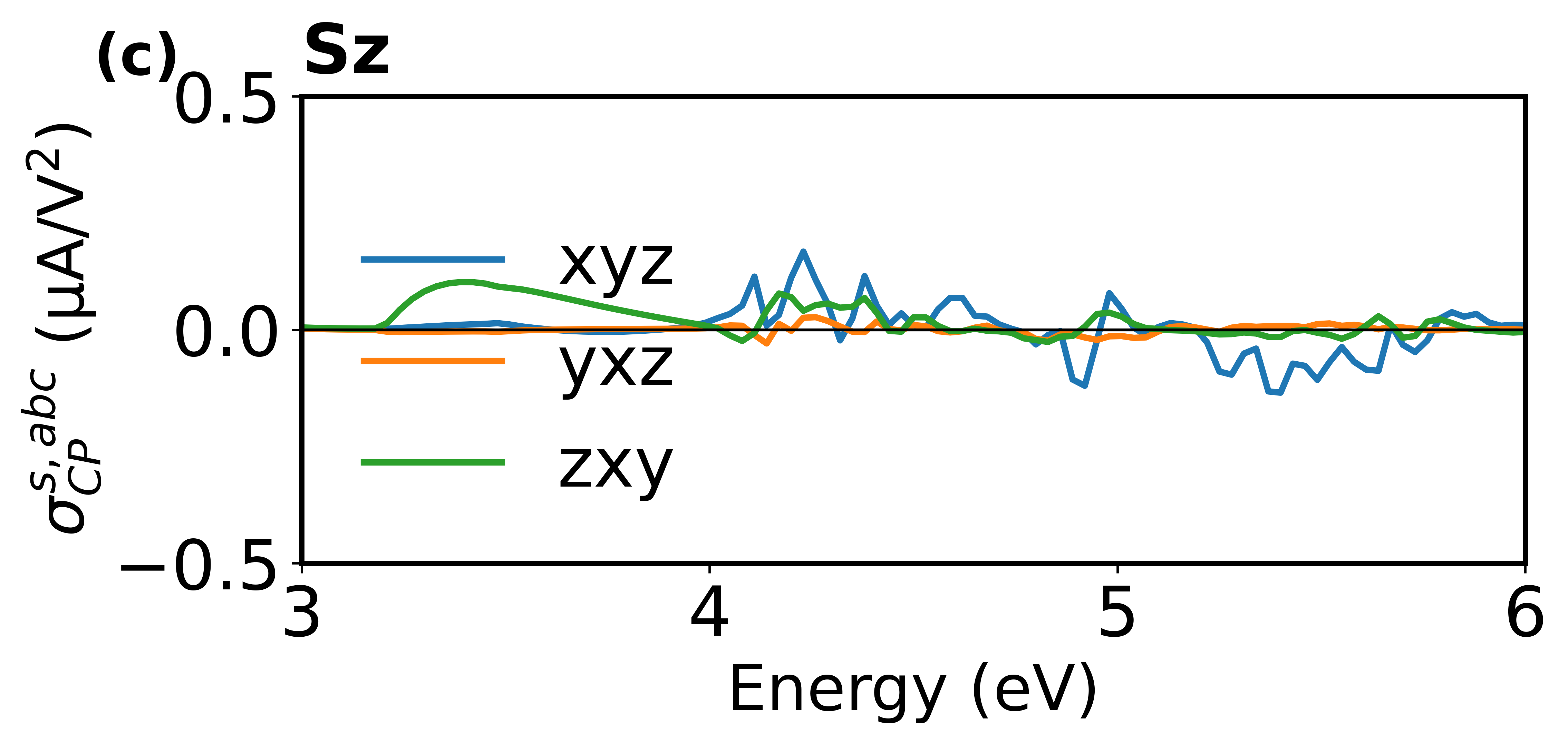}%
}
\vspace{1mm}
\parbox{\textwidth}{%
  \centering
  \includegraphics[width=0.32\textwidth]{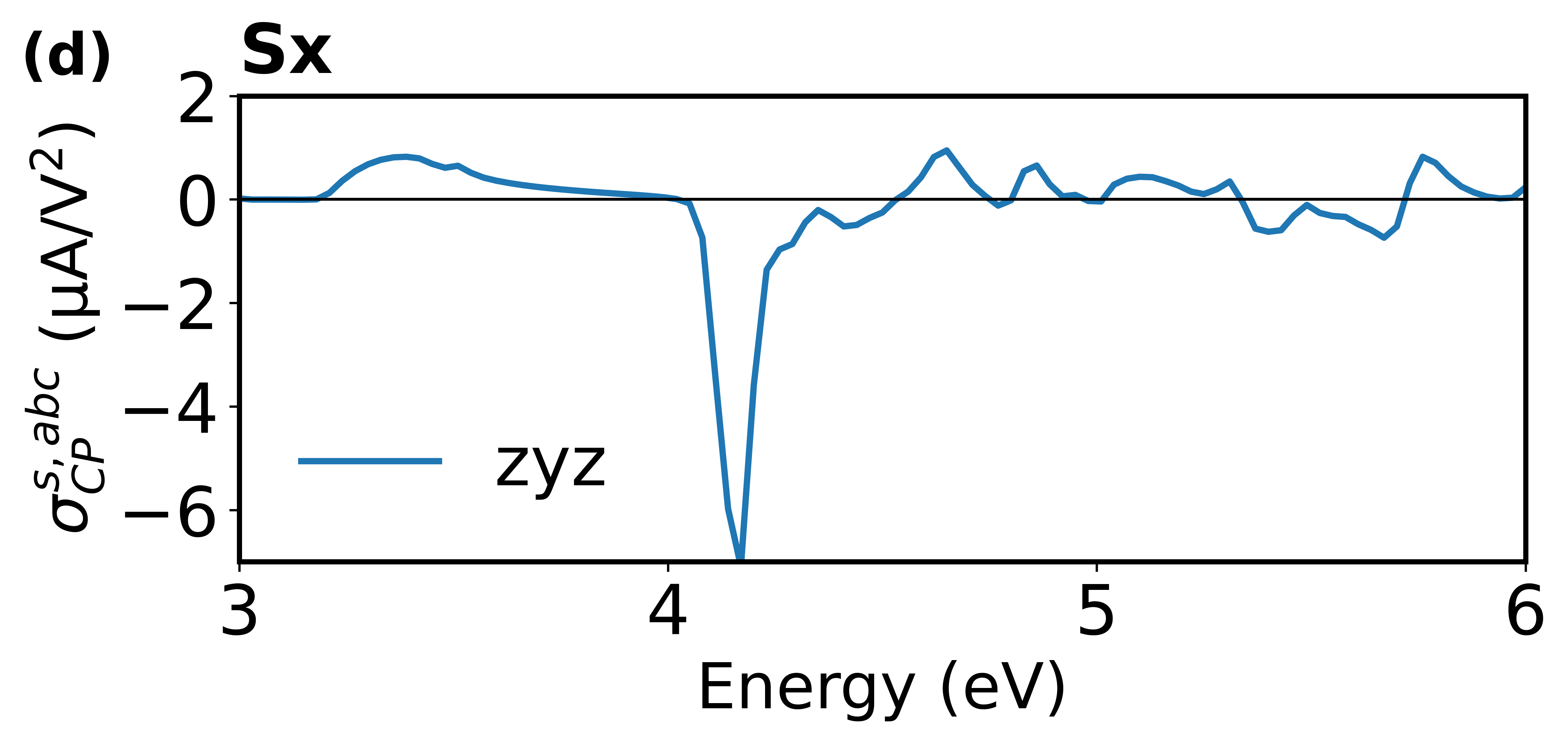}\hfill
  \includegraphics[width=0.32\textwidth]{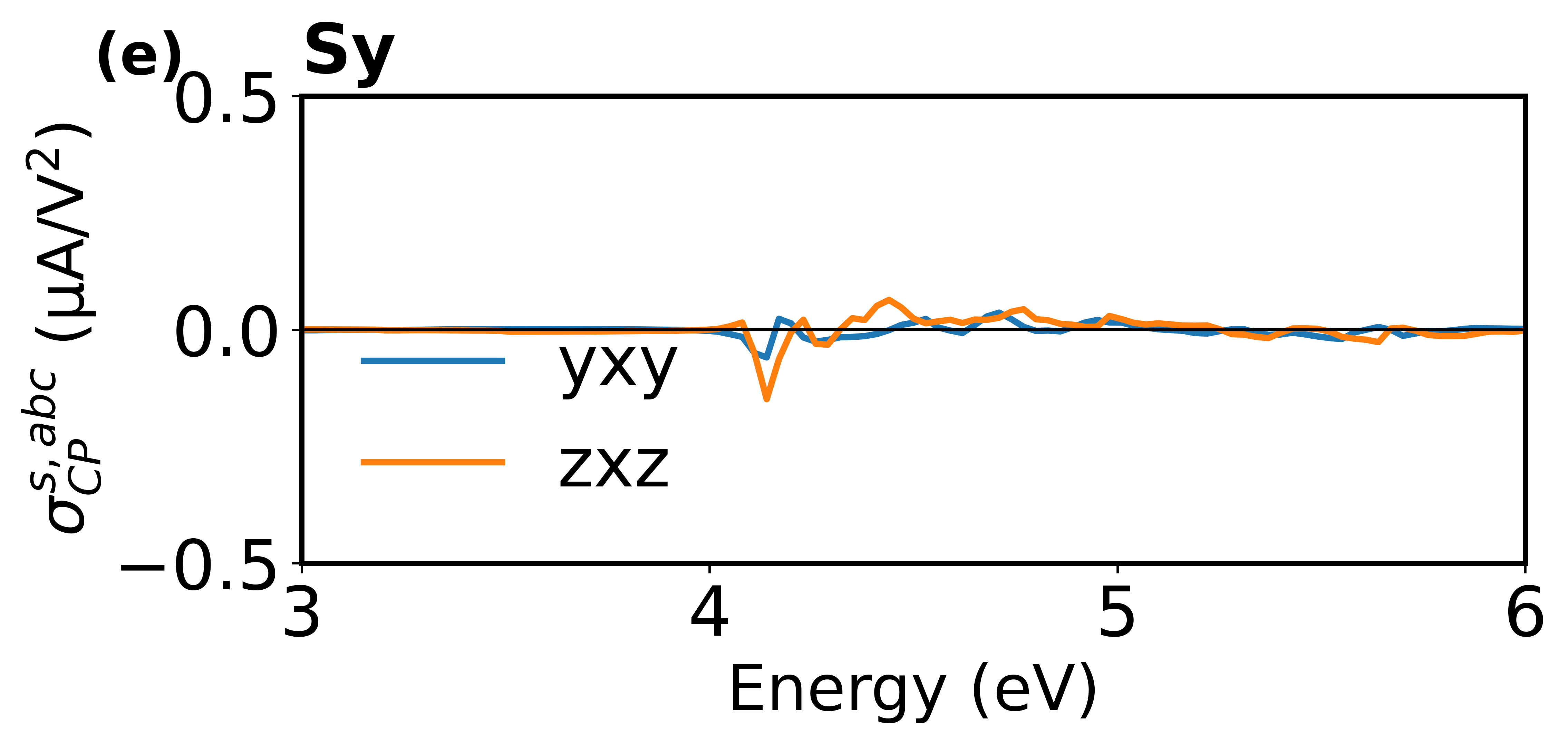}\hfill
  \includegraphics[width=0.32\textwidth]{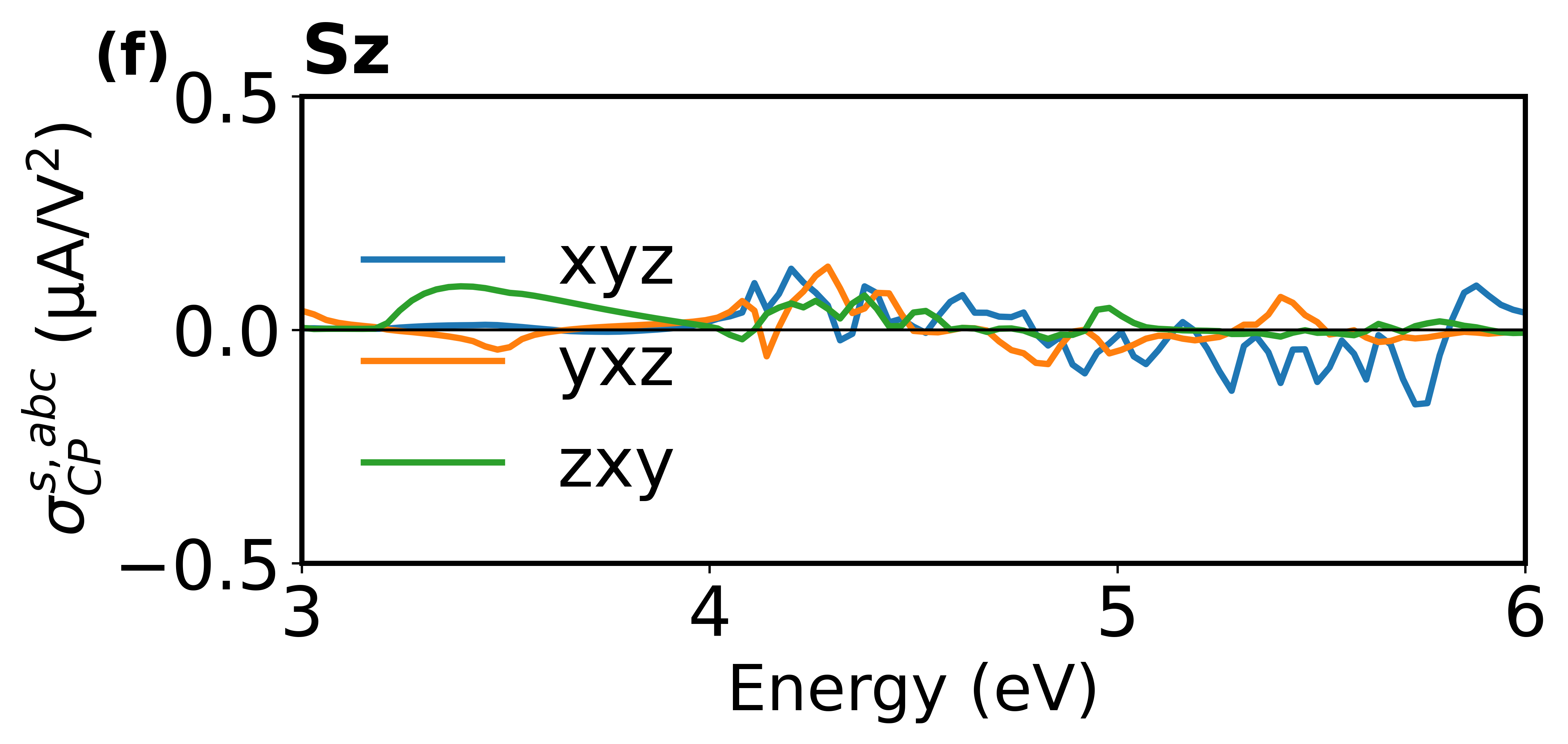}%
}
\vspace{1mm}
\parbox{\textwidth}{%
  \centering
  \includegraphics[width=0.32\textwidth]{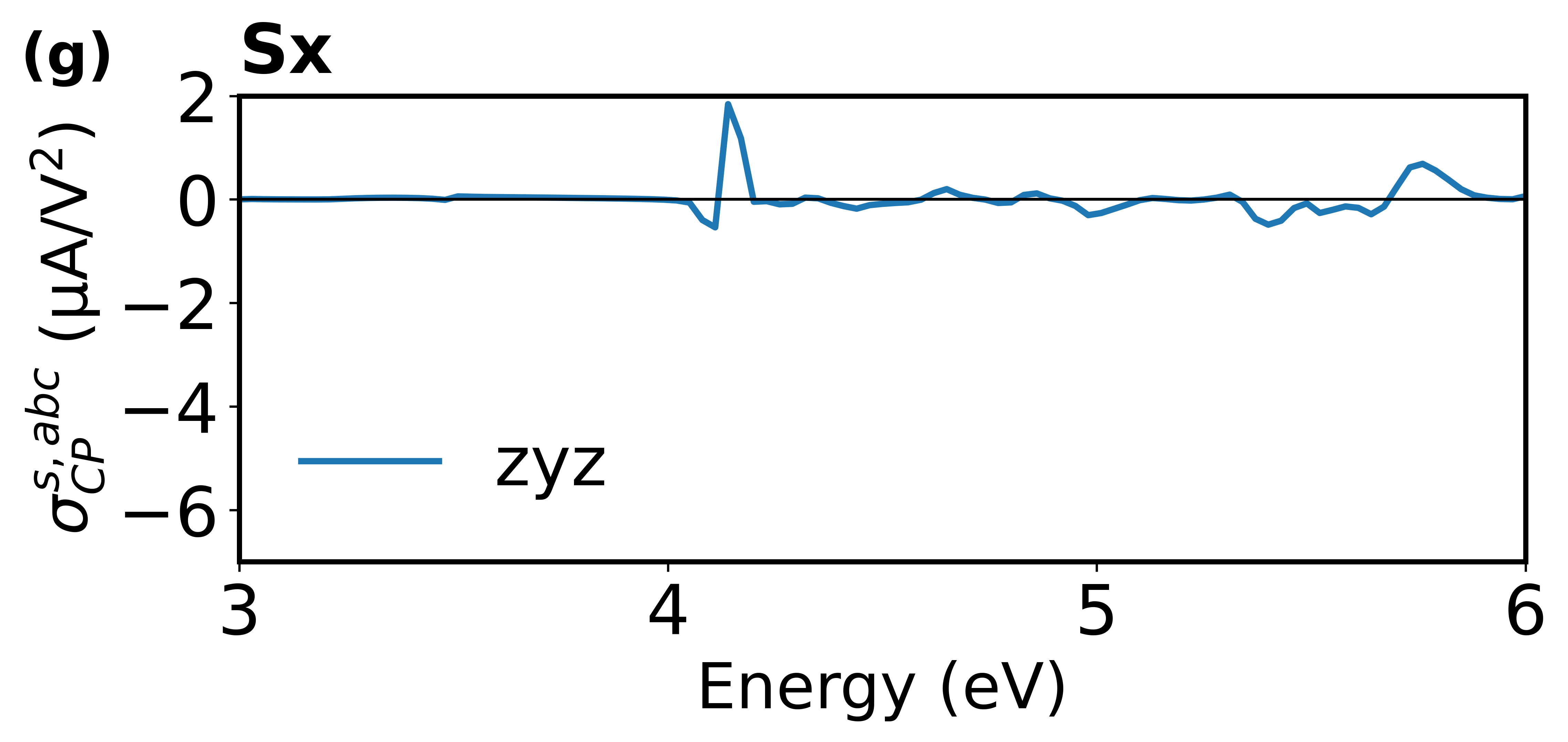}
  \includegraphics[width=0.32\textwidth]{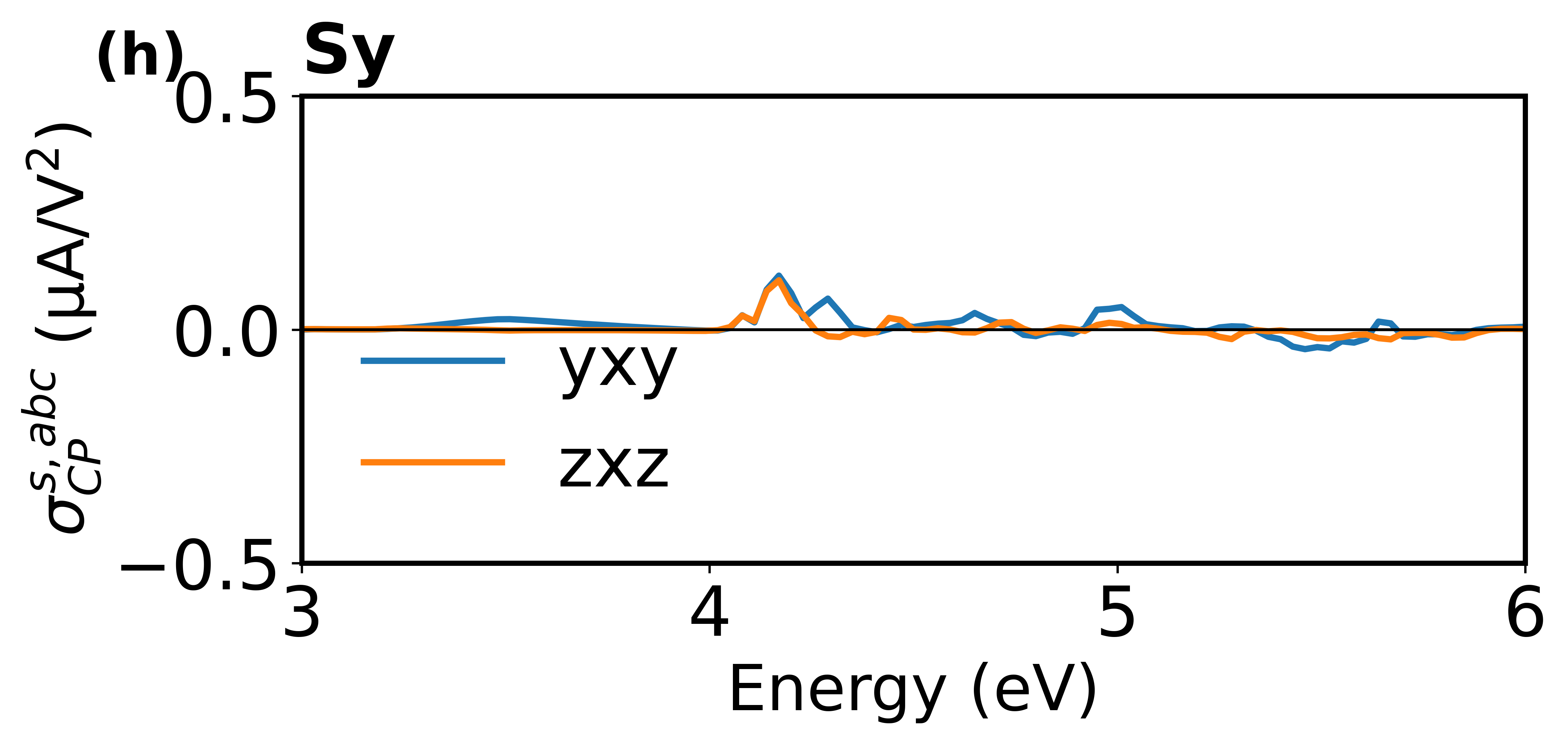}\hfill
  \includegraphics[width=0.32\textwidth]{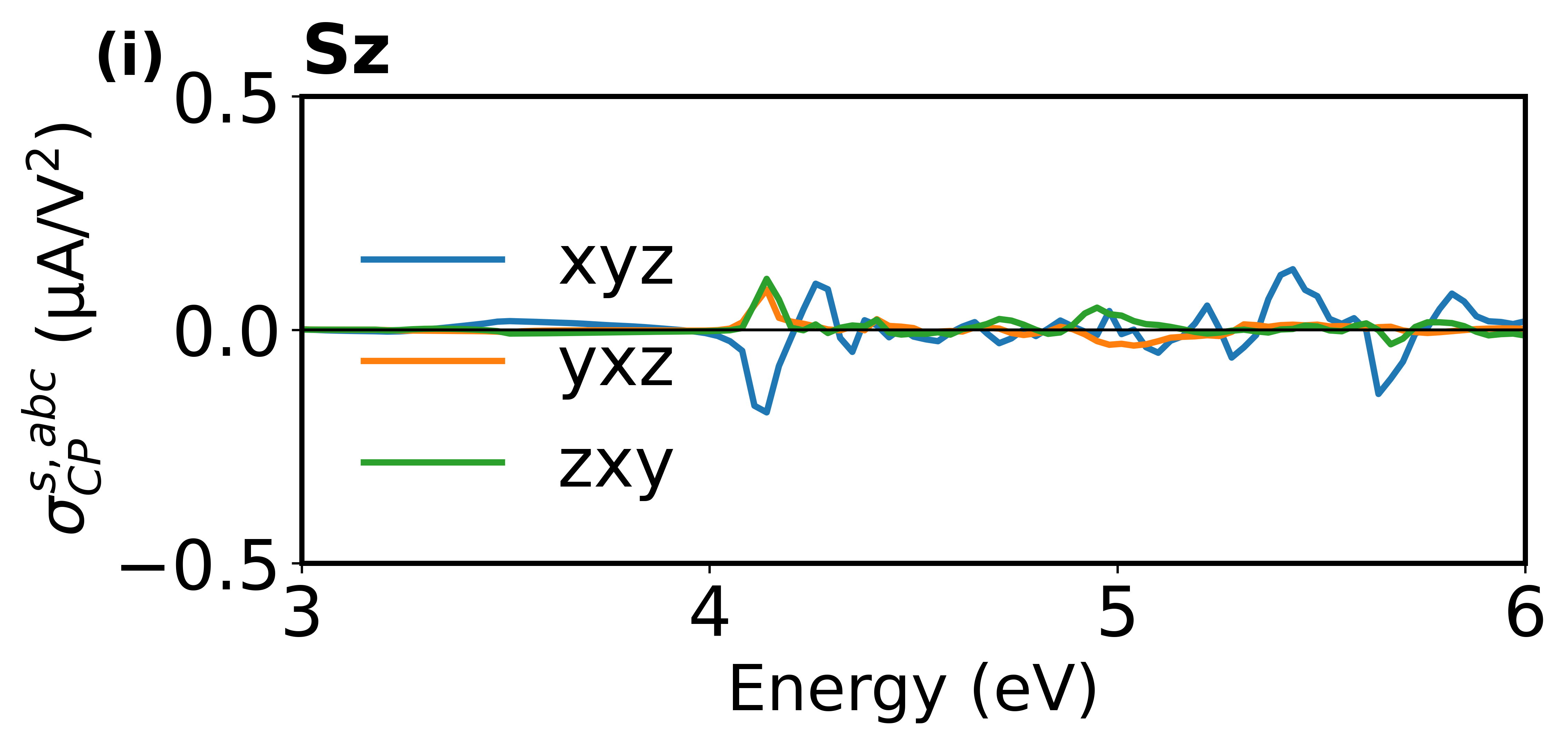}%
}
\caption{Main components of the CP spin shift photoconductivity for (a–c) $\mathbf N\parallel x$, (d–f) $\mathbf N\parallel y$, and (g–i) $\mathbf N\parallel z$, resolved into spin components $S_x$, $S_y$, and $S_z$. Components of comparable magnitude are plotted using the same $y$-axis range, while different scales are used for components of different orders of magnitude.}
\label{fig:spin_circular_shift_9panel}
\end{figure*}

In Fig. \ref{fig:spin_circular_shift_9panel} we report the main elements of the CP spin shift conductivity for the Néel vector orientations $\mathbf{N}\parallel x$ (a-c), $\mathbf{N}\parallel y$ (d-f), and $\mathbf{N}\parallel z$ (g-i) and for the spin components $S_x$, $S_y$ and $S_z$.
The CP shift conductivity is even under time reversal as the LP injection, therefore it shows the same behaviour, namely the same elements are allowed for all three Néel vector orientations, although their classification within the group--subgroup hierarchy depends on $\mathbf N$. It has no contribution in the collinear spin-group limit. It therefore separates into a coplanar sector and a purely SOC-induced sector. 
The elements related to the coplanar sector are the dominant ones, namely those for $\mathbf{N}\parallel$ $y$ and  $\mathbf{N}\parallel$ $z$ with spin component $S_x$.
$\sigma_{\mathrm{CP}}^{x,zyz}$ reachs around $6~\mu\text{A}/\text{V}^2$ and $2~\mu\text{A}/\text{V}^2$ near $\hbar\omega=4.2$~eV for $\mathbf{N}\parallel$ $y$ and  $\mathbf{N}\parallel$ $z$, respectively, as shown in panels (d) and (g) of Fig. \ref{fig:spin_circular_shift_9panel}.
The purely relativistic contributions are shown in the other panels and the largest peaks are one order of magnitude smaller.

\clearpage

\subsection{CP injection conductivity}

\begin{figure*}[h!]
\centering
\parbox{\textwidth}{%
  \centering
  \includegraphics[width=0.32\textwidth]{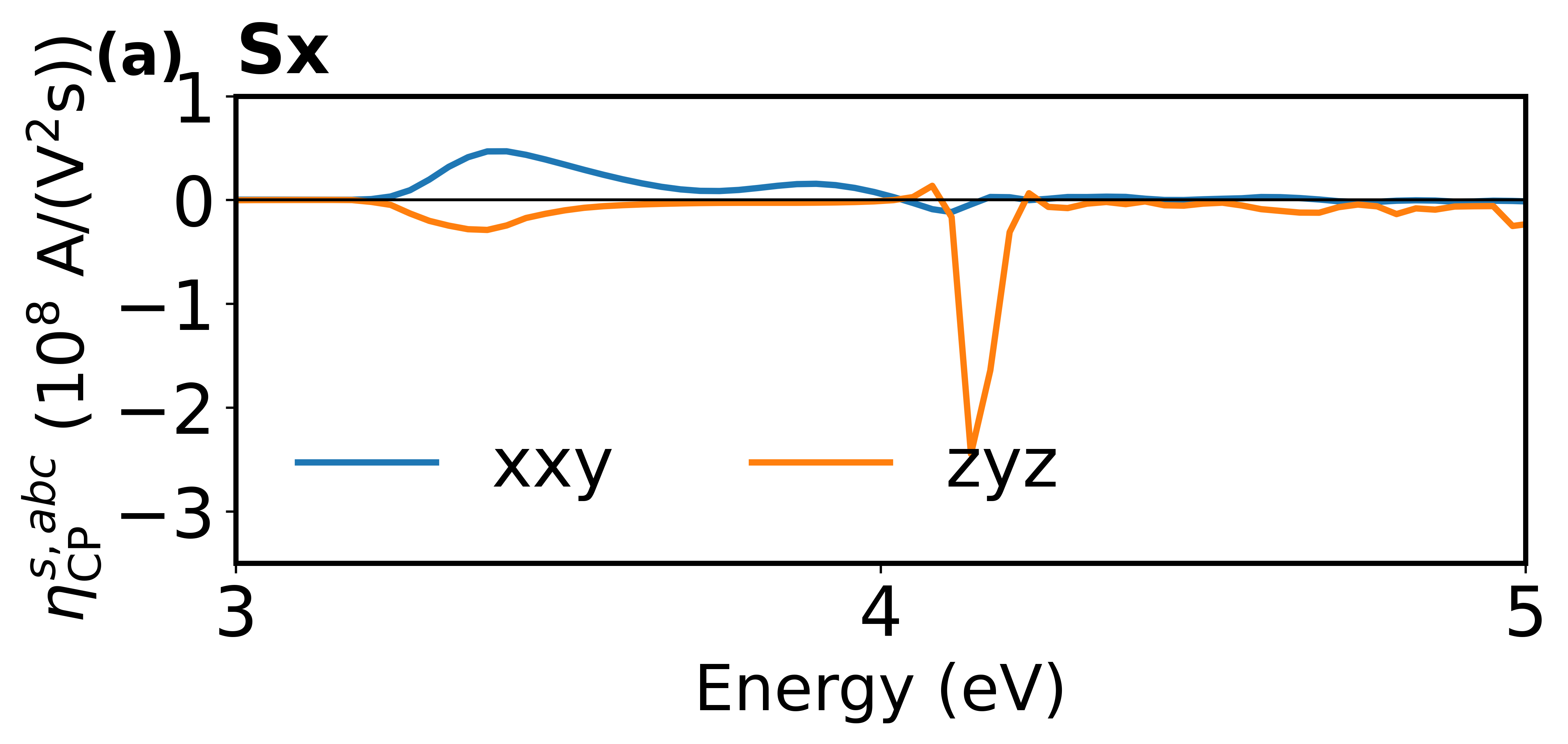}\hfill
  \includegraphics[width=0.32\textwidth]{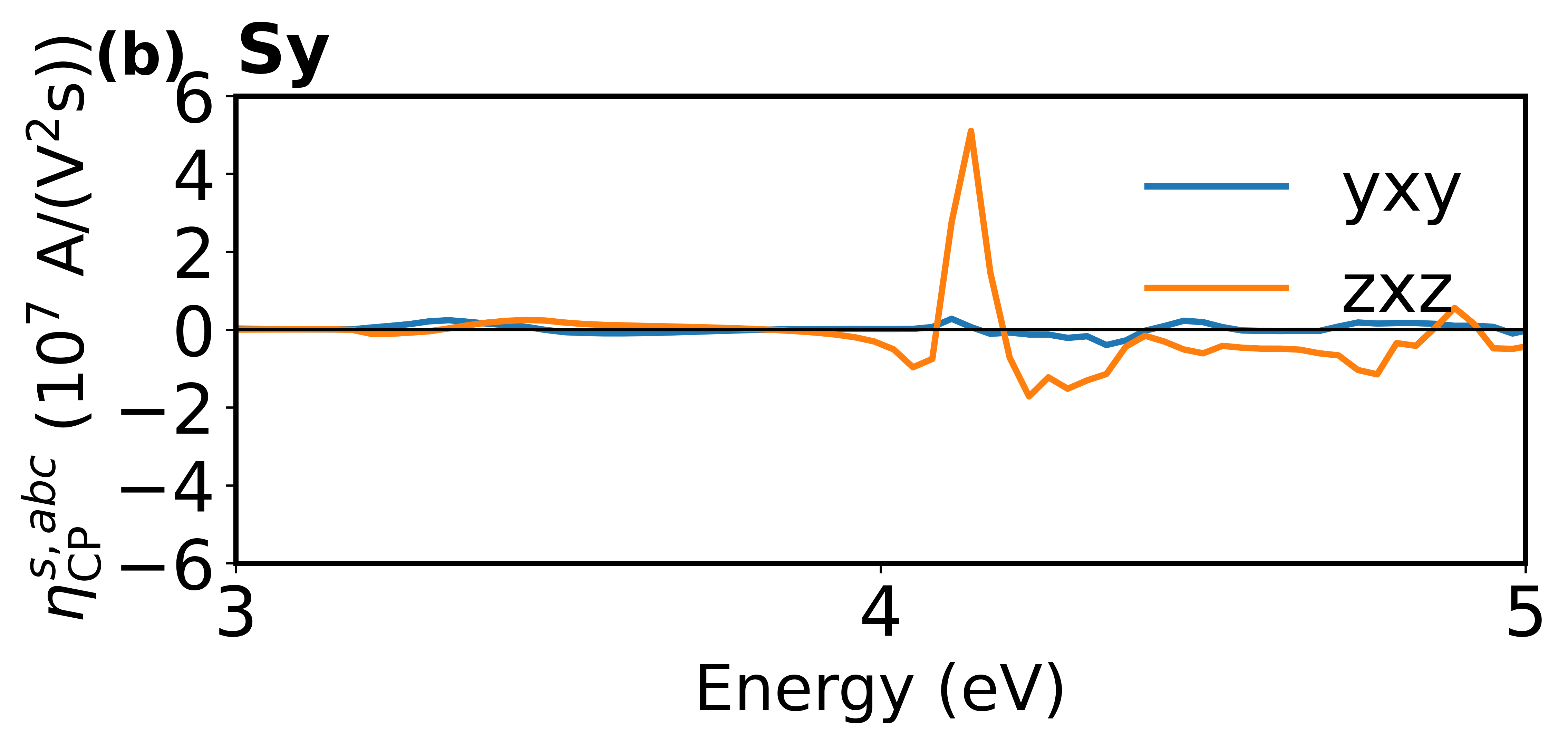}\hfill
  \includegraphics[width=0.32\textwidth]{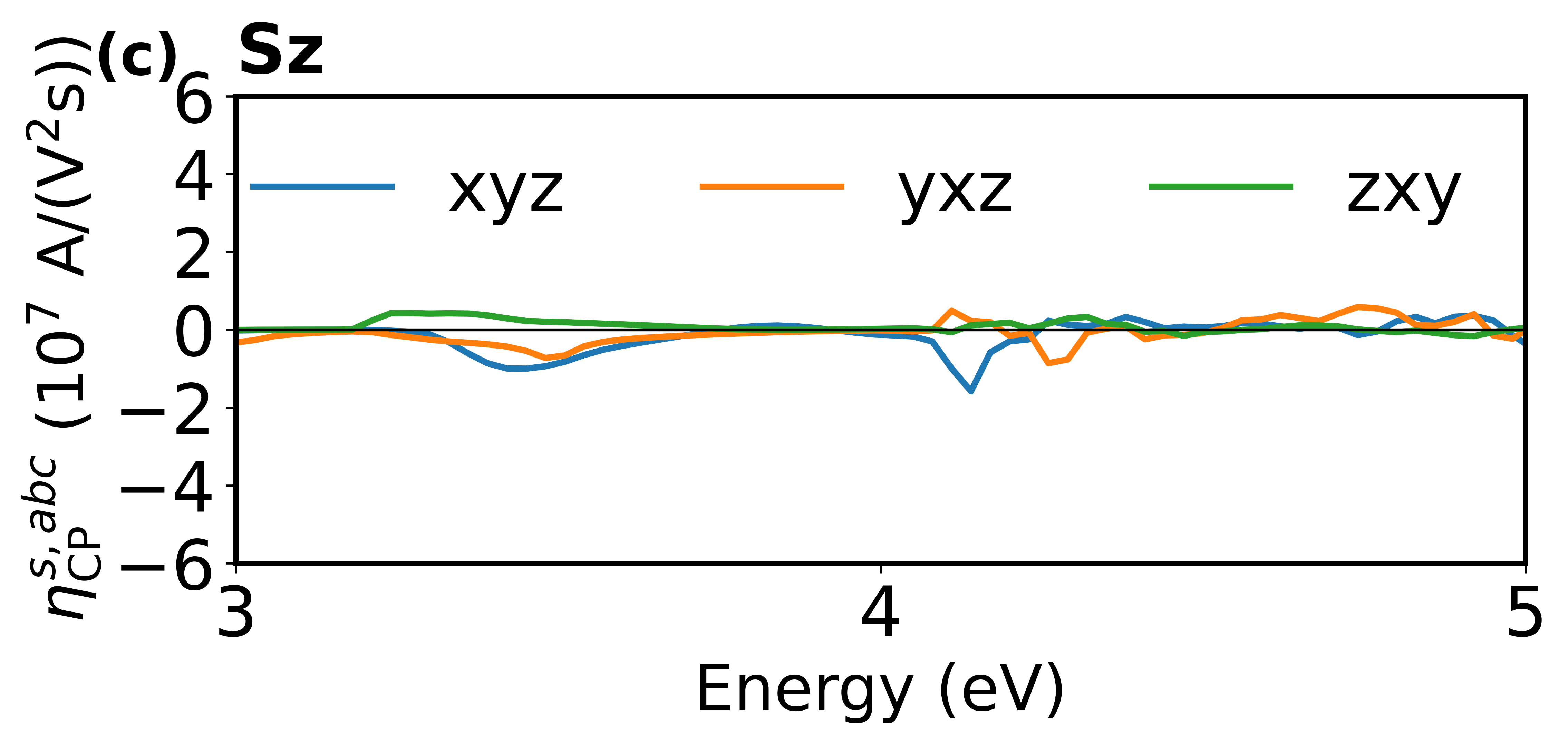}%
}
\vspace{1mm}
\parbox{\textwidth}{%
  \centering
  \includegraphics[width=0.32\textwidth]{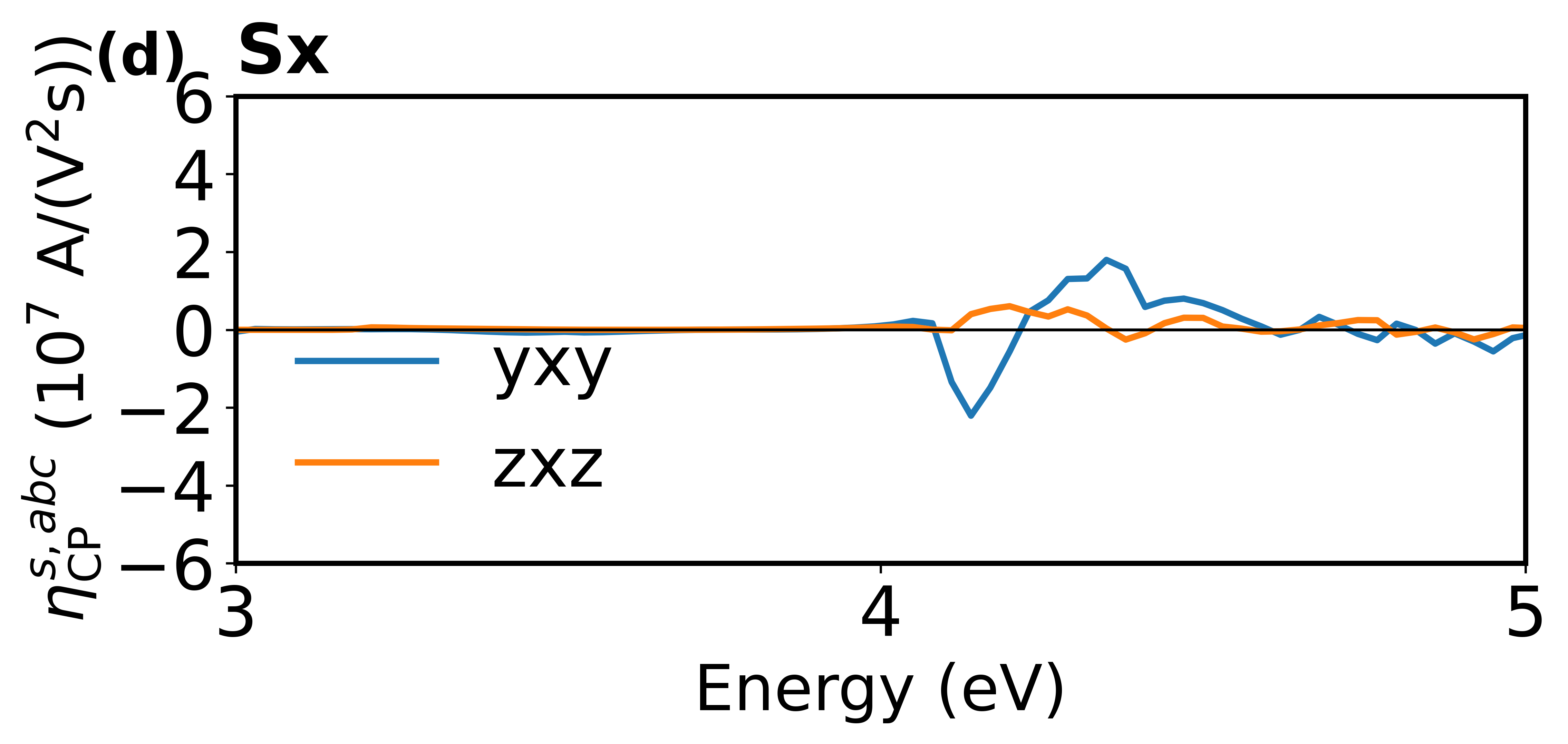}\hfill
  \includegraphics[width=0.32\textwidth]{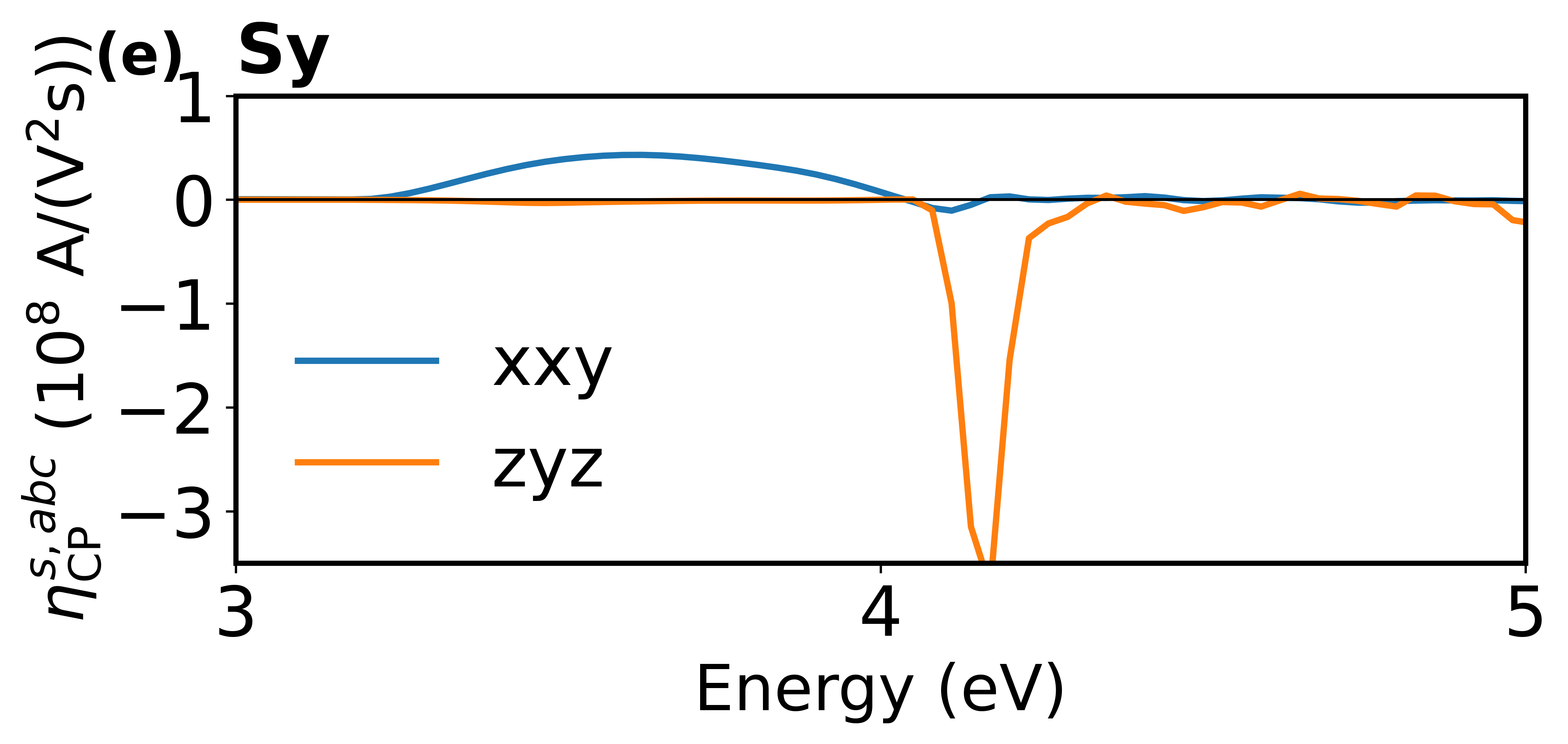}\hfill
  \includegraphics[width=0.32\textwidth]{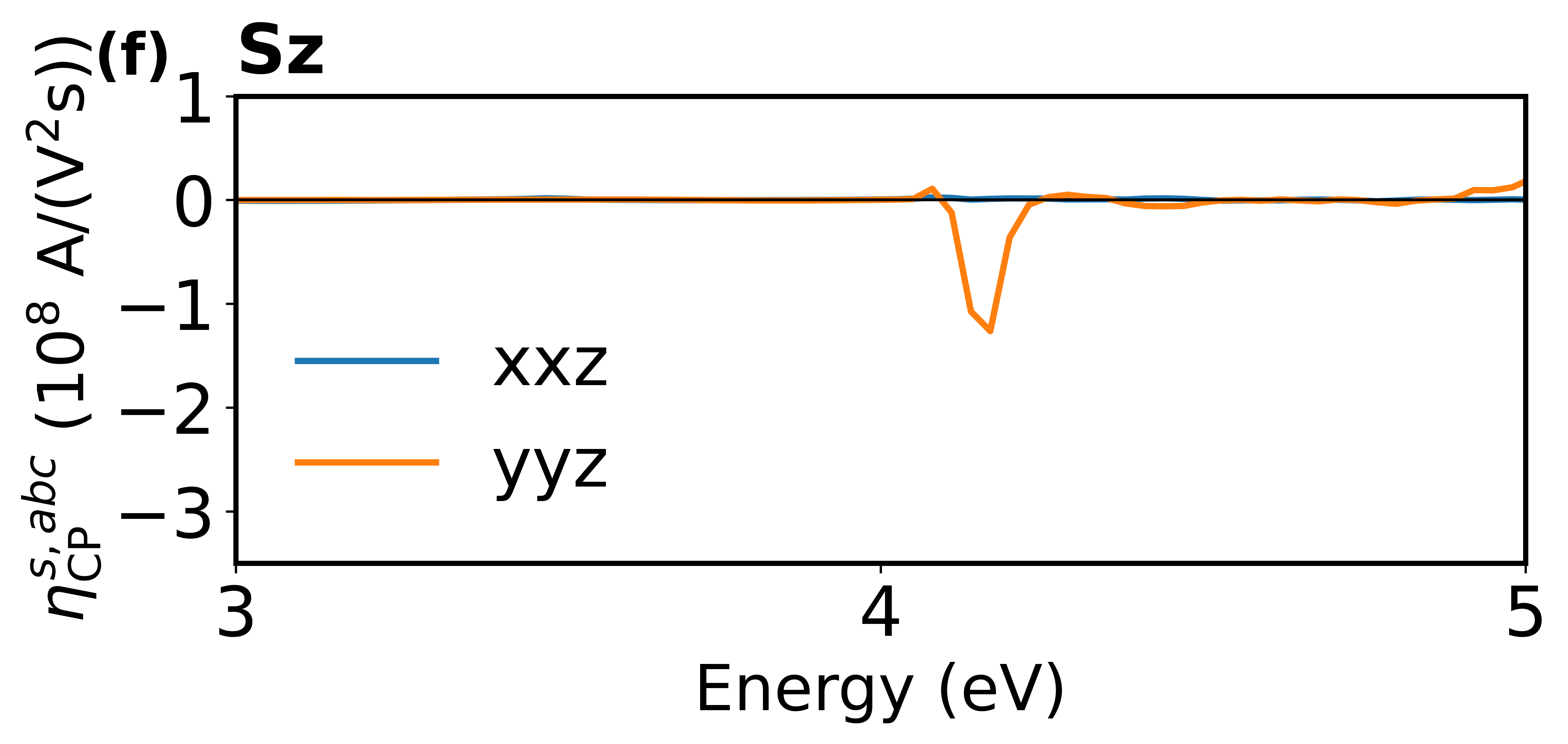}%
}
\vspace{1mm}
\parbox{\textwidth}{%
  \centering
  \includegraphics[width=0.32\textwidth]{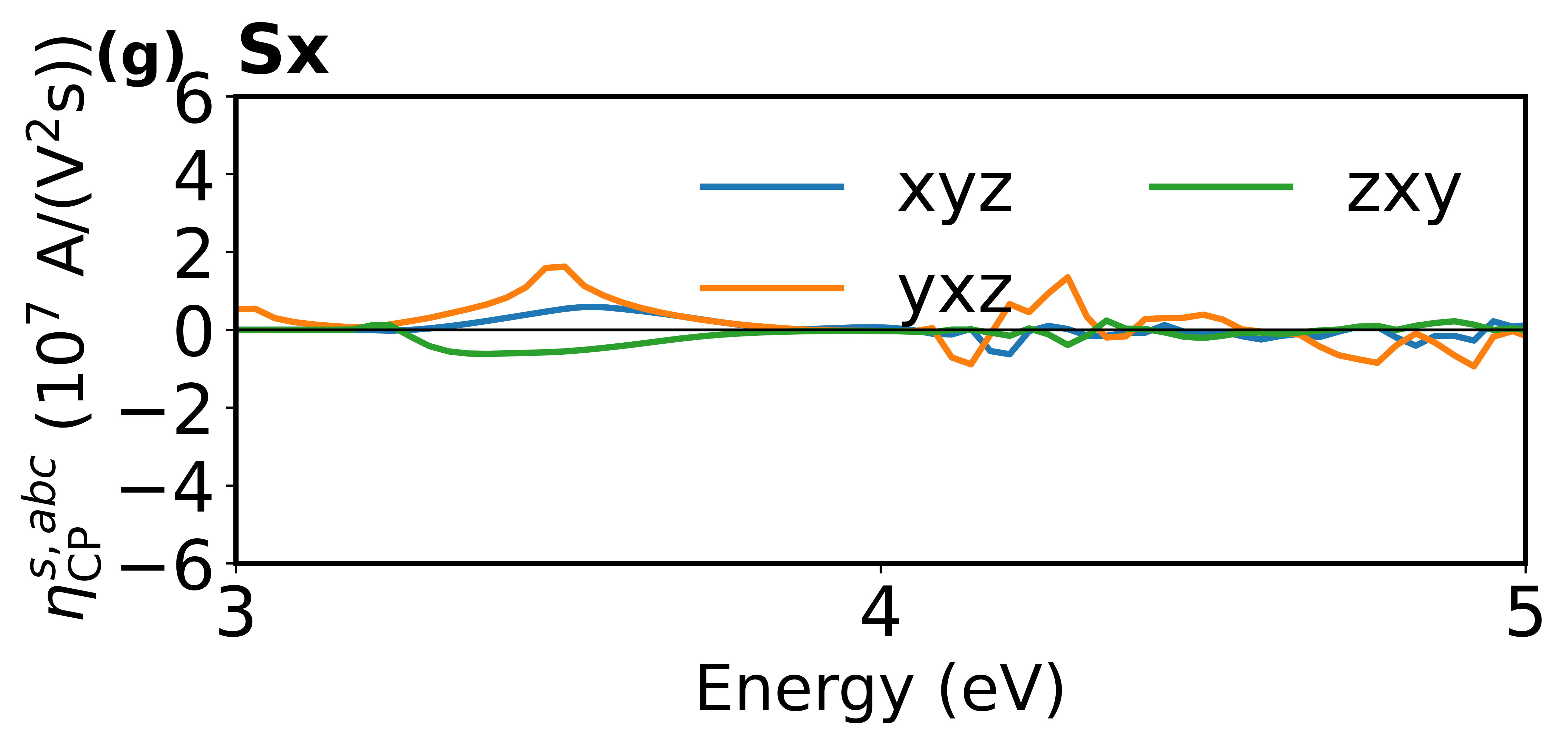}\hfill
  \includegraphics[width=0.32\textwidth]{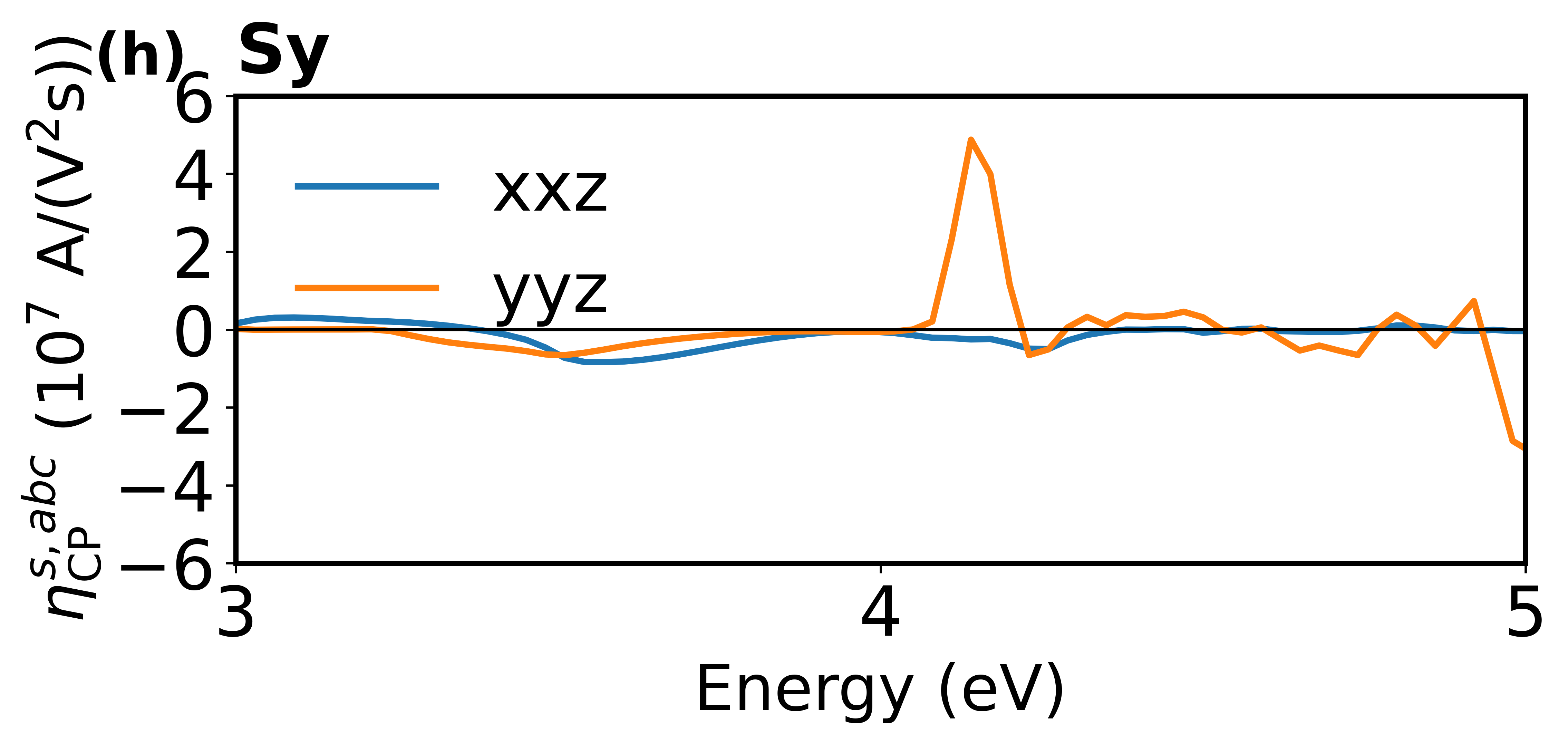}\hfill
  \includegraphics[width=0.32\textwidth]{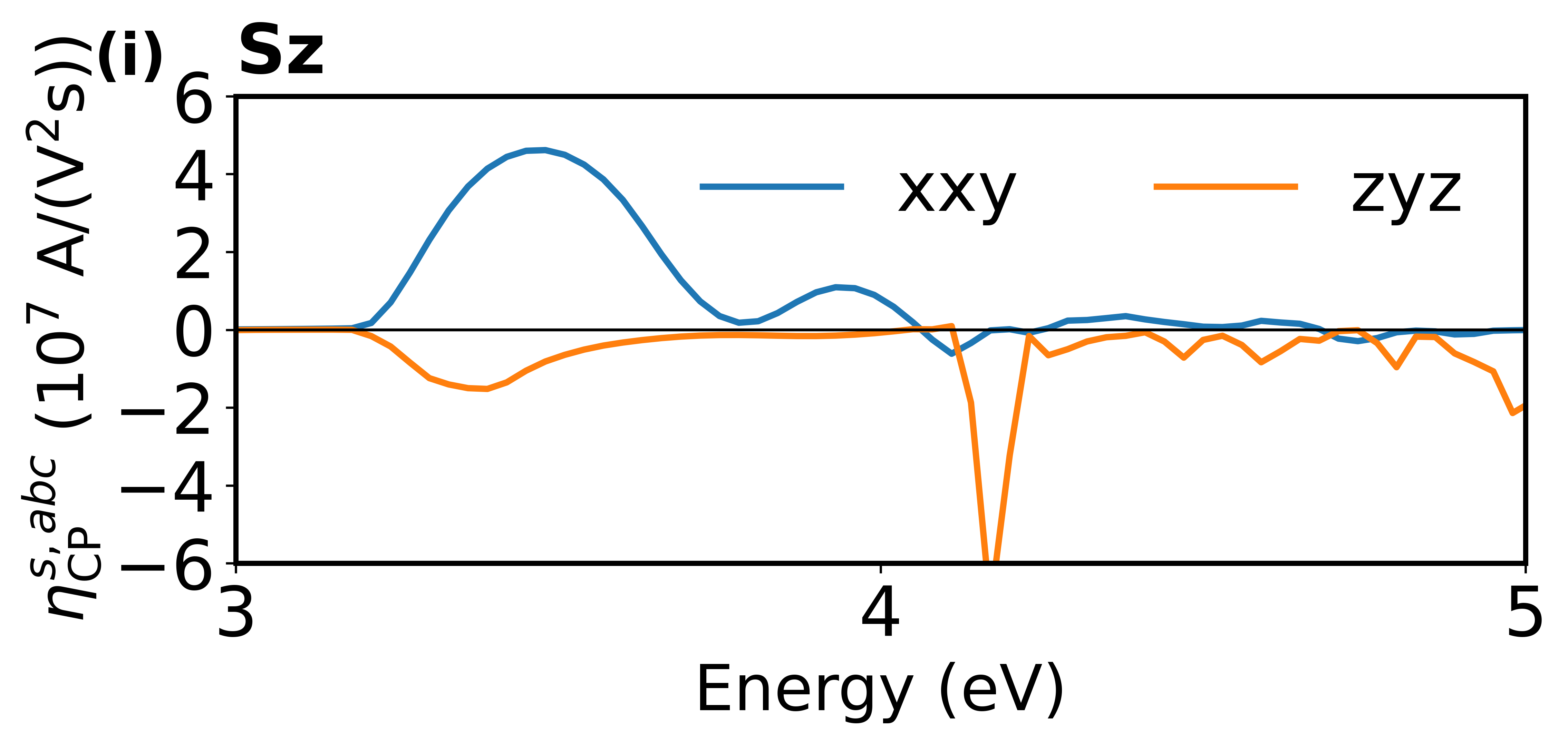}%
}
\caption{Main components of the CP spin injection photoconductivity for (a–c) $\mathbf N\parallel x$, (d–f) $\mathbf N\parallel y$, and (g–i) $\mathbf N\parallel z$, resolved into spin components $S_x$, $S_y$, and $S_z$. Components of comparable magnitude are plotted using the same $y$-axis range, while different scales are used for components of different orders of magnitude.}
\label{fig:spin_circular_injection_9panel}
\end{figure*}

In Fig. \ref{fig:spin_circular_injection_9panel} we report the main elements of the CP spin injection conductivity for the Néel vector orientations $\mathbf{N}\parallel x$ (a-c), $\mathbf{N}\parallel y$ (d-f), and $\mathbf{N}\parallel z$ (g-i) and for the spin components $S_x$, $S_y$ and $S_z$.
This response is odd under time-reversal and it behaves as the LP shift conductivity.
It shows different elements depending on the Néel vector orientations and presents nonrelativistic contributions and elements related to weak ferromagnetism and to relativistic effects.
By looking at Fig. \ref{fig:spin_circular_injection_9panel} (a), (e), (i), we can see that the elements along the Néel vector direction are inherited from the nonrelativistic case, and the main elements are $\eta_{\mathrm{CP}}^{s,xxy}$ and $\eta_{\mathrm{CP}}^{s,zyz}$, with $s=x,y,z$ for $\mathbf N\parallel x,y,z$, respectively, reaching peak absolute values of the order of $10^{8}~\mathrm{A}/(\mathrm{V}^{2}\mathrm{s})$. 
The coplanar and purely SOC-induced elements are approximately one order of magnitude smaller, with maximum absolute values of the order of $10^{7}~\mathrm{A}/(\mathrm{V}^{2}\mathrm{s})$, as we can see from the other panels.

\clearpage

\end{document}